\documentclass[english,floatfix,twocolumn,showpacs, aps, prx, reprint, superscriptaddress, longbibliography]{revtex4-2}

\usepackage{amsmath}
\usepackage{amssymb}
\usepackage{bbm}
\usepackage{graphicx}
\usepackage[shortlabels]{enumitem}
\usepackage{booktabs, footnote}
\usepackage{hyperref}
\hypersetup{colorlinks=true, citecolor=blue, linkcolor=blue, urlcolor=blue, breaklinks=true}
\usepackage[capitalize,noabbrev]{cleveref}
\usepackage{physics}

\crefname{figure}{Fig.}{Figs.}
\Crefname{figure}{Figure}{Figures}
\crefname{section}{Sec.}{Secs.}
\Crefname{section}{Section}{Sections}
\crefname{subsection}{Sec.}{Secs.}
\Crefname{subsection}{Section}{Sections}
\crefname{equation}{Eq.}{Eqs.}
\Crefname{equation}{Equation}{Equations}
\crefformat{figure}{Fig.~#2#1#3}
\Crefformat{figure}{Fig.~#2#1#3}
\crefformat{subfigure}{Fig.~#2#1#3}
\Crefformat{subfigure}{Fig.~#2#1#3}
\crefformat{section}{Sec.~#2#1#3}
\Crefformat{section}{Sec.~#2#1#3}
\crefformat{subsection}{Sec.~#2#1#3}
\Crefformat{subsection}{Sec.~#2#1#3}
\crefformat{equation}{Eq.~#2(#1)#3}
\Crefformat{equation}{Eq.~#2(#1)#3}

\makeatletter
\def\frontmatter@title@above{\addvspace{-82pt}}
\makeatother

\begin{document}


\title{Probing emergent prethermal dynamics and resonant melting on a programmable quantum simulator}

\author{Siva~Darbha}
~\thanks{These authors contributed equally to this work.}
~\affiliation{National Energy Research Scientific Computing Center, Lawrence Berkeley National Laboratory, Berkeley, CA 94720, USA}

\author{Alexey~Khudorozhkov}
~\thanks{These authors contributed equally to this work.}
~\affiliation{QuEra Computing Inc., 1284 Soldiers Field Road, Boston, MA, 02135, USA}
~\affiliation{Department of Physics, Boston University, Boston, MA 02215, USA}

\author{Pedro~L.~S.~Lopes}
~\affiliation{QuEra Computing Inc., 1284 Soldiers Field Road, Boston, MA, 02135, USA}

\author{Fangli~Liu}
~\affiliation{QuEra Computing Inc., 1284 Soldiers Field Road, Boston, MA, 02135, USA}

\author{Ermal~Rrapaj}
~\affiliation{National Energy Research Scientific Computing Center, Lawrence Berkeley National Laboratory, Berkeley, CA 94720, USA}

\author{Jan~Balewski}
~\affiliation{National Energy Research Scientific Computing Center, Lawrence Berkeley National Laboratory, Berkeley, CA 94720, USA}

\author{Majd~Hamdan}
~\affiliation{QuEra Computing Inc., 1284 Soldiers Field Road, Boston, MA, 02135, USA}

\author{Pavel~E.~Dolgirev}
~\affiliation{QuEra Computing Inc., 1284 Soldiers Field Road, Boston, MA, 02135, USA}
~\affiliation{Department of Physics, Harvard University, Cambridge, Massachusetts 02138, USA}

\author{Alexander~Schuckert}
~\affiliation{QuEra Computing Inc., 1284 Soldiers Field Road, Boston, MA, 02135, USA}

\author{Katherine~Klymko}
~\affiliation{National Energy Research Scientific Computing Center, Lawrence Berkeley National Laboratory, Berkeley, CA 94720, USA}

\author{Sheng-Tao~Wang}
~\affiliation{QuEra Computing Inc., 1284 Soldiers Field Road, Boston, MA, 02135, USA}

\author{Mikhail~D.~Lukin}
~\affiliation{QuEra Computing Inc., 1284 Soldiers Field Road, Boston, MA, 02135, USA}
~\affiliation{Department of Physics, Harvard University, Cambridge, Massachusetts 02138, USA}

\author{Daan~Camps}
~\email[Corresponding author: ]{dcamps@lbl.gov}
~\affiliation{National Energy Research Scientific Computing Center, Lawrence Berkeley National Laboratory, Berkeley, CA 94720, USA}

\author{Milan~Kornja\v{c}a}
~\email[Corresponding author: ]{mkornjaca@quera.com}
~\affiliation{QuEra Computing Inc., 1284 Soldiers Field Road, Boston, MA, 02135, USA}


\begin{abstract}

The dynamics of isolated quantum systems following a sudden quench plays a central role in many areas of material science, high-energy physics, and quantum chemistry. Featuring complex phenomena with implications for thermalization, non-equilibrium phase transitions, and Floquet phase engineering, such far-from-equilibrium quantum dynamics is challenging to study numerically, in particular, in high-dimensional systems. Here, we use a programmable neutral atom quantum simulator to systematically explore quench dynamics in spin models with up to 180 qubits. By initializing the system in a product state and performing quenches across a broad parameter space, we discover several stable, qualitatively distinct dynamical regimes. We trace their robustness to Floquet-like prethermal steady states that are stabilized over long emergent timescales by strong dynamical constraints. In addition, we observe sharp peaks in the dynamical response that are quantitatively explained by the structured melting of prethermalization through resonances. In two dimensions, we uncover a sharp dynamical response change that converges with increased system size, that is linked to the proliferation of N{\'e}el-order defects and indicative of a dynamical phase transition with no equilibrium analogs. Uncovering an intricate interplay between quantum prethermalization and emergent dynamical phases, our results demonstrate the use of quantum simulators for revealing complex non-equilibrium quantum many-body phenomena.

\end{abstract}
\maketitle



\begin{figure*}[htb]
\centering
\includegraphics[width=\textwidth]{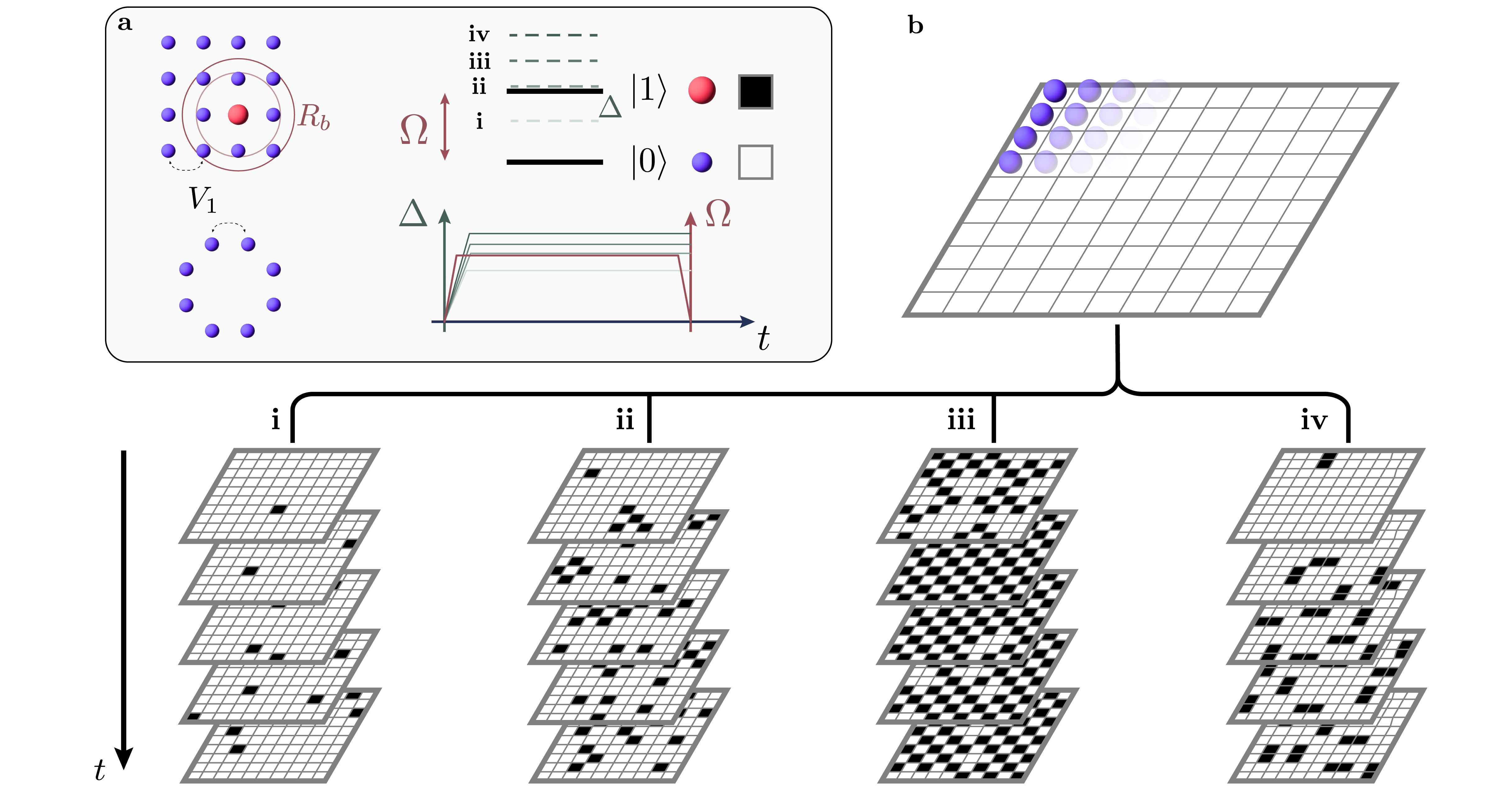}
\caption{
\textbf{The emergence of distinct dynamical regimes in neutral atom quench experiments.} 
\textbf{a.} Rubidium atoms are arranged in square grids (2D) or square chains with flattened corners (1D), and they are driven by constant quenches in the detuning field $\Delta$ and the Rabi frequency $\Omega$. Dynamics couples ground states (purple spheres, white tiles) and excited states (red spheres, black tiles) with a range of interaction strengths as measured by the blockade radius $R_\mathrm{b}$. \textbf{b.} The atoms are initialized in the all-zero state, with post-quench excitations proliferating until the system reaches a non‑equilibrium steady state. Steady-state sampling reveals a sequence of dynamical regimes, each defined by the stable long-time averages of key local observables. As the detuning quench strength increases, spatial excitation patterns evolve: from \textbf{i.} isolated excitations, to \textbf{ii.} mixed-excitation domains, to \textbf{iii.} large, defect-rich checkerboards, to abrupt regime changes, and \textbf{iv.} at still higher detunings, to the emergence of isolated multi-excitation island patterns. Comparable behavior is observed in the one-dimensional chain.
}
\label{fig:overview}
\end{figure*}

Understanding far-from-equilibrium quantum dynamics is an outstanding challenge in quantum science~\cite{Polkovnikov:2011, Mitra:2018,Rudner:2020, Vasseur:2016, Eisert:2015}.
Quantum quenches, involving rapid changes of system parameters, offer a simple yet powerful setting to uncover the fundamental mechanisms of dynamical phenomena such as thermalization, non-equilibrium phase transitions, and ergodicity breaking~\cite{Rigol:2008, Srednicki:1994, Deutsch:2018, Abanin:2019, Heyl:2018, Papic:2022}. While established theoretical paradigms~\cite{Landau:1937obd,Sachdev:2011} have fruitfully guided experiments~\cite{Ebadi:2021, Greiner:2002} for systems in equilibrium, by contrast, a general framework that classifies the rich non-equilibrium quench dynamics of isolated quantum systems remains elusive~\cite{Eisert:2015}. Analytical and numerical tools typically address near-integrable models, one-dimensional geometries, small systems, or specific critical points~\cite{Sengupta:2004, Verstraete:2004, Cazalilla:2006, Schollwock:2011, Heyl:2013, Titum:2019, Titum:2020, Lin:2022, Begusic:2025, Park:2025}, limiting their predictive power in more general settings. As a result, the range of phenomena accessible in large, non-integrable quantum systems far from equilibrium remains poorly understood. A central open question is how large, qualitative changes in post‑quench dynamics --- particularly dynamical phase transitions (DPTs)~\cite{Heyl:2013,Karrasch:2013,Zvyagin:2016,Halimeh:2017,Zunkovic:2018,Karch:2025,Hashizume:2025} --- interplay with prethermal states that can persist over emergently long timescales~\cite{Moeckel:2008,Berges:2004,Abanin:2017,Mori:2018,Haghshenas:2025}.

Here we use a programmable neutral-atom quantum simulator~\cite{aquila:2023,Bernien:2017,Bluvstein:2021,GonzalezCuadra:2024} 
to explore systematic quenches in spin models with up to 180 atomic qubits,  experimentally uncovering a rich set of dynamical regimes stabilized by prethermalization physics. Among these, we observe a broad region in parameter space where the system rapidly mixes within a kinetically constrained sector of the Hilbert space, alongside sharp features arising from selective resonant excitations. A Floquet-like prethermalization mechanism, yielding exponentially long non‑thermal dynamics~\cite{Abanin:2017,Mori:2018}, quantitatively captures the constrained dynamics. In addition, the resonant excitations provide pathways for the breakdown of prethermalization, offering unique insight into the structured melting of prethermal states. In two spatial dimensions, we observe a pronounced change in dynamical behavior, associated with a proliferation of N{\'e}el-order defects, that converges with increasing system size. 
The defect proliferation resembles an equilibrium criticality mechanism but cannot be traced to either the ground-state or thermal phase diagrams~\cite{Keesling:2019,Ebadi:2021}. Unlike previously observed dynamical regimes and transitions~\cite{Daley:2022,Smale:2019,Muniz:2020,Chu:2020,Karch:2025,Zhang:2017,De:2025,Xu:2020}, it indicates a prethermal dynamical phase transition without a near-integrable or equilibrium analog.

\section*{Quantum simulations of quench dynamics}
\label{sec:setup}

\begin{figure*}[htb]
\centering
\includegraphics[width=\textwidth]{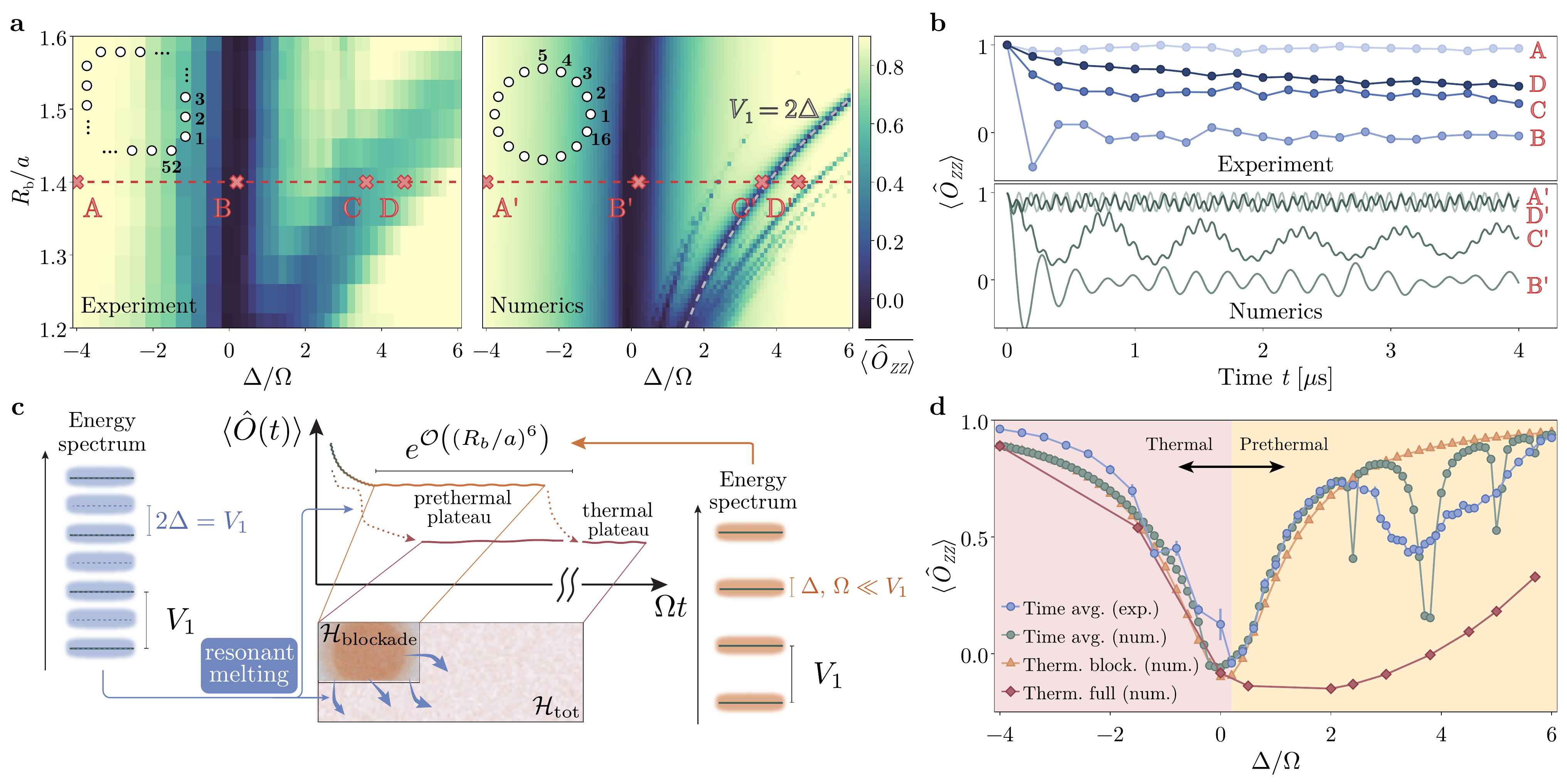}
\caption{
\textbf{Dynamical regimes of the 1D chain.} 
Distinct dynamical regimes are extracted from long-time quench dynamics in experiments (up to 4$\, \mathrm{\mu s}$ time evolution) with $52$ atoms and numerics with $16$ atoms (up to 10$\, \mathrm{\mu s}$ time evolution in \textbf{a} and up to 100$\, \mathrm{\mu s}$ in \textbf{d}). 
\textbf{a.} Dynamical phase diagrams (experiment left, numerics right) displaying the time-averaged expectation value of the two-site observable $\hat{O}_{ZZ}$, revealing a central region and side resonances. The atom schematics in the upper left corners show the geometries.
The red dashed horizontal line represents the cut $R_\mathrm{b} / a = 1.4$, while the gray dashed line satisfies $V_1 = 2\Delta$. 
\textbf{b.} The time evolution of $\langle\hat{O}_{ZZ}\rangle$ for characteristic points on the cut. 
\textbf{c.} Schematic overview of the phenomenology of the different dynamical regimes. The integer structure of the energy spectrum at small $\Delta$ and $\Omega$ results in an effective Floquet-like drive that induces prethermalization in the blockade subspace. In the side peaks, the spectral gaps become commensurate with the drive frequency, providing heating pathways for prethermalization melting via resonances on specific excitation patterns. 
\textbf{d.} The time-averaged expectation values of the two-site observable $\hat{O}_{ZZ}$ on the $R_\mathrm{b} / a = 1.4$ cut (experiment and numerics), and the thermal expectation values of the observable for the effective temperature of the quench using the nearest-neighbor blockade subspace (thermal blockade) and the full Hilbert space (thermal full).
}
\label{fig:phases_1d}
\end{figure*}

We perform quantum dynamics experiments using a programmable, analog neutral atom quantum simulator~\cite{aquila:2023}.  
A single atom $j$ encodes a qubit, where the ground state is $\ket{0_j}$ and a Rydberg state is $\ket{1_j}$. The Hamiltonian governing the dynamics of the $N$-qubit system is:
\begin{equation}
\hat{H} = \frac{\Omega (t)}{2} \sum_j \hat{X}_j - \Delta (t) \sum_j \hat{n}_j + \sum_{j<k} V_{jk} \hat{n}_j \hat{n}_k \, ,
\label{eq:hamiltonian}
\end{equation}
where $\hat{X}_j = \outerproduct{0_j}{1_j} + \outerproduct{1_j}{0_j}$ is the Pauli $X$ operator, $\hat{n}_j = \outerproduct{1_j}{1_j}$ is the Rydberg density operator, $\Omega (t)$ is the Rabi frequency, $\Delta (t)$ is the detuning, and $V_{jk} = C_6 / r_{jk}^6$ is the Rydberg-Rydberg interaction in terms of the atom separation $r_{jk}$ and an interaction constant $C_6$. The atoms experience the Rydberg blockade, which suppresses simultaneous excitations for separations $r_{jk} \lesssim R_\mathrm{b}= (C_6 / \Omega)^{1/6}$, where $R_\mathrm{b}$ is the blockade radius~\cite{Jaksch:2000,Lukin:2001}. We initialize the system in the $\ket{0}^{\otimes N}$ state, evolve it under the time-constant Hamiltonian $\hat{H}$, and measure it in the computational basis. Such an experiment represents a hardware-native quench, with the all-zero state directly prepared during state initialization and the constant $\Delta$ and $\Omega$ pulses serving as quench parameters.

\Cref{fig:overview} sketches our data collection process and observed phenomenology; experiments are performed in 1D closed chains of up to $52$ atoms and 2D square lattices with up to $180$ atoms. Guided by the known equilibrium phase diagrams~\cite{Ebadi:2021}, we quench to a constant detuning $\Delta / \Omega \in [-4.0, 6.0]$ and sweep the overall energy scales within the nearest-neighbor blockade regime $R_\mathrm{b} / a \in [1.2, 1.6]$ by varying the atom separation $a$ (see Methods).

To probe the quench dynamics, we examine the expectation values of diagonal local observables $\langle\hat{O}\rangle$ that are products of $\hat{Z} = \outerproduct{0}{0} - \outerproduct{1}{1}$ or $\hat{n}$ as well as their long-time averages
\begin{equation} 
\overline{\langle \hat{O} \rangle} \equiv \frac{1}{T} \int_{t_0}^{t_0 + T} \langle \hat{O}(t) \rangle \, \mathrm{d}t. 
\label{eq:observables}
\end{equation}
These diagonal local observables are efficient to sample in the experiment and characterize the structure of Rydberg excitations. In 1D, the most important example is
\begin{align}
\hat{O}_{ZZ} = \frac{1}{N} \sum_j \hat{Z}_j \hat{Z}_{j+1}, \label{eq:Onn}
\end{align}
which reflects the short-range correlations in the system. Excitation patterns are further resolved via \emph{$k$-island observables}, which we define as 
\begin{equation}
\hat{O}^{(k)}_L = \frac{1}{N} \sum_j (1-\hat{n}_{j}) \hat{n}_{j+1} \hdots \hat{n}_{j+k} (1-\hat{n}_{j+k+1}). \label{eq:OLk}
\end{equation}
These observables count the number of isolated clusters of exactly $k$ consecutive excitations.

\section*{Prethermal dynamics in 1D systems}
\label{sec:prethermal}

We probe the long-time behavior of the two-site $\hat{O}_{ZZ}$ observable in a 1D chain of atoms, and observe rich features in the long-time averages (\Cref{fig:phases_1d}a), indicating several distinct dynamical regimes characterized by qualitative changes in the correlations. We define a ``dynamical phase diagram'' by using the long-time value of the time-averaged observable as an order parameter. Naively, when $\Delta$ is quenched to a large negative value, one expects a small number of Rydberg excitations; as $\Delta$ is increased, naturally, the number of such excitations should grow. We observe that this intuition breaks down around $\Delta/\Omega \sim 0$, as the long-time expectation value $\langle \hat{O}_{ZZ} \rangle$ displays a non-monotonic behavior, resulting in a distinct central region in the phase diagram that remains fixed with increasing atom-atom interaction strength. Furthermore, for larger detuning, we observe a broad side peak that shifts to higher $\Delta / \Omega$ with increasing $R_\mathrm{b} / a$, as seen in \Cref{fig:phases_1d}a.

To elucidate the experimental results, we perform exact diagonalization simulations on a 16-atom ring. The numerics qualitatively recover the experimental results, albeit with higher resolution, due to longer evolution times, no sampling limitations, no variability in atom positions and measurements, and no decoherence (\Cref{fig:phases_1d}a). On the representative cut $R_\mathrm{b} / a = 1.4$, we observe stable oscillations at frequencies characteristic of the underlying physics (\Cref{fig:phases_1d}b), such as fast oscillations at point A$'$ with a frequency approximate to $\Delta$. In the regimes where local correlations differ significantly from their initial values, the oscillations slow down and become irregular. Examining the time averages of the signals across the dynamical phase diagram in \Cref{fig:phases_1d}a offers a clearer picture of the line shapes: the broad experimental side peak resolves into a fan of peaks in simulation, with one dominant and several weaker neighbors. We attribute the broadening and coalescence of the peaks to atom position disorder in the experiment~\cite{aquila:2023}, as reproduced in numerical simulations with position fluctuations (see Supplementary Information).

\begin{figure*}[htb]
\centering
\includegraphics[width=\textwidth]{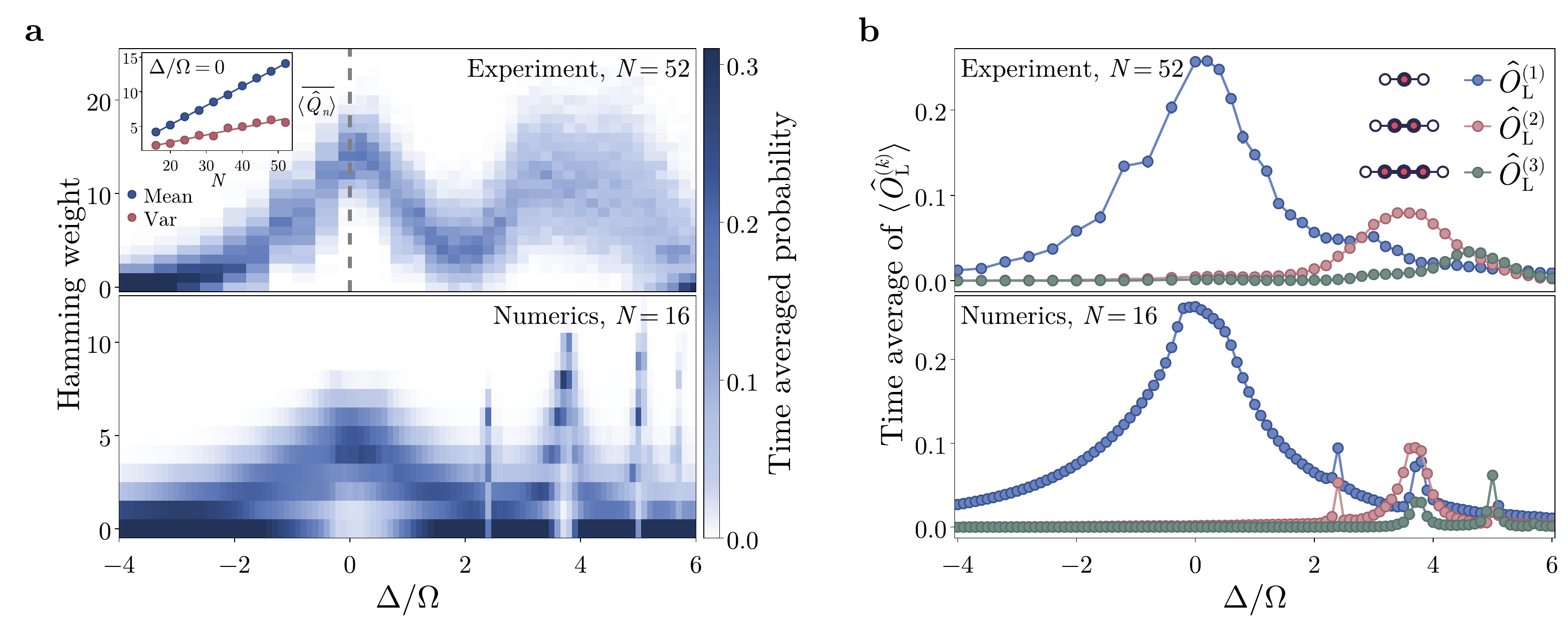}
\caption{
\textbf{Experimental characterization of the central region and resonances for the 1D chain.} 
\textbf{a.} The Hamming weight ($\hat{Q}_n$) distribution of experimentally and numerically sampled post-quench states along the representative cut $R_\mathrm{b} / a = 1.4$ at a fixed system size $N$. At $\Delta/\Omega = 0$ (inset), the excitation distribution exhibits a mean of $N/4$ and a variance of $N/8$ across experimentally probed system sizes, consistent with equal excitation probability across the blockade subspace.
\textbf{b.} The time-averaged expectation values of the $k$-island observables for $k = 1,2,3$ (illustrated in the legend) resolve the fan of side peaks in the experiment. The island operator peaks coincide with resonances between the energy of island excitations and the initial state. The sub-peaks indicate sub-dominant contributions of the island mixtures to a given resonance.
}
\label{fig:characterization_phases_1d}
\end{figure*}

We now examine the emergence and nature of the dynamics across different regimes, beginning with the broad central region near $\Delta / \Omega \approx 0$. Unlike integrable or near-integrable systems, where dynamical signatures align closely with ground-state phases~\cite{Heyl:2013,Titum:2019,Haldar:2021}, this region does not map onto any equilibrium phase~\cite{Keesling:2019}. For large negative detunings, the all-zero initial state approximates an isolated eigenstate, leading to a low density of Rydberg excitations and thermal behavior as one enters from the left ($\Delta / \Omega \lesssim 0$). Near the central peak, the system transitions to non-thermal behavior ($\Delta / \Omega \approx 0$), as manifested by comparing the response of the $\hat{O}_{ZZ}$ local observable, both measured and simulated, to its thermal expectation value with respect to the system Hamiltonian (\Cref{fig:phases_1d}d). Thermal quantities are estimated via a canonical ensemble with the same mean energy as the initial state. The persistence of non-thermal dynamics over a wide parameter range and the distinct character of this central region call for a theoretical framework capturing the underlying mechanism. Given the blockaded structure of the Hilbert space, illustrated in~\Cref{fig:phases_1d}c, it is natural to interpret the transient state observed at experimentally relevant times as a prethermal regime~\cite{Berges:2004, Moeckel:2008, Abanin:2017, Mori:2018}.

In this picture (\Cref{fig:phases_1d}c), the dynamics of the central region corresponds to a non-equilibrium steady state constrained to the blockade subspace but exploring it broadly. At timescales longer than those accessible experimentally, the system is expected to thermalize in the full Hilbert space. This interpretation follows naturally from the discrete structure of the Rydberg Hamiltonian $\hat{H}$ in \Cref{eq:hamiltonian}, which enables a direct application of Floquet prethermalization theory~\cite{Abanin:2017, Mori:2018, Haghshenas:2025}. We decompose $\hat{H}$ according to energy scales: $\hat{H} = \hat{H}_0 + \lambda \hat{V}'$, where $\hat{H}_0 = V_1 \sum_{\langle jk \rangle } \hat{n}_j \hat{n}_k$ captures the dominant nearest-neighbor Rydberg interaction, while $V'$ combines the remaining terms of typical strength $\lambda = \max{\{\Delta, \Omega, V_2 \}}$, where $V_2$ is the second-neighbor sub-dominant interaction and $\lambda \ll V_1$ in a wide portion of the explored parameter space. The spectrum of $\hat{H}_0$ contains only integer multiples of $V_1$ that correspond to the number of blockade violations in a configuration. Consequently, $\hat{V}'$ in the interaction picture is effectively driven periodically with the period $T = 2\pi \lambda/V_1$. Under these conditions, the system relaxes to a prethermal state governed by an approximately conserved truncated Floquet Hamiltonian, with an exponentially long thermalization time $\mathcal{O} \left(e^{(V_1/\lambda)} \right)$~\cite{Abanin:2017}. This contrasts with the polynomial relaxation times typical of quenched systems with dynamical constraints (via the Fermi golden rule) and arises from the structure of the Rydberg Hamiltonian and the time-independent nature of the quench~\cite{Mori:2018}. As a consequence, the steady states we observe are expected to survive for uncharacteristically long times, supporting the robustness of the experimental dynamical regimes.

We directly test the Floquet prethermalization picture in numerics by noting that the leading order truncated Floquet Hamiltonian is equivalent to the projection of the Rydberg Hamiltonian to the blockade subspace~\cite{Bernien:2017, Bluvstein:2021}. We expect that the energy under the Floquet Hamiltonian is conserved in the prethermal regime, and we thus proceed to calculate ``prethermal'' expectation values for our quench (see Supplementary Information) by constraining the canonical ensemble within the blockade subspace, finding excellent agreement between the long-time expectation values and the prethermal ones (\Cref{fig:phases_1d}d).

\begin{figure*}[htb]
\centering
\includegraphics[width=\textwidth]{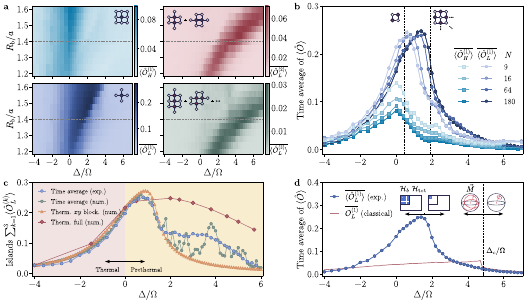}
\caption{
\textbf{Dynamical regimes of the 2D lattice.} 
\textbf{a.} Decomposed experimental dynamical phase diagrams for $180$ atoms displaying the time-averaged expectation values of several $k$-island observables. $\hat{O}_H^{(k)}$ and $\hat{O}_L^{(k)}$ measure islands with and without diagonal blockade restrictions, respectively, according to shapes and sizes shown in the schematics. The grey dashed horizontal lines show the representative cut $R_\mathrm{b} / a = 1.4$, the focus of panels \textbf{b}-\textbf{d}. 
\textbf{b.} The time-averaged expectation values of the $1$-island observables on the cut  for different lattice sizes $N$. The black dashed vertical lines show the locations $\Delta / \Omega = c (R_\mathrm{b} / a)^6$ for $c = 1/16 , 1/4$, which correspond to the smallest and largest resonant N{\'e}el order patches in the classical limit, shown by the atom schematics. 
\textbf{c.} The time-averaged expectation values of the island operator sum on the cut $R_\mathrm{b} / a = 1.4$ for $16$ atoms, and the thermal expectation values of the observable for the effective temperature of the quench using the nearest-neighbor blockade subspace (thermal $xy$ blockade) and the full Hilbert space (thermal full). 
\textbf{d.} The time-averaged response of the classical spin dynamics model, obtained from the $S \rightarrow \infty$ limit of the quantum spins, and the quantum counterpart for $180$ atoms, on the cut $R_\mathrm{b} / a = 1.4$. The classical response exhibits a dynamical phase transition (black dashed vertical line) given by the approximate solution to the Stoner-Wohlfarth astroid (Supplementary Information). A schematic shows representative dynamics on either side of the classical transition, and similarly for the quantum analog.
}
\label{fig:phases_experiments_2d}
\end{figure*}

To further probe blockade-subspace prethermalization experimentally, we examine the diagonal Hamming weight probability distribution, as presented in \Cref{fig:characterization_phases_1d}a. This distribution counts the number of Rydberg excitations and, for the $52$-atom system at $\Delta/\Omega \approx 0$, is peaked near $13$--$14$ atoms, corresponding to the most populated Hamming weight subspaces consistent with blockade constraints. Expanding the system from $16$ to $52$ atoms (see also Supplementary Information) reveals that the mean Hamming weight at $\Delta = 0$ scales as $N/4$, with variance $N/8$ (\Cref{fig:characterization_phases_1d}a, inset). The Hamming weight statistics is thus consistent with a state that explores the whole blockade subspace homogeneously. We note that the effective blockade subspace Hamiltonian at $\Delta=0$ (the PXP Hamiltonian) has the form of a 1D lattice gauge theory (LGT) related to the Schwinger model~\cite{Surace:2020}, and hosts a manifold of many-body scarred eigenstates~\cite{Bernien:2017, Turner:2018}. The all-zero state in the LGT framework corresponds to a charge-full state, and the blockade subspace dynamics can be related to damped collective plasma oscillations between effective charges and electric fields~\cite{Mark:2025}.

\section*{Resonant melting}
\label{sec:melting}

We next consider the region $\Delta / \Omega \gtrsim 2$ featuring multiple sharp side peaks. The peaks signal the ``melting'' of prethermalization, which manifests as deviations from the prethermal blockade subspace averages. While the prethermal steady states are ultimately expected to melt for a thermodynamically large system, the melting at the observed peaks is drastically hastened when the all-zero state becomes resonant to a specific manifold of blockade-violating states (namely $k$-island states). This yields non-oscillatory terms in the time dynamics that survive in long-time averages. In the Floquet-like prethermalization picture, the resonances correspond to drive frequencies commensurate with the system excitation energies, which leads to heating and prethermalization breakdown.

The lowest-order resonance arises between the all-zero state and a state predominantly containing isolated ``$0110$'' ($k=2$) excitation islands. In the classical (product state) limit, the resonance condition is reached when $2\Delta = V_1 = \Omega (R_\mathrm{b}/a)^6$ (see also \Cref{fig:phases_1d}a), which indeed yields the dominant observed resonance at $\Delta/\Omega \approx 3.7$ on the $R_\mathrm{b}/a = 1.4$ cut. The $(R_\mathrm{b}/a)^6$ dependence directly accounts for the characteristic resonance shape being observed. Pairwise excited $1$- and $2$-islands (``$0110 \hdots 010$''), become resonant at $3 \Delta \approx V_1$, yielding $\Delta / \Omega \approx 2.5$ on the cut; excited $3$-islands (``$01110$'') arise when $3 \Delta \approx  2 V_1$, yielding $\Delta / \Omega \approx 4.9$; and so on. We directly probe the excitation islands by measuring corresponding island observables, as defined in \Cref{eq:OLk}. The results are presented in \Cref{fig:characterization_phases_1d}b and confirm the physical picture of the resonances given above, with full dynamical phase diagrams in terms of the island observables presented in the Supplementary Information. 

The resonance-selective nature of the island observables allows us to resolve at least two distinct side peaks (``$0110$'' and ``$01110$'') in the experimental data, albeit broadened by atom position fluctuations. The central peak is characterized by a large number of $k=1$ islands, consistent with the dynamics that explores large regions of the blockade subspace. A time-dependent perturbation theory analysis (see Supplementary Information) qualitatively recovers other experimentally observed trends of the resonances, including the relative peak magnitudes. The resonances, while primarily governed by lowest-order excitation processes, exhibit richer structures. For example, the main $2\Delta = V_1$ resonance also contains pairwise excited mixtures of $k=3$ and $k=1$ islands (``$01110...010$''), detectable as subpeaks of the corresponding operators in both numerics and experiment (\Cref{fig:characterization_phases_1d}b). The effective Hamiltonian near the resonance (see Supplementary Information) supports conversion between island types, notably mobile $k=1$ islands and immobile $k=2,3$ islands, hinting at exotic fracton-like behavior.

\section*{Dynamical phase transition in 2D systems}
\label{sec:2D_results}

We extend our experiment to the 2D square lattice with up to $180$ atoms. The dynamical phase diagram is presented in \Cref{fig:phases_experiments_2d}a through several relevant island observables, uncovering intricate dynamical regimes. Owing to higher connectivity, the 2D set of $k$-island observables expands beyond the 1D case: we distinguish excitation patches isolated along the $x$ and $y$ directions, and ``$xyd$'' islands additionally isolated along diagonals. Similar to the 1D chain, the dynamical phase diagrams show a broad central region characterized by extensive $1$-island excitations in which the system explores the blockade subspace. Several experimentally resolvable side peaks stem from resonances between the all-zero state and various $k$-islands. Novel to the 2D lattice, the two $1$-island observables ($\hat{O}_H^{(1)}$ and $\hat{O}_L^{(1)}$) reveal an apparent two-peak structure in the central region. The peaks diverge and broaden rapidly with increasing $R_\mathrm{b} / a$: the $xy$ $1$-island peak shifts to larger $\Delta / \Omega$, while the $xyd$ $1$-island peak with additional diagonal connections remains steady at $\Delta / \Omega \approx 0$ and broadens slowly. Comparing experiments between $9$ and $180$ atoms (see also Supplementary Information), increasing the system size or $R_\mathrm{b}/a$ shifts the $xy$ $1$-island peak to larger $\Delta / \Omega$ values and shifts the peak weight to produce a distinct, sharp exit edge (\Cref{fig:phases_experiments_2d}b).

The emergence of the central dynamical regime in 2D is consistent with the blockade-subspace prethermalization picture, as the Hamiltonian structure drives Floquet-like prethermalization for $V_1 \gg \Delta, \Omega$. This is evident from the 16-atom numerics in \Cref{fig:phases_experiments_2d}c, where nearest-neighbor blockade-subspace prethermal averages coincide with long-time averages of local island observables (see also Supplementary Information). In contrast, the full Hilbert-space thermal average deviates strongly for $\Delta / \Omega \gtrsim 0$, with resonant melting driving the system away from prethermalization at specific $\Delta / V_1$ ratios.

Distinct from the 1D chain, the diverging two-peak structure in the central region arises from the 2D connectivity shaping the blockade subspace. The nearest-neighbor blockade subspace becomes weakly fractured due to the energy cost of second-neighbor interactions, $V_2 = C_6 / (\sqrt{2} a)^6 = V_1 / 8$, where fracturing increases with $R_\mathrm{b} / a$. Floquet-like prethermalization can occur in the second-neighbor blockade subspace owing to the integer spectrum of  $H_0' = 8 V_2 \sum_{\langle ij \rangle} n_i n_j + V_2 \sum_{\langle\langle ij \rangle \rangle} n_i n_j$ on shorter timescales, $\mathcal{O}(e^{(V_2/\lambda)})$. This could yield a second prethermal plateau, though the condition $V_2 \sim \max\{\Delta, \Omega\}$ prevents its full realization here. Nevertheless, the dynamics begin to transition from a one- to a two-step prethermal plateau in the central regime, thus presenting a path for exploring both the emergence and breakdown of prethermalization in intermediate, theoretically unexplored regimes.

The sharp right edge at the exit of the central region is a characteristic feature of 2D dynamics and opens a distinct possibility of a dynamical phase transition at prethermal times. Microscopically, the edge appears due to the sharp decay in the number of accessible microscopic states in the blockade subspace. In the perturbative limit, at large $R_\mathrm{b} / a$ and significant $\Delta/\Omega$, the energy density in the blockade subspace drops sharply when the all-zero state becomes approximately resonant with the state containing the maximum number of excitations in the blockade subspace, i.e., the N{\'e}el state. The trends observed experimentally for the exit edge of the central region now follow: the classical N{\'e}el state in the large system limit gives $\Delta_{\text{N{\'e}el}} = 2 V_2$, as shown in \Cref{fig:phases_experiments_2d}b, which is close to the $180$-atom observations. At smaller system sizes, the classical maximum N{\'e}el cluster energy is shifted by the boundary correction $\Delta_{\text{N{\'e}el}} / V_2 \approx 2 - 2(1/N_x + 1/N_y)$, describing how the right edge converges to the large system limit in the experiment.

Immediately to the left of the edge, the dynamics is dominated by defect-proliferating N{\'e}el states within the blockade subspace, whose classical density decays exponentially with both $|\Delta - \Delta_{\text{N{\'e}el}}|$ and the system size. Similarly, the transition matrix elements exhibit exponential decay with system size as the Hamming distance between the N{\'e}el cluster and the all-zero state increases. These trends suggest a sharp transition at the large system and large $R_\mathrm{b}/a$ limits, that survives up to exponentially long times --- a proximate quantum dynamical phase transition driven by prethermalization. The underlying mechanism, defect proliferation during the evolution of the N{\'e}el configuration, resembles equilibrium phase transitions~\cite{Sachdev:2011}. However, the parameter space of our observed transitions lies well beyond the known boundaries of the equilibrium N{\'e}el phase identified in prior experiments~\cite{Ebadi:2021}. Furthermore, while the entrance to the central regime is thermal, no thermal transition equivalent~\cite{Halimeh:2017, Zunkovic:2018} is possible at the exit regardless of quench temperature, as the ground state transition~\cite{Ebadi:2021} lies at significantly higher detunings and the thermal dome is expected to shrink further with temperature. This dynamical transition, while expected to be present in thermodynamically large systems at finite (prethermal) times, likely does not persist in long (thermal) times. This is analogous to the observed finite-size, finite-time proximate phase transitions in ground state preparations, which does not preclude accurate probes of the underlying proximate criticalities~\cite{Keesling:2019, Ebadi:2021}.

The qualitative picture of a crossover at the entrance and a DPT at the exit of the central region has a suggestive classical analog related to the theory of magnetic hysteresis (see Supplementary Information). In the classical limit of the Rydberg Hamiltonian, where spins are promoted to vectors $S \rightarrow \infty$, the classical equations of motion for the zero-momentum spectrum decouple, and the total magnetization $\vec{M}$ fully governs the equivalent collective dynamics. The classical system exhibits limited precession away from the initial state at large $|\Delta|/\Omega$, and shows a smooth crossover around $\Delta \approx 0$ to the collective state where the system explores extensive regions of the magnetization sphere, analogous to the quantum central region. The magnetization changes abruptly with a dynamical phase transition point at the location given by the Stoner-Wohlfarth astroid~\cite{Stoner:1948}. The classical equivalent of the 1-island observable is presented on the representative $R_\mathrm{b}/a$ cut in \Cref{fig:phases_experiments_2d}d, and agrees quantitatively with quantum dynamics at large detunings. In the limit $V_i \gg \Omega$, the approximate position of the transition is given by $\Delta_c / \Omega \approx (3/2) (\sum_i z_i V_i / \Omega)^{1/3} \propto (R_\mathrm{b}/a)^2$, where $z_i$ is the coordination number corresponding to $V_i$. This strongly deviates from that of the potential quantum DPT line close to $\Delta_{\text{N{\'e}el}}/\Omega \propto (R_\mathrm{b}/a)^6$. 

Therefore, while the prospective quantum DPT at prethermal times may represent a quantum analog of the Stoner–Wohlfarth transition, quantum fluctuations substantially alter the classical picture. The classical Stoner-Wohlfarth transition turns into a crossover for sufficiently small anisotropy, i.e., $V_i$. An analogous scenario likely occurs in the quantum system: as Floquet-like prethermalization breaks down near $R_\mathrm{b}/a \sim 1$, the system transitions into a single dynamical regime without accessible thermal transitions, yielding only a crossover at the exit. The experimental data in \Cref{fig:phases_experiments_2d}a support this interpretation, showing a gradual loss of sharpness in the exit edge around $R_\mathrm{b}/a=1.2$.

\section*{Outlook}
\label{sec:outlook}

Our experiments in both one and two dimensional systems uncover qualitatively distinct dynamical regimes of the Rydberg Hamiltonian. They reveal Floquet-like prethermal dynamics at long emergent timescales and probe the structured melting of the prethermalization through resonant excitations, resulting in a dynamical phase diagram that stands in sharp contrast to the corresponding ground-state phase diagram~\cite{Ebadi:2021}. Prethermal dynamics stabilizes a wide dynamical regime within a constrained subspace, which in 2D exhibits a sharp change in dynamical behavior converging with increasing system size, indicating a  possibility for a proximate dynamical phase transition at prethermal times for large systems. Interestingly, the transition appears to be an analog of the classical Stoner-Wohlfarth transition with clear quantum-driven features. 

Several immediate directions emerge from our findings. Our results motivate the need for a systematic classification of the observed prethermal quantum dynamical regimes, particularly the possible quantum Stoner-Wohlfarth transition at the exit of the central region in two dimensions. Theoretical frameworks behind currently known DPTs usually rely on thermal and quantum equilibrium criticality~\cite{Heyl:2013, Halimeh:2017, Titum:2019, Zunkovic:2018}, and thus do not apply to our observations. The prethermal behavior we observe also spans diverse parameter regimes, providing a possibility for further studies of the intermediate regimes that are theoretically difficult to describe, including the emergence of the Floquet-like prethermal plateau and appearance of multiple plateaus at distinct though exponentially long timescales. The resonance structure, with its fine-grained hierarchy of dynamical peaks, suggests a deep correspondence between quench dynamics and the geometric organization of the many-body spectrum. Investigating the effective dynamics within each resonance could reveal novel Hamiltonians of interest beyond the commonly studied PXP Hamiltonian in the blockade subspace. The robustness of our resonance regime observations shows, by precedent, the experimental viability of such studies~\cite{Liu:2022}. Expanded experimental capabilities, such as two-copy entanglement entropy probes~\cite{Bluvstein:2022}, or mid-circuit perturbations that allow measurement of dynamical spectral functions and out-of-time-order correlators~\cite{Swingle:2016}, could provide complementary information about spatio-temporal correlations and the mechanism behind observed dynamical responses, particularly the potential proximate DPT. Extending such studies to distinct initial states~\cite{Daniel:2023}, programmable disorder~\cite{Hashizume:2025}, periodic driving~\cite{Bluvstein:2021, Hudomal:2022}, or open system dynamics~\cite{Chertkov:2023, Iadecola:2025} could also reveal how non-equilibrium phenomena manifest under more general conditions. Finally, several regimes we uncover, notably in two dimensions, may not be easily accessible directly using existing numerical techniques, underscoring the unique potential of quantum simulators to probe non-equilibrium many-body physics~\cite{Daley:2022}.


\begin{acknowledgments}

We acknowledge helpful discussions with Soonwon Choi, Manuel Endres, Daniel K. Mark, Carolyn Zhang, and Phillip Weinberg. This research was supported by the U.S. Department of Energy (DOE) under Contract No. DE-AC02-05CH11231, through the National Energy Research Scientific Computing Center (NERSC), an Office of Science User Facility located at Lawrence Berkeley National Laboratory. 
Work at Harvard was supported by the U.S. Department of Energy (DOE Quantum Systems Accelerator Center, grant number DE-AC02-05CH11231, and the QUACQ program, grant number DE-SC0025572),  the Center for Ultracold Atoms (a NSF Physics Frontiers Center, PHY-1734011), the National Science Foundation (grant numbers PHY-2012023 and  CCF-2313084).  Experiments were conducted on the neutral atom analog quantum simulator Aquila, built and operated by QuEra Computing \cite{aquila:2023}. We ran our numerical simulations with the Bloqade software package developed by QuEra Computing \cite{BloqadeJulia:2023, bloqadepython:2024} on the Perlmutter supercomputer at the National Energy Research Scientific Computing Center (NERSC) based at Lawrence Berkeley National Laboratory \cite{perlmutter:2023}.

\end{acknowledgments}

%

\clearpage
\onecolumngrid
\begin{center}
{\bf Supplementary Information: ``Probing emergent prethermal dynamics and resonant melting on a programmable quantum simulator''}
\end{center}

In this Supplementary Information, we present experimental details and supporting calculations for the results in the main text. In particular, we start by describing experimental and numerical methods in \Cref{supp:sec:methods}, \Cref{supp:sec:DPTs_1D_exp} provides a deeper look into the 1D experiments, while \Cref{supp:sec:DPTs_1D_num} provides supplemental numerical data supporting the observations in the main text, including the details of the thermal ensemble calculations. \Cref{supp:sec:resonances} presents perturbation theory calculations of the resonances, covering the dynamical emergence of the resonances and the effective Hamiltonian at the main resonance. Additional details of the 2D experiments are presented in \Cref{supp:sec:DPTs_2D_exp}, while the accompanying 2D numerics follow in \Cref{supp:sec:DPTs_2D_num}. Finally, the Stoner-Wohlfarth classical dynamics is covered in \Cref{supp:sec:classical_spin_models}.

\section{Methods}
\label{supp:sec:methods}

\subsection{Experimental protocols and data analysis}
\label{supp:subsec:methods:experiments}

\begin{figure*}[htb]
\centering
\includegraphics[width=1\textwidth]{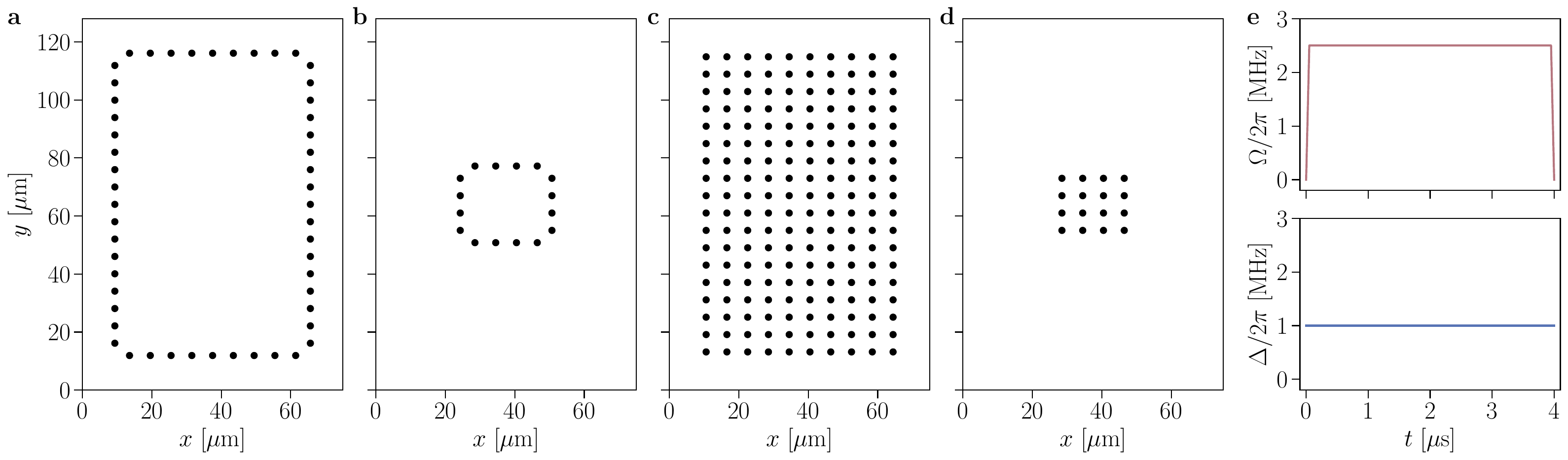}
\caption{\textbf{Experimental geometries and waveforms.} The lattices are shown for $R_\mathrm{b} / a = 1.4$. 1D lattice with \textbf{a.} 52 atoms; and \textbf{b.} 16 atoms, where the field of view can accommodate 2 parallel copies with separation $b = 5a$. 2D square lattice with \textbf{c.} 180 atoms; and \textbf{d.} 16 atoms, where the field of view can accommodate 6 parallel copies with separation $b = 5a$. \textbf{e.} Pulse sequences. (\textit{Top}) The Rabi frequency $\Omega$ is set to a constant value $\Omega / 2\pi = 2.5$ MHz. (\textit{Bottom}) The detuning $\Delta$ is set in the range $\Delta / \Omega \in [-4.0, 6.0]$, and shown here for $\Delta / \Omega = 0.4$.}
\label{fig:experimental_setup}
\end{figure*}

We perform experiments using the neutral atom analog quantum simulator Aquila, with its operating principles described in detail in Ref.~\cite{aquila:2023}. The platform uses $^{87}$Rb atoms, with each atom $j$ encoding a qubit. In the analog mode, the qubit states are the ground state $\ket{g} = \ket{5s_{1/2}} = \ket{0}$ and the highly-excited Rydberg state $\ket{r} = \ket{70 s_{1/2}} = \ket{1}$, yielding the qubit Hilbert space $\mathcal{H}_{(2)}$. The array of $N$ atoms has the Hilbert space $\mathcal{H} = \mathcal{H}_{(2)}^{\otimes N}$ spanned by the computational basis $\ket{x}$ where $x \in \{0,1\}^{\otimes N}$ are bitstrings. The Hamiltonian for the neutral atom array is given in \Cref{eq:hamiltonian}. We initialize the state in the $\ket{0}^{\otimes N}$ configuration, evolve it under the Hamiltonian, and measure the final state in the computational basis. The measurement thus returns a bitstring of length $N$. 

We examine two geometries. First, a 1D chain with periodic boundary conditions, arranged in a rectangle with flattened corners (\Cref{fig:experimental_setup}a,b). Such a geometry is determined by the accessible operation region and the row distance constraints of the experiment. We perform extensive runs for $52$ atoms to examine the dynamics across a broad parameter space and to map a landscape of steady dynamical regimes, which we equivalently refer to as a dynamical phase diagram, as well as more limited runs for $16$ atoms and intermediate sizes to probe system size effects. Secondly, a 2D lattice with open boundary conditions is arranged in a rectangle (\Cref{fig:experimental_setup}c,d), where similarly we perform extensive runs for $180$ atoms and limited runs for $9$, $16$, and $64$ atoms. 

In each case, we quench the Rabi frequency from its initial zero value to the constant value $\Omega / 2\pi = 2.5$~MHz, and set the detuning to a constant value $\Delta / \Omega$ (\Cref{fig:experimental_setup}e). The Rabi frequency sets the energy scale of the Hamiltonian and the blockade radius $R_\mathrm{b} = (C_6 / \Omega)^{1/6}$, where $C_6 = 2\pi \cdot 862690~\mathrm{MHz} \cdot \mu\mathrm{m}^6$ is the van der Waals coefficient. We set the dimensionless ratio $R_\mathrm{b} / a$, and thus the blockade strength, by setting the atom separation $a$. We map the dynamical regimes over the following parameter points: $\Delta / \Omega \in [-4.0, 6.0]$, with $\Delta / \Omega \in [-4.0, -0.4]$ in steps of $0.4$ and $\Delta / \Omega \in [0.0, 6.0]$ in steps of $0.2$; and $R_\mathrm{b} / a \in [1.2, 1.6]$ in steps of $0.04$. For each point on the dynamical phase diagram, we simulate the dynamical evolution for end time points $t \in [0.0, 4.0]$ $\mu$s in steps of $0.4$ $\mu$s (see also \Cref{fig:time_series_2d}). 

For the 1D chain with $52$ atoms, we run $120$ shots per time point; in the smaller system of $16$ atoms, we replicate $2$ parallel copies in the field of view, and run $50$ shots per time point, for $100$ effective shots. For the 2D lattice with $180$ atoms, we run $100$ shots per time point; in the smaller system of $16$ atoms, we replicate $6$ parallel copies in the field of view, and run $20$ shots per time point, for $120$ effective shots. 

The measurement results are processed as follows. We first filter out shots in which any atoms are lost during the filling procedure, thereby reducing to the subset of successful shots. It is crucial to take this filtering step for the 1D chain, though less so for the 2D lattice due to its increased multidimensional connectivity, allowing one to retain shots with lost atoms to improve statistics. For each diagonal observable, we compute the expectation value of the observable on each measured bitstring, and collect the results from the ensemble of measurements to obtain the spectral probability distribution. For the island observables, in lieu of this approach, we equivalently perform an iterative depth-first search on the geometrically arranged bitstring to compute the number and size of islands. The expectation value of an observable is simply the probability-weighted average of its bitstring expectation values. We compute the time average of the expectation value over the range $t_i = 0.8$ $\mu$s to $t_f = 4.0$ $\mu$s, to exclude the initial transient. We use the standard error of a given statistic to estimate its measurement error bars. 

\subsection{Numerical calculations}
\label{supp:subsec:methods:numerics}

The numerical simulations of the Rydberg Hamiltonian [\Cref{eq:hamiltonian}] are performed using exact diagonalization (ED), implemented in Julia via the open-source package Bloqade.jl~\cite{BloqadeJulia:2023}. All simulations are executed on the Perlmutter supercomputer \cite{perlmutter:2023} at the National Energy Research Scientific Computing Center (NERSC), hosted at Lawrence Berkeley National Laboratory. This ED-based approach enables us to directly access the full many-body wavefunction and compute the extensive exact spectra and time evolution datasets efficiently for system sizes of up to $N \sim 20$ atoms. Crucially, the simulations are performed in the full Hilbert space of dimension $|\mathcal{H}| = 2^N$, without restriction to the blockade subspace. This allows us to capture phenomena that arise from blockade violations, including the emergence of resonance peaks in the dynamical regimes landscape (\Cref{fig:phases_1d}a).

Time evolution is simulated under a time-independent Hamiltonian, with parameters chosen to closely match the experimental setup. The Rabi frequency is fixed at $\Omega / 2\pi = 2.5$ MHz, and the van der Waals coefficient is taken to be $C_6 = 2\pi \cdot 862690~\mathrm{MHz} \cdot \mu\mathrm{m}^6$, setting the blockade radius via $R_\mathrm{b} = (C_6/\Omega)^{1/6}$. We scan the detuning over $\Delta / \Omega \in [-4.0, 6.0]$ in steps of $0.1$, and vary the interatomic spacing $a$ to sweep the dimensionless parameter $R_\mathrm{b} / a \in [1.2, 1.6]$ in steps of $0.00625$. 

All simulations are initialized in the product state $\ket{0}^{\otimes N}$ and are evolved to a final time of $10$ $\mu$s using a time step of $0.01$ $\mu$s. At every time step, the full wavefunction is stored and observables are evaluated during post-processing. We consider two atomic geometries: a 1D circle (chain with periodic boundary conditions) and a 2D square lattice with open boundary conditions (see \Cref{fig:phases_numerics_2d}). Most simulations are performed at system size $N = 16$, with select cases run for $N = 10, 12, 14, 18,$ and $20$ atoms to probe finite-size effects. 

Additionally, in 1D, for $N=16$, $R_\mathrm{b}/a = 1.4$, and $\Delta / \Omega \in [-4.0, 6.0]$, we perform longer time evolution simulations, with the final time of $100$ $\mu$s (keeping the time step at $0.01$ $\mu$s). We observe no significant difference in time averages compared to $10$ $\mu$s runs (\Cref{fig:phases_1d}d), as further discussed in the Supplementary Information.

\section{Experimental characterization of dynamical regimes in 1D} 
\label{supp:sec:DPTs_1D_exp}

\begin{figure*}[htb]
\centering
\includegraphics[width=0.667\textwidth]{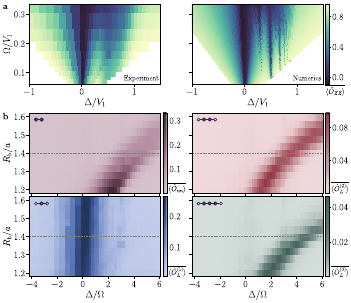}
\caption{\textbf{Dynamical regimes of the 1D chain in terms of alternative variables and observables.} 
\textbf{a.} The phase diagrams from \Cref{fig:phases_1d}a with the axes rescaled to $(\Delta/V_1, \Omega/V_1)$. The diagrams are presented in terms of the $\hat{O}_{ZZ}$ observable, and show (\textit{left}) the experimental results with $52$ atoms and (\textit{right}) the numerical results with $16$ atoms in a circular geometry.
\textbf{b.} Experimental phase diagrams for $52$ atoms displaying the time-averaged expectation values of the $\hat{O}_{nn}$ observable [\Cref{eq:Onn}] and $k$-island observables $\hat{O}_L^{(k)}$ for $k = 1,2,3$ (\Cref{eq:OLk} and surrounding text). 
The atom schematics in the upper corners show the types and sizes of Rydberg excitations that the observables measure. The grey dashed horizontal line shows the cut $R_\mathrm{b} / a = 1.4$.
}
\label{fig:phases_1d_alternative}
\end{figure*}

In our quench experiments, we systematically explore a rectangular grid of points in the input parameter space $(\Delta/\Omega, R_\mathrm{b}/a)$, and thus present most of our results natively in terms of these variables. We select this grid to enable direct comparisons with previous experiments characterizing the ground state equilibrium phases of the Rydberg Hamiltonian~\cite{Bernien:2017,Keesling:2019,Ebadi:2021}. However, alternative sets of variables can highlight different features of the results. In \Cref{fig:phases_1d_alternative}a, we present the phase diagrams for the 1D chain in terms of the rescaled variables $(\Delta/V_1, \Omega/V_1)$, where $V_1$ is the nearest-neighbor interaction energy, which satisfies $V_1/\Omega = (R_\mathrm{b}/a)^6$. The two rescaled variables remove the $(R_\mathrm{b}/a)^6$ scaling of the side peak resonances, and the second variable $\Omega/V_1$ offers a natural classical limit of $\Omega/V_1 \rightarrow \infty$. In the resulting phase diagram, the central region and resonances remain fixed at particular values of $\Delta/V_1$.

In the main text, we present the dynamical phase diagram for the 1D chain in terms of the $\hat{O}_{ZZ}$ observable (\Cref{fig:phases_1d}a). In light of the resonance structure discussed in the Results section, it is instructive to also examine the phase diagrams constructed from the $\hat{O}_{nn}$ observable and the island observables, as was done in the latter case for the 2D lattice (\Cref{fig:phases_experiments_2d}a). These phase diagrams follow in \Cref{fig:phases_1d_alternative}b, and confirm that the features observed on the $R_\mathrm{b} / a = 1.4$ cut indeed extend throughout the parameter space. In particular, we observe that the broad central region is characterized by the excitation of $1$-islands and the side peaks are characterized by nearest-neighbor blockade violations. Importantly, the side peaks, largely smeared in the $\hat{O}_{ZZ}$ and $\hat{O}_{nn}$ signals, become effectively resolved when using size-$k$ island observables for $k = 2, 3$, in line with their origin as energy resonances from specific Rydberg excitation patterns. 

\begin{figure*}[htb]
\centering
\includegraphics[width=1\textwidth]{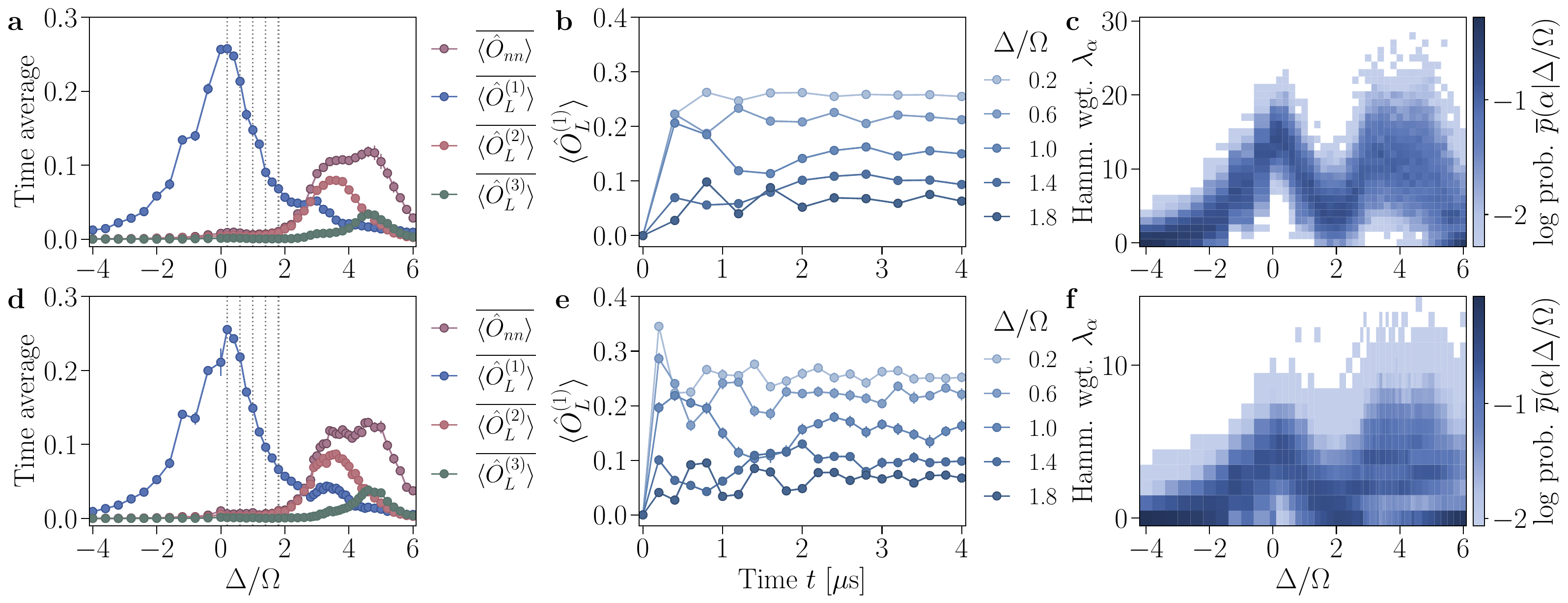}
\caption{\textbf{System size dependence in 1D chain experiments.} Experimental results on the cut $R_\mathrm{b} / a = 1.4$ for \textbf{a}-\textbf{c.} $N = 52$ atoms and \textbf{d}-\textbf{f.} $N = 16$ atoms. The geometries are shown in \Cref{fig:experimental_setup}. 
\textbf{a},\textbf{d.} The time-averaged expectation values of the $\hat{O}_{nn}$ observables and the $k$-island observables $\hat{O}_L^{(k)}$ for $k = 1,2,3$. 
\textbf{b},\textbf{e.} The time evolution of the expectation value of the 1-island observable for several values of $\Delta / \Omega$ on the right exit from the central region, shown as dotted vertical lines in \textbf{a},\textbf{d}. 
\textbf{c},\textbf{f.} The time-average $\bar{p} ( \alpha | \Delta / \Omega ) = \frac{1}{T} \sum_i p ( \alpha | \Delta / \Omega , t_i )$ of the Hamming weight probability distribution $p(\alpha | \Delta/\Omega , t) = \operatorname{Tr}{(\rho(t) \hat{P}_\alpha)}$. Projectors $\hat{P}_\alpha$ arise from the spectral decomposition of the (unnormalized) Hamming weight operator $\hat{Q}_n \equiv N \hat{O}_n = \sum_\alpha \lambda_\alpha \hat{P}_\alpha$, where $\lambda_\alpha = 0, \hdots, N$ denote the possible Hamming weights. The time average is equivalent to marginalization over $t_i$ with $p (t_i) = \frac{1}{T}$. The distribution in panel \textbf{c} is the same as in \Cref{fig:characterization_phases_1d}a.
}
\label{fig:system_size_experiments_1d}
\end{figure*}

As outlined in Sec.~\ref{supp:sec:methods}, we perform experiments on the 1D chain from $52$ to $16$ atoms to investigate the system size dependence of the observed dynamical regimes, focusing on the representative $R_\mathrm{b} / a = 1.4$ cut through the parameter space. On this cut, the features of the size-normalized island observables for $52$ atoms (\Cref{fig:characterization_phases_1d}b, repeated in \Cref{fig:system_size_experiments_1d}a for a clear comparison) are largely unchanged compared to those for $16$ atoms (\Cref{fig:system_size_experiments_1d}d), indicating that the mechanisms underlying each dynamical region are qualitatively similar across different system sizes in the 1D case. The size-normalized $\hat{O}_{nn}$ observable also remains largely unchanged with system size, further corroborating the resonance picture of the side peaks.

In addition to time averages, the time evolution of size-normalized local observables is generically unchanged with increasing system size, suggesting that their dynamics are not strongly influenced by finite-size effects. For instance, the time series of the $1$-island observable are similar at different system sizes for a series of points on the right exit of the central region (\Cref{fig:system_size_experiments_1d}b,e), a particularly volatile segment since the state evolution shifts from exploring most of the nearest-neighbor blockade subspace to a local subregion of it. Importantly, they quickly converge to a steady value in approximately $t \sim 0.5$ $\mu$s. 

In contrast to expectation values of size-normalized observables, full spectral probability distributions encode richer dynamical information than simple averages, since unnormalized observables are more sensitive to system size effects. The unnormalized Hamming weight probability distribution is of particular interest, since it directly reflects the total number of excited Rydberg atoms and thus effectively captures the spread of Rydberg excitations. Since the observables rapidly converge to steady values, the \emph{time-averaged} Hamming weight probability distribution can be used as a condensed description of the quench dynamics. These are shown in \Cref{fig:system_size_experiments_1d}c,f for $52$ and $16$ atoms, where panel c also appears in \Cref{fig:characterization_phases_1d}a. The distribution becomes sharper with increasing system size. 

Notably, at $\Delta / \Omega = 0$, which corresponds to the central peak in the dynamical regimes landscape, we find that the time-averaged distribution shifts to higher Hamming weight with increasing system size. We presented the scaling of its mean and variance in the inset of \Cref{fig:characterization_phases_1d}a, where linear fits yield the slope of $\sim 0.275 \pm 0.002$ for the mean and $\sim 0.106 \pm 0.006$ for the variance. Larger systems support higher Hamming weight configurations in the nearest-neighbor blockade subspace, and thus enable the post-quench state to traverse more expansive configuration spaces. The distribution of Hamming weights in this state is given approximately by the binomial distribution $B \left( N/2 , 1/2 \right)$, since the largest Hamming weight state in the blockade subspace has $N/2$ Rydberg excitations, and in the limit of equal blockade subspace exploration a single atom excitation has probability $1/2$. The mean of this distribution is $N/4$ and the variance is $N/8$. The experimental scalings given above yield broadly consistent prefactors and are thus consistent with comprehensive (ergodic) evolution in the blockade subspace.

\section{Numerical characterization of dynamical regimes in 1D} 
\label{supp:sec:DPTs_1D_num}

\subsection{Non-diagonal observables and dynmical response convergence}
\label{supp:sec:DPTs_1D_convergence}
\begin{figure*}[htb]
\centering
\includegraphics[width=1.0\textwidth]{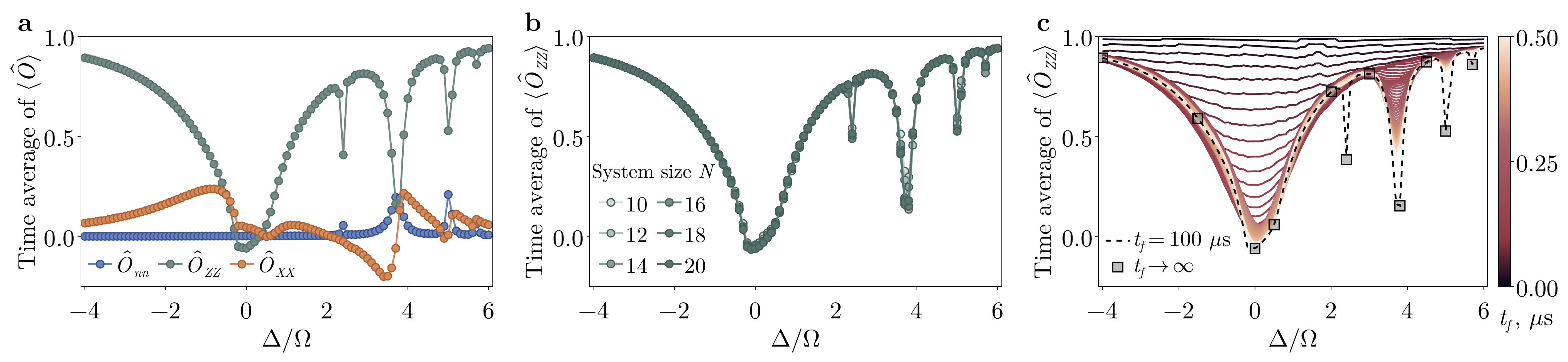}
\caption{
\textbf{Characterization of dynamical properties for the 1D chain.} Numerical studies of multiple observables, scaling, and averaging for the 1D chain in a circular geometry on the cut $R_\mathrm{b} / a = 1.4$. The time-averaged expectation values are computed as $\overline{\expval*{\hat{O}}} = \frac{1}{t_f} \int_{0}^{t_f} \dd{t} \langle \hat{O}(t) \rangle$. \textbf{a.} The time averages of the Rydberg density $\hat{O}_{nn}$ operator, the diagonal $\hat{O}_{ZZ}$ operator (also presented in the main text \Cref{fig:phases_1d}), and the off-diagonal $\hat{O}_{XX}$ operator. The $\hat{O}_{nn}$ response vanishes at the main peak, but captures the resonances, suggesting their origin from blockade violations; the $\hat{O}_{XX}$ response showcases non-analyticities at points where the dynamics is expected to change abruptly. \textbf{b.} The time averages for different system sizes $N = 10, \hdots, 20$ and final time $t_f = 10~\mu$s. The curves are nearly identical, indicating a weak system size dependence. \textbf{c.} The time averages for $N=16$ atoms and different final times $t_f$. The curves rapidly converge in $t_f \approx 0.5~\mu$s. The dashed line denotes the curve for $t_f = 100~\mu$s, which is the longest time examined in the numerics, while the squares denote the diagonal ensemble averages, which are equal to the time averages in the strictly infinite-time limit.  
}
\label{fig:characterization_observables_1d}
\end{figure*}

We further interrogate the observed experimental trends with numerical simulations at small system sizes, which offer finer resolution in the input parameters and time steps. In \Cref{fig:characterization_observables_1d}a, we present observables beyond $\hat{O}_{ZZ}$, including the diagonal Rydberg density $\hat{O}_{nn}$ operator, as well as the off-diagonal $\hat{O}_{XX}$ operator. The $\hat{O}_{nn}$ observable time averages indicate the physics underlying the main peak versus the side peaks. This observable assumes non-vanishing values only outside of the first-neighbor blockade subspace. In our data, we observe a strong match between finite values of the time-averaged $\hat{O}_{nn}$ observable and the side peaks only, suggesting the stabilization of resonant clustered excitations on these peaks. We also leverage the numerical time-averaged observables for small finite-size scaling analyses, finding that the sharp features remain nearly unchanged across system between $N=10$ and $20$ atoms (\Cref{fig:characterization_observables_1d}b). 
The numerical results also corroborate the rapid temporal convergence of the time-averaged observables, and permit finer tracking of the time-averaged value as a function of final time (\Cref{fig:characterization_observables_1d}c), explicitly revealing a transient timescale of approximately $t \approx 0.5~\mu\mathrm{s}$. We confirm numerically the stability of the pre-thermalized oscillations to $100$ $\mu$s, very long times relative to the natural scales of our system dynamics. In addition, we compute diagonal ensemble averages by explicitly diagonalizing the full Hamiltonian at selected values of $\Delta / \Omega$:
\begin{equation}
    \langle \hat{O} \rangle_{\mathrm{DE}} \equiv \sum_i |c_i|^2 \langle \psi_i | \hat{O} | \psi_i \rangle \, ,
\end{equation}
where $\ket{\psi_i}$ are the eigenstates of the Hamiltonian and $c_i \equiv \braket{\psi_i}{0}$ denotes the overlap between the initial state and the corresponding eigenstate. The diagonal ensemble average $\langle \hat{O} \rangle_{\mathrm{DE}}$ corresponds exactly to the observable expectation value in the infinite-time limit, assuming no degeneracies in the eigenspectrum. In \Cref{fig:characterization_observables_1d}c, we compare these values (grey squares) to the numerically obtained 100~$\mu$s time averages (dashed line). The excellent agreement between the two confirms the convergence of our simulations. While one might expect a separation between prethermal and infinite-time thermal values, the observed convergence of time averages and diagonal ensemble values in our simulations is likely an artifact of small system size. In finite systems, diagonal ensemble observables deviate from ETH by terms that scale as the inverse square root of the many-body density of states \cite{Beugeling:2014}. The density of states is significantly suppressed at large detuning outside of the resonances themselves, thus requiring a larger system size for convergence between the diagonal and thermal ensembles. In the large-scale experimental systems --- such as the 52-atom chain --- we expect that the deviations between thermal and diagonal ensembles are strongly suppressed, due to the exponential growth of the many-body density of states with system size.

\subsection{Quench dynamics and many-body spectrum}
\label{supp:sec:DPTs_1D_spectrum}

\begin{figure*}[htb]
\centering
\includegraphics[width=1\textwidth]{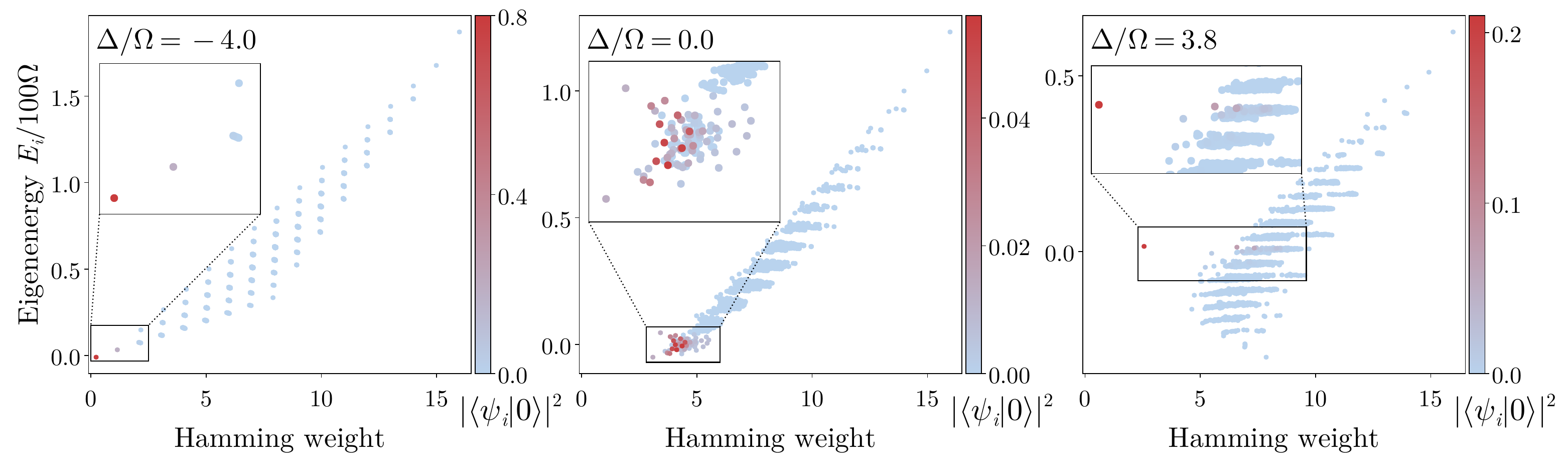}
\caption{
\textbf{Characterization of the eigenspectrum for the 1D chain.} 
Scatter plots of eigenenergies $E_i$ and mean Hamming weights for energy eigenstates $\ket{\psi_i}$ with zero momentum, from numerics with $16$ atoms in a circular geometry. The color density characterizes the overlap probability of each eigenstate with the all-zero initial state. Following the cut $R_\mathrm{b} / a = 1.4$, for $\Delta / \Omega = -4.0, 0.0, 3.8$, we observe a concentration of high overlaps among states in the lowest energy cluster as we center ourselves in the main peak of the phase diagram. This suggests that quenches into this regime will lead to state evolutions exploring sectors of the blockaded subspace. Increasing the detuning allows for the resonant crossing of the all-zero state with sectors with varying blockade violation count.
}
\label{fig:eigenspectrum_1d}
\end{figure*}

The nature of the state evolution in each dynamical regime is reflected in the underlying energy eigenspectrum, which can be explicitly computed in our numerical setups. To this aim, in \Cref{fig:eigenspectrum_1d} we analyze the eigenenergies at several quench points that each represent a different dynamical regime, for $16$ atoms on the cut $R_\mathrm{b} / a = 1.4$. Deep in the disordered (equilibrium) phase at $\Delta/\Omega = -4.0$ (left panel), the eigenstates are approximately product states, and thus the all-zero initial state approximately remains in this state at future times. On a side peak resonance at $\Delta/\Omega = 3.8$ (right panel), the eigenstates are arrayed in a structured tower; the initial state overlaps with a few dominant blockade-violating eigenstates in an energy-resonant manifold, and thus evolves in a limited part of the Hilbert space outside of the blockade subspace. At the peak $\Delta / \Omega = 0$ (middle panel) of the central region, the all-zero initial state overlaps with many energy eigenstates in the nearest neighbor blockade subspace, and thus the state coherently oscillates throughout the subspace. This last result accords with our experimentally obtained system-size scaling for the mean and variance of the Hamming weight probability distribution (\Cref{fig:characterization_phases_1d}a, inset), supporting the full exploration of the accessible blockade subspace.

\subsection{Thermal and prethermal expectation values}
\label{supp:sec:DPTs_1D_thermal}

\begin{figure*}[htb]
\centering
\includegraphics[width=1.0\textwidth]{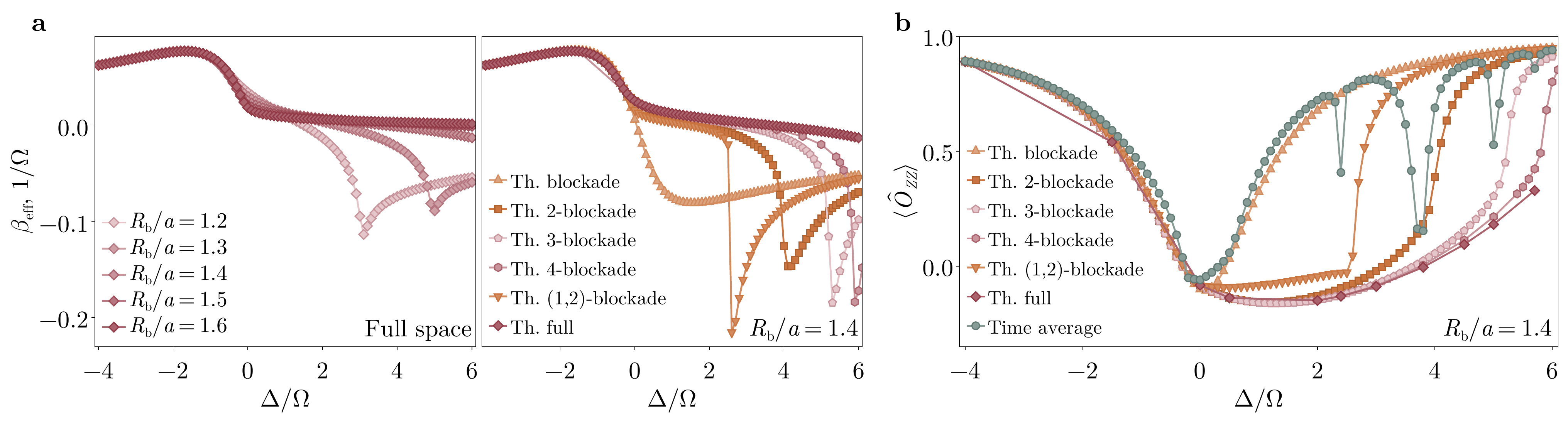}
\caption{
\textbf{Thermal observables for the 1D chain.} Comparative analysis of the effective inverse temperature and thermal expectation values of $\hat{O}_{ZZ}$ for $16$ atoms, different subspaces of the full Hilbert space, and different values of $R_\mathrm{b}/a$. \textbf{a.} The effective temperature increases with $\Delta/\Omega$ until an elbow-like turn at zero; temperatures become negative, with non-analyticities in the temperature correlated with resonance peaks when thermal averages are computed within specific subspaces of the Hilbert space. \textbf{b.} The time average (green dots) and thermal average coincide over progressively larger overlapping subspaces for detuning values up to $\Delta/\Omega=0$, after which a separation occurs. Constraining the thermal mean value across the specific subspaces leads to a sequential convergence between thermal and time averages, with jumps correlated with resonant peaks. Thermal averages constrained to the blockade subspace (light orange up-triangles) and time averages match everywhere except for the side resonances. The label ``Th.~$k$-blockade'' denotes a subspace spanned by all configurations that consist of islands up to size $k$. ``Th.~(1,2)-blockade'' denotes a subspace that is a union of the blockade subspace and configurations that include an \emph{equal} number of 1- and 2-islands. 
}
\label{fig:subspaces_1d}
\end{figure*}

To probe whether the quench dynamics lead to thermal behavior -- as would be expected under the eigenstate thermalization hypothesis (ETH) \cite{Deutsch:1991, Srednicki:1994, Rigol:2008, Deutsch:2018}  -- we compute the effective inverse temperature $\beta_\mathrm{eff}$ associated with the energy of the initial state $\ket{0}^{\otimes N}$. This is defined by matching the initial energy to that of a thermal ensemble with inverse temperature $\beta_\mathrm{eff}$: 
\begin{equation}
\mel{0}{\hat{H}}{0} = \expval*{\hat{H}}_\mathrm{th} \equiv \frac{\Tr(\hat{H} e^{-\beta_\mathrm{eff} \hat{H}})}{\Tr(e^{-\beta_\mathrm{eff} \hat{H}})} \, .
\end{equation}
By solving this equation, we obtain $\beta_\mathrm{eff}$ as a function of $\Delta / \Omega$, shown in \Cref{fig:subspaces_1d}a, left for several values of $R_\mathrm{b} / a$. These effective temperatures are then used to compute corresponding thermal expectation values of local observables, such as $\expval*{\hat{O}_{ZZ}}_\mathrm{th}$ shown in \Cref{fig:phases_experiments_2d}c (labeled ``Th. full (num.)'', red diamonds). We also consider similar evaluations of the temperature and observables while projecting the trace sums into specific subspaces according to specific excitation subsectors in an onion-shell scheme around the blockade subspace, e.g. as only the blockade subspace, or that plus all 2-island excitations (labeled ``2-blockade'' in \Cref{fig:subspaces_1d}), and so on. The ``(1,2)-blockade'' label denotes the subspace consisting of the blockade subspace plus the subspace spanned by configurations that have an equal number of 1- and 2-islands.

This thermodynamic mapping provides a valuable reference point for assessing whether the long-time dynamics of the system can be captured by thermal equilibrium behavior. At first, let us consider thermal averages over the whole Hilbert space. At large negative detuning $\Delta/\Omega \ll 0$, the effective inverse temperature $\beta_\mathrm{eff}$ is large and positive, corresponding to a low-temperature thermal ensemble. In this regime, the quench energy is small, and the system remains close to the ground state of the post-quench Hamiltonian, while the time-averaged observables closely match the thermal predictions, consistent with thermalization under the ETH.

Near the center of the dynamical phase diagram, around $\Delta/\Omega \approx 0$, the effective temperature diverges ($\beta_\mathrm{eff} \to 0$), signaling a transition to an infinite-temperature-like regime. For positive detunings, $\beta_\mathrm{eff}$ even becomes negative. In this region, the canonical ensemble fails to describe the long-time dynamics, and the time-averaged observables significantly deviate from thermal expectations, at least on the observed timescale. 

The origin of the non-thermal behavior in the broader region to the right of the central peak --- including between the resonances --- becomes clearer when thermal expectations are evaluated within constrained subspaces, defined by specific excitation patterns. In particular, when thermal averages are restricted to the blockade subspace, we find an excellent agreement with the “backbone” of the time-averaged observables (\Cref{fig:subspaces_1d}b, green circles vs. light orange up-triangles). This provides a clear indication of prethermalization of the post-quench state within the subspace obeying the nearest-neighbor blockade constraint.

As we relax these constraints and consider progressively larger subspaces that permit additional types of blockade-violating excitations, the thermal and time-averaged values begin to diverge. This discrepancy persists until the detuning reaches the location of the next resonance, which corresponds to the dominant type of newly allowed excitation. The resonances coincide with sharp, discontinuous jumps in the effective temperature.

We interpret this structure as a sequence of prethermalization melts, in which the resonant pathways for heating outside of the initial prethermal subspace open. The fact that time averages at the resonance points do not fully match the thermal predictions of the full Hilbert space leaves an open question regarding the ultimate thermalization dynamics at these points. Similarily, discrepancies between the island-subspace thermal values and the respective resonance shapes show that the resonance content is richer than the leading island contributions would suggest. As an example, in \Cref{supp:sec:resonances_effective}, we discuss the complex physics of the main resonance in detail.

\subsection{Dynamical phase diagram under disorder}
\label{supp:subsec:DPTs_1D_disorder}

The dynamical phase diagram of the 1D chain shows a prominent central feature around $\Delta / \Omega = 0$ in the experimental data, consistent with numerical simulations. In contrast, the sharp resonance peaks at positive detuning are not observed in the experimental $\hat{O}_{ZZ}$ data, though clearly visible in the numerical phase diagram. Instead, the experimental side features are smeared and appear as a single broad peak. We conjecture that random imperfections in atom positions in the experiment may broaden or obscure the sharp resonance features in the numerics, and thereby produce this discrepancy. To test this, we repeat our numerical simulations with added positional disorder and examine its effect on the dynamical observables.

\begin{figure*}[htb]
\centering
\includegraphics[width=1.0\textwidth]{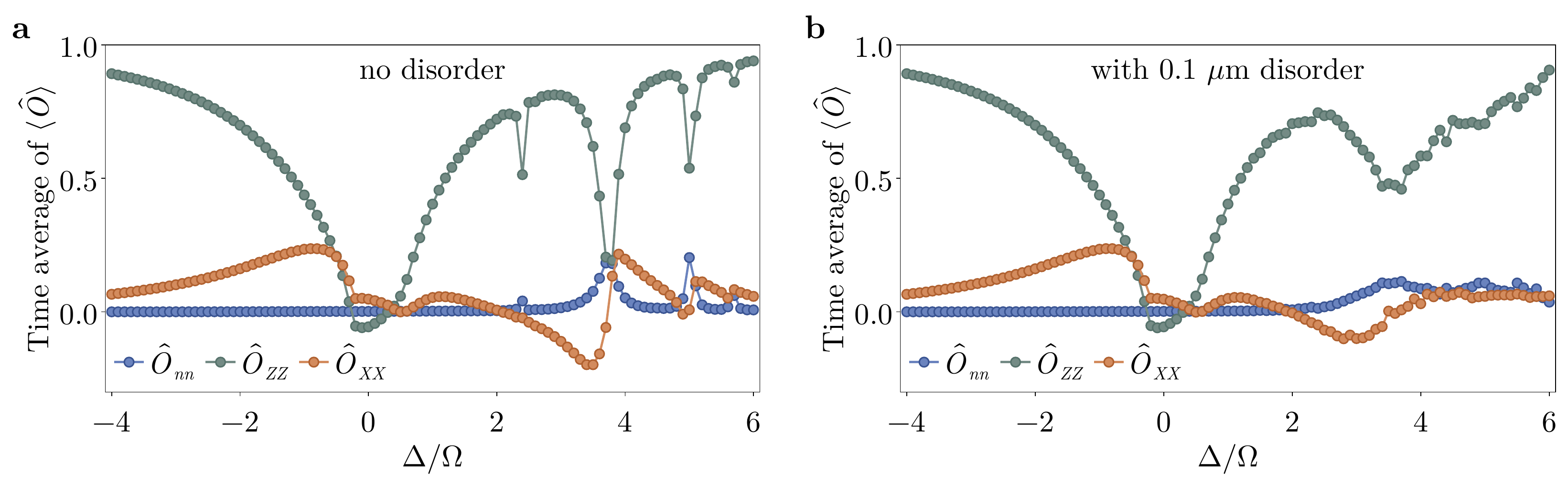}
\caption{
\textbf{Effect of position disorder on the observables in the 1D chain.} 
Numerical studies of the time-averaged expectation values of the two-site observables $\hat{O}_{ZZ}$, $\hat{O}_{nn}$, and $\hat{O}_{XX}$ (\Cref{eq:Onn} and surrounding text), for the 1D chain with circular geometry and $16$ atoms, and on the cut $R_\mathrm{b} / a = 1.4$. The atoms have separation $a = 6.0$ $\mu$m with: \textbf{a.} no position disorder; and \textbf{b.} hardware-relevant position disorder, where every atom position $(x_j,y_j)$ is shifted by $(\delta x_j, \delta y_j)$, where $\delta x, \delta y$ are randomly chosen from a normal distribution with mean zero and standard deviation $0.1$ $\mu$m, and then averages are taken over $10$ disorder realizations.
}
\label{fig:disorder_1d}
\end{figure*}

We simulate a $16$ atom circular chain at the fixed value of $R_\mathrm{b} / a = 1.4$, and compute the time-averaged expectation values of local two-site observables. The clean, disorder-free results are shown in \Cref{fig:disorder_1d}a (identical to \Cref{fig:phases_1d}d), where several sharp resonance peaks appear at positive values of $\Delta / \Omega$. We then introduce positional disorder by randomly shifting each atom’s position $(x_j, y_j)$ by $(\delta x_j, \delta y_j)$ drawn from a Gaussian distribution with zero mean and standard deviation $0.1~\mu$m -- a realistic estimate for the hardware platform \cite{aquila:2023}. The observables are then averaged over $10$ independent disorder realizations. 

As shown in \Cref{fig:disorder_1d}b, this modest level of disorder is sufficient to broaden the sharp resonances into a smoother feature, leading to a dynamical phase diagram that more closely resembles experimental observations. In each realization, positional disorder slightly changes the positions of each resonance peak. Upon averaging, this results in a single broadened resonance centered at the clean-system peak, with a width set by the disorder strength. Notably, while the peak sharpness is lost, the total integrated amplitude (peak area) is approximately preserved.

Importantly, the central region of the dynamical phase diagram near $\Delta / \Omega = 0$ remains essentially unaffected by disorder. It remains robust because the dynamics in this region are governed primarily by the blockade constraint itself, which prohibits nearest-neighbor excitations and is thus largely unaffected by positional disorder.

\section{Perturbation theory of the resonances}
\label{supp:sec:resonances}

The observed resonances provide structured melting pathways for a blockade subspace-constrained prethermal state. In the Floquet-like prethermalization picture, the resonances emerge from enhanced heating at frequencies commensurate with the effective drive frequencies. In this supplement, we provide a detailed view of the emergence of resonances through time-dependent perturbation theory (\Cref{supp:sec:resonances_emergence}), qualitatively explaining the relative magnitudes of the resonances. Furthermore, the observation of a robust resonant response opens a unique possibility for in-depth studies of distinct effective Hamiltonian physics with neutral atom arrays. As an example of possible exotic emergent physics, we derive an effective Hamiltonian that is valid in the vicinity of the 1D 2-island resonance in \Cref{supp:sec:resonances_effective}.

\subsection{Emergence of the resonances}
\label{supp:sec:resonances_emergence}

In this subsection, we examine the qualitative features of the resonances, which can be readily explained using time-dependent perturbation theory. Our experimental setup starts with the all-zero state, labeled $\ket{0}$ here for simplicity, and evolves it with a time-independent Hamiltonian. Since the resonances appear in the regime $\Delta, V_1 \gg \Omega$, we can straightforwardly apply time-dependent perturbation theory by splitting the Hamiltonian into a diagonal (``classical'') part and an off-diagonal perturbation (``quantum fluctuation''):
\begin{equation}
\hat{H} = \hat{H}_c + \hat{V}_q \, , \qquad
\hat{H}_c = - \Delta (t) \sum_j \hat{n}_j + \sum_{j<k} V_{jk} \hat{n}_j \hat{n}_k \, , \qquad
\hat{V}_q = \frac{\Omega}{2}\sum_j\hat{X}_j \, .
\label{eq:ptham}
\end{equation}
The eigenstates of $H_c$ are product states in the computational basis; we label these as $\{ \ket{m}\}$ with associated (degenerate) eigenvalues $E_m$, which describe the classical energies. 
The Dyson series~\cite{Sakurai:1994} yields an expansion for the transition matrix elements $c_n (t) = \!\mel{n}{U(t)}{0}$, and thus their long-time average:
\begin{align}
\overline{c}_n &= \lim_{T \rightarrow \infty, \delta \rightarrow 0} \frac{1}{T}\int_{t_0}^{t_0 + T} dt \left[ \delta_{n0} + c_n^{(1)} (t) + c_n^{(2)} (t) + \hdots \right] , \label{eq:longtimeaverage_pt} \\
c_n^{(1)} (t) &= \frac{1}{i} \int_{t_0}^{t} dt_1 \!\mel{n}{\hat{V}_q}{0}\! e^{i(E_n-E_0)t_1} e^{-\delta t_1} , \notag \\
c_n^{(2)} (t) &= \frac{1}{i^2} \sum_{m} \int_{t_0}^{t} dt_1 \int_{t_0}^{t_1} dt_2 \!\mel{n}{\hat{V}_q}{m}\! \!\mel{m}{V_q}{0}\! e^{i(E_n-E_m)t_1} e^{i(E_m-E_0)t_2} e^{-\delta t_1} e^{-\delta t_2} , \notag
\end{align}
where we used a regularization parameter $\delta>0$ to evaluate the time averages, with $\delta \ll \Omega, \Delta, V_i$. In practice, a finite $\delta$ will more closely match the exact behaviour in which all resonances have finite width. No qualitative features from the perturbation theory are affected by the choice of $\delta$. The long-time averages of the matrix elements recover the eigenstate corrections obtained in time-independent perturbation theory~\cite{Sakurai:1994}. If we apply the Dyson series to estimate the expectation value $d(t) = \langle \tilde{O} \rangle$ of an operator, properly shifted $\tilde{O} = \hat{O} - \!\mel{0}{\hat{O}}{0}$, then we obtain the following expression up to second order in the Rabi frequency:
\begin{align}
\overline{d} &= 2\mathrm{Re} \left[ d^{(1)} (t) + d^{(2)} (t) + \hdots \right] , \label{eq:operator_pt} \\
d^{(1)} (t) &= -\frac{\Omega}{2} \sum_{m \neq 0} \frac{\!\mel{m}{\sum_i\hat{X_i}}{0}\! \!\mel{0}{\tilde{O}}{m}\!}{E_m-E_0-i\delta_m} , \notag \\
d^{(2)} (t) &= \frac{1}{2} \left(\frac{\Omega}{2}\right)^2 \sum_{m, m'\neq 0} \frac{\!\mel{m}{\sum_i\hat{X_i}}{0}\! \!\mel{0}{\sum_j\hat{X_j}}{m'}\! \!\mel{m'}{\tilde{O}}{m} + \!\mel{m}{\sum_i\hat{X_i}}{m'}\! \!\mel{m'}{\sum_j\hat{X_j}}{0}\! \!\mel{0}{\tilde{O}}{m}\!}{\left(E_m-E_0-i\delta_m\right) \left(E_{m'}-E_0-i\delta_{m'}\right)} . \notag
\end{align}

The emergence of resonances can be directly inferred from the structure of the denominators in the perturbative expansion. In general, resonances in the phase diagram occur when $E_m = E_0$. The amplitude of the resonance is qualitatively related to the lowest order $p$ in the perturbation theory at which matrix elements in the numerator become non-zero. Since the initial state and the form of the perturbation are simple, we can find $p$ for all Pauli strings by analyzing the Hamming distances between contributing states. At $p$-th order, the perturbative expansion involves $p$ insertions of the off-diagonal term $\sum_i \hat{X}_i$, such that the set of intermediate states forms a ladder starting from the all-zero initial state and leading to the state with Hamming weight at most $p$. For a diagonal operator $\hat{O}$, a state with Hamming weight $k$ can produce a non-vanishing resonance only if the ladder is sufficiently long to traverse from this state to the all-zero state and back, and thus $p \geq 2k$. However, if $\hat{O}$ is a Pauli string that contains $l$ instances of $\hat{X}$, then the ladder length can be shortened, to at most $p \geq 2k-l$. More precisely, for an off-diagonal observable $\hat{O}$, the minimum order $p$ for a non-vanishing contribution is set by the lowest Hamming weight state reached when $\hat{O}$ acts on the resonance state:
\begin{equation}
p = \min_{\{\ket{m},\, E_m=E_0\}} \left[ 2\, \mathrm{hdist}(\ket{0}, \ket{m}) - \mathrm{hdist}(\ket{0}, \hat{O}\ket{m}) \right] \, .
\label{eq:perturbation_result}
\end{equation}

We apply this study to several of the prominent resonances. For all diagonal operators, including the commonly measured $\hat{Z}_i$ and $\hat{Z}_i\hat{Z}_{i+1}$, the perturbation theory order of the $k$-island resonance is $p=2k$. The main $k=2$ resonance is thus the strongest, with each side resonance suppressed by an additional prefactor of $(\Omega/ \Delta)^2$. This directly agrees with the numerical results of \Cref{fig:phases_1d}c and the experiments. On the other hand, the lowest-order response for the main $k=2$ island resonance arises from the $\hat{X}_i \hat{X}_{i+1}$ operator. In general, off-diagonal operators tailored to the resonance will show enhanced response. In contrast, island operators project out the individual island contributions, but being diagonal, still have $p=2k$. Finally, for the simplest case of the $k=2$ resonance in the 1D system and the $\hat{O}_{XX}$ observable, the expression from \Cref{eq:operator_pt} evaluates to:
\begin{equation}
\overline{\langle \hat{O}_{XX}\rangle}^{(2)} = \left(\frac{\Omega}{2}\right)^2 \mathrm{Re} \left[ \frac{1}{\left(2\Delta-V_1-i\delta\right)\left(\Delta-i\delta\right)} + \frac{1}{\left(\Delta-i\delta\right)^2} \right] .
\label{eq:operator_ptXX}
\end{equation}

\begin{figure*}[htb]
\centering
\includegraphics[width=0.667\textwidth]{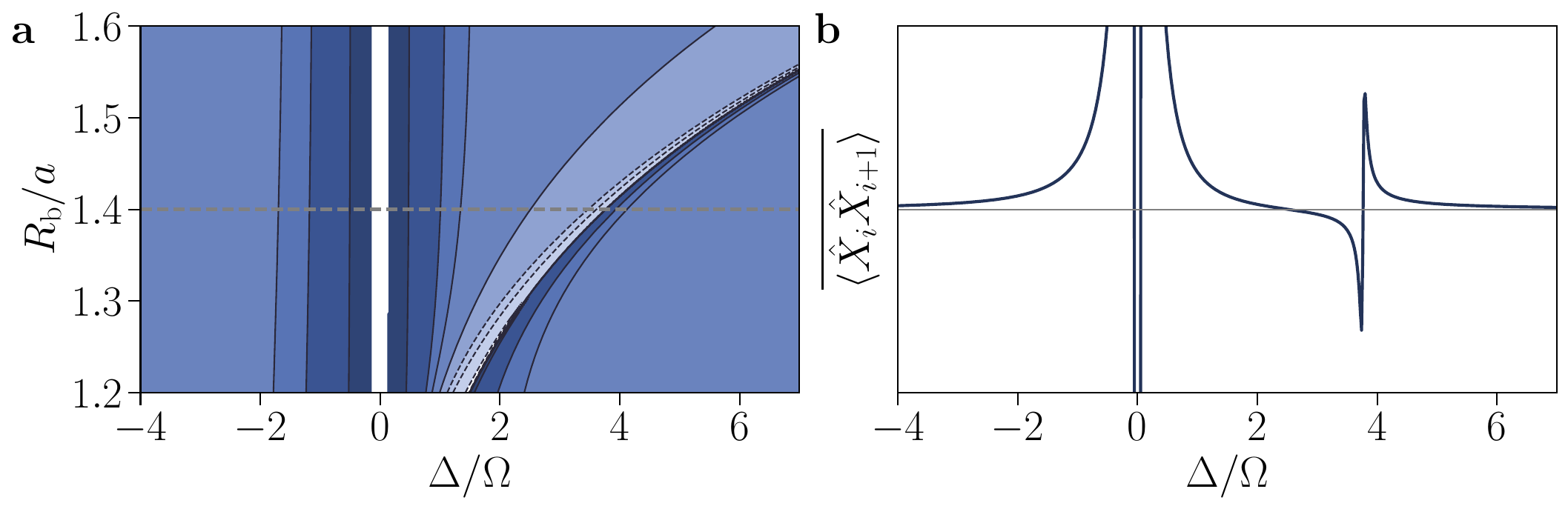}
\caption{\textbf{Perturbation theory of the resonances.} The long-time average of the $\hat{X}_i \hat{X}_{i+1}$ operator [\Cref{eq:operator_ptXX}] from second order perturbation theory \textbf{a.} on a contour plot over the parameter space of the detuning and blockade radius, and \textbf{b.} on the cut $R_\mathrm{b} / a = 1.4$. The regularization parameter is set to $\delta = 0.05 \Omega$. The expectation value amplitude is not shown since it is unphysical (regularization dependent); for visual clarity, the amplitude range is restricted and the contours have nonuniform amplitude spacing.}
\label{fig:resonances_perturbation}
\end{figure*}

This equation is plotted in \Cref{fig:resonances_perturbation}. It is strictly valid only in the perturbative regime $\Delta, V_1 \gg \Omega$, though we plot it over the full extent of the phase diagram shown in \Cref{fig:phases_1d}. The expansion should apply for most of the $k=2$ resonance parameters. We use a finite regularization $\delta$. The amplitude is unphysical, since it is regularization-dependent in perturbation theory. In contrast, the total area under the peaks is a more physical quantity for comparing the strength of the resonances. Qualitatively, the perturbative estimate for the position and shape of the $k=2$ resonance peak matches the numerical and experimental results presented in the main text reasonably well. In particular, the two-peaked structure with the sign change is recovered at the $k=2$ resonance. This structure results from the simple pole structure at $2\Delta = V_1$, which is specific for the resonance and the operator chosen. Generically, diagonal operator resonances will lead to even-order poles due to the symmetric nature of the perturbation theory ladder. Their shape is thus similar to the second-order pole driven peak structure at $\Delta=0$ in \Cref{fig:resonances_perturbation}, arising from the second term in \Cref{eq:operator_ptXX}.

\subsection{Effective Hamiltonian of the 2-island resonance} 
\label{supp:sec:resonances_effective}

The effective PXP Hamiltonian for the central region has been explored extensively and has led to predictions and observations of exotic phenomena, such as quantum many-body scars, lattice gauge theory dynamics, false vacuum decay, and quantum coarsening~\cite{Bernien:2017, Bluvstein:2021, Surace:2020, GonzalezCuadra:2024, Darbha:2024a, Darbha:2024b, Manovitz:2025}. In contrast, the study of effective physics in non-PXP regimes, notably the antiblockade regime characterized by $\Delta = V_1$, has been hindered by the lack of robustness to disorder~\cite{Mattioli:2015, Liu:2022, Marcuzzi:2017}. The robustness of the observed resonances in our experiments opens up the possibility of probing the detailed microscopic dynamics of the respective effective Hamiltonians. As an example of the exotic phenomena that might be accessible, we turn to explore the effective Hamiltonian at the dominant $2$-island resonance in the 1D chain.

The set of relevant, classically configurations at the two island resonance, $V_1 = 2\Delta$ includes configuration states with any number of $2$-islands ($0110$), but also mixed island configurations. The lowest Hamming distance configurations among these are $3+1$ island mixtures ($01110...010$). We again split the Rydberg Hamiltonian into a diagonal part ($\hat{H}_c$) and an off-diagonal perturbation ($\hat{V}_q$), according to \Cref{eq:ptham}. The space of all classical configurations at the $V_1 = 2\Delta$ resonance defines the projector $\hat{P}$, while all the other states are energetically well separated with respect to the $\hat{H}_c$ Hamiltonian since $V_1, \Delta \gg \Omega, V_2$, with projector $\hat{Q} = \hat{I} - \hat{P}$. The effective dynamics in the $\hat{P}$-defined $2$-island subspace is induced by virtual fluctuations to $\hat{Q}$ states, with the effective Hamiltonian in the lowest (second) order of the perturbation theory given by:
\begin{equation}
\hat{H}_{\mathrm{eff}}^{(2)} 
= \hat{P} \hat{H}_{c} \hat{P}
+ \hat{P} \hat{V}_q \hat{P} 
+ \hat{P} \hat{V}_q \hat{Q} \frac{1}{E_P^{(0)}\hat{I} - \hat{Q} \hat{H}_c \hat{Q}} \hat{Q} \hat{V}_q \hat{P} \, 
\end{equation}
where $E_P^{(0)}$ denotes the (degenerate, up to $V_2$), classical energy of the $\hat{P}$-subspace. 

The effective Hamiltonian can now be directly evaluated:
\begin{align}
H_{\mathrm{eff}}^{(2)} 
&=  \frac{\Omega^2}{2\Delta} \sum_{i} (1-\hat{n}_{i-1}) (1-\hat{n}_{i+2}) \left( \outerproduct{11}{00} + \mathrm{h.c.} \right)_{(i,i+1)} (1-\hat{n}_{i-1}) (1-\hat{n}_{i+2}) \label{eq:Heff_2islands_t1}\\
& + \frac{\Omega^2}{6\Delta} \sum_{i}  \left( \outerproduct{11011}{11101} + \outerproduct{11011}{10111} \right)_{(i,\ldots,i+4)} \label{eq:Heff_2islands_t2} \\
& - \frac{\Omega^2}{2\Delta} \sum_{i} (1-\hat{n}_{i-1}) (1-\hat{n}_{i+3}) \left( \outerproduct{010}{100} + \outerproduct{010}{001} \right)_{(i,i+1,i+2)} (1-\hat{n}_{i-1}) (1-\hat{n}_{i+3})\label{eq:Heff_2islands_t3}\\
& + V_1 \sum_i \hat{n}_i \hat{n}_{i+1} - \Delta\sum_i\hat{n}_i  + \mathcal{O} \left(V_2, \frac{\Omega^2V_2}{\Delta^2}, \frac{\Omega^2(V_1-2\Delta)}{\Delta^2}, \frac{\Omega^4}{\Delta^3} \right) \label{eq:Heff_2islands_t4}.
\end{align}
The terms in the Hamiltonian describe the following dynamical processes:
\begin{enumerate}[$\bullet$, labelindent=0pt, leftmargin=*, itemsep=0pt, topsep=0pt, parsep=0pt]
\item Two-island creation, \Cref{eq:Heff_2islands_t1}. The two islands can be created and destroyed in a second-order process, provided that the island's neighboring sites are not excited. On its own, this term describes PXXP dynamics for two-islands, analogous to PXP blockade subspace dynamics; however, in contrast to blockade subspace, this term does not exhaust all leading-order processes.
\item $3+1$ mixed island conversion from the neighboring two islands, \Cref{eq:Heff_2islands_t2}. While the direct creation of mixed islands is a higher-order process, island mixtures can generally be created in a lower-order process by conversion from neighboring islands. The most important contribution stems from two neighboring $2$-islands converting to a $3+1$ island mixture through a $5$-site shuffle, $11011 \rightarrow 11101$. The $5$-site shuffle process is not conditioned on the environment, as long as it operates within allowed resonant configurations and induces further conversions. For example, the $3+2+1$ neighboring island configuration can be converted to a $4+1+1$ island mixture, $11101101 \rightarrow 11110101$; similarly $11110110101 \rightarrow 1111101010$, etc. All higher-order shuffles beyond the $5$-site shuffle vanish in the leading perturbation theory order due to non-locality.
\item One-island hopping, \Cref{eq:Heff_2islands_t3}. While hopping for two-to three- and higher-weight islands vanishes at second order, as expected for a non-local process, one islands are mobile, conditioned on neighboring excitions. This induces exotic physics for $3+1$ and other island mixtures. While $3+1$ islands can be created as a local mixture from two-island conversions, the $1$-island mobility allows for generically non-local excitation mixtures. However, due to the resonance condition, the presence of one part of the $3+1$ island mixture always implies the presence of the second part somewhere in the system.
\item Off-resonant detuning, \Cref{eq:Heff_2islands_t4}. For small deviations of $V_1 - 2\Delta$ from the exact resonance, all allowed classical configurations become detuned, with higher Hamming-weight configurations being proportionally more detuned. 
\end{enumerate}

The effective two-island resonance Hamiltonian thus hosts physics well beyond simple PXP-like two-island creation. In particular, the local interconversion between different island mixtures, the mobility of single-island excitations, and the immobility of higher-weight islands draw a strong analogy to fracton physics~\cite{Nandkishore:2019}, with the potential to generate emergent glassy dynamics~\cite{Chamon:2005, Hirsbrunner:2025}. Moreover, the non-local mixed-island constraints, such as the $3+1$ mixed excitations, are reminiscent of spin-ice physics~\cite{Castelnovo:2012}, where the presence of a monopole necessarily entails a compensating partner. The key difference, however, is that here the constituents display starkly different mobilities, highlighting the fracton-like character of the excitations.

It is important to note that while the processes summarized in the effective Hamiltonian of the resonances are all equally relevant within the subspace, our dynamics, starting from an all-zero state, predominantly probes two-island creation. Indeed, as covered in the preceding section, only the two-island creation processes appear in the leading order of time-dependent perturbation theory, with $3+1$ and higher weight mixtures coming in as higher-order contributions. Experimental data clearly show that most of the two-island resonance weight that we observe originates from isolated two-islands, but the subleading $3+1$ mixture contributions are also explicitly detected in experiments. The details of the effective two-island resonance Hamiltonian could be probed more effectively for carefully chosen initial product states, readily preparable by local detuning~\cite{aquila:2023, Kornjaca:2024}.

\section{Experimental characterization of dynamical regimes in 2D} 
\label{supp:sec:DPTs_2D_exp}

\begin{figure*}[htb]
\centering
\includegraphics[width=1\textwidth]{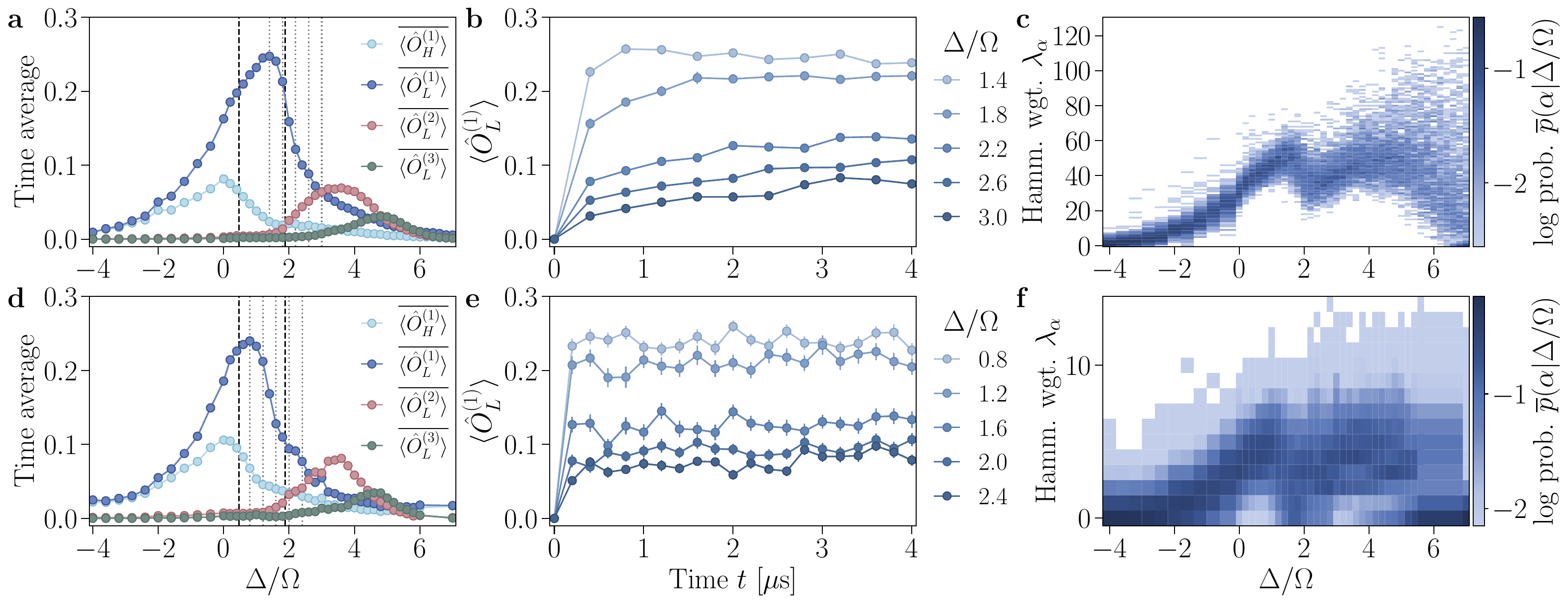}
\caption{
\textbf{System size dependence in 2D square lattice experiments.} Experimental results on the cut $R_\mathrm{b} / a = 1.4$ for \textbf{a}-\textbf{c.} $180$ atoms and \textbf{d}-\textbf{f.} $16$ atoms. The geometries are shown in \Cref{fig:experimental_setup}. 
\textbf{a,d.} The time-averaged expectation values of several $k$-island observables, namely $xyd$ $1$-islands and $xy$ $k$-islands for $k = 1,2,3$. 
\textbf{b,e.} The time evolution of the expectation values of the $xy$ 1-island observable for several values of $\Delta / \Omega$ on the right exit from the central region, shown as dotted vertical lines in \textbf{a} and \textbf{d}, respectively. 
\textbf{c,f.} The time-average $\bar{p} ( \alpha | \Delta / \Omega ) = \frac{1}{N_T} \sum_i p ( \alpha | \Delta / \Omega , t_i )$ of the Hamming weight probability distribution $p(\alpha | \Delta/\Omega , t) = \operatorname{Tr}{(\rho(t) \hat{P}_\alpha)}$. Projectors $\hat{P}_\alpha$ arise from the spectral decomposition of the (unnormalized) Hamming weight operator $\hat{Q}_n \equiv N \hat{O}_n = \sum_\alpha \lambda_\alpha \hat{P}_\alpha$, where $\lambda_\alpha = 0, \hdots, N$ denote the possible Hamming weights. The time average is equivalent to marginalization over $t_i$ with $p (t_i) = \frac{1}{N_t}$.}
\label{fig:system_size_experiments_2d}
\end{figure*}

As described in Sec.~\ref{supp:sec:methods}, we perform experiments on the 2D square lattice with a range of system sizes up to $180$ atoms on the  $R_\mathrm{b} / a = 1.4$ cut. To examine finite-size effects in this geometry, we analyze the size-normalized island observables and the time-averaged unnormalized Hamming weight probability distribution, in analogy with the 1D chain analysis (\Cref{fig:system_size_experiments_1d}). The results are presented in \Cref{fig:system_size_experiments_2d}. 

The 2D system shows notable size-dependent effects, in contrast to the 1D case. 
The features of the $xy$ 1-island observables change with system size (\Cref{fig:system_size_experiments_2d}a,d), as seen and explained in the Results section (\Cref{fig:phases_experiments_2d}b). 
The time series are similar for both system sizes (\Cref{fig:system_size_experiments_2d}b,e). 
The time-averaged Hamming weight probability distribution is much sharper at 180 atoms than 16 (\Cref{fig:system_size_experiments_2d}c,f). 
Notably, the central peak lies in the range $0 \lesssim \Delta / \Omega \lesssim 2$, and at 180 atoms it becomes much sharper and shifts to $\Delta / \Omega \approx 1.9$. 
Indeed, the difference in sharpness of the hamming weight distributions is most pronounced near the conjectured critical point, in the vicinity of the classical resonance with large N{\'e}el cluster, $\Delta / \Omega = 1/4 (R_\mathrm{b} / a )^6$, as shown in \Cref{fig:phases_experiments_2d}b. 
The probability distributions thus change with system size for the same reasons as the $xy$ $1$-island observable discussed in the Results section, namely because larger systems can support larger N{\'e}el-ordered domains, resulting in the faster decay of both the many-body density of states and the relevant matrix elements.

\begin{figure*}[htb]
\centering
\includegraphics[width=1\textwidth]{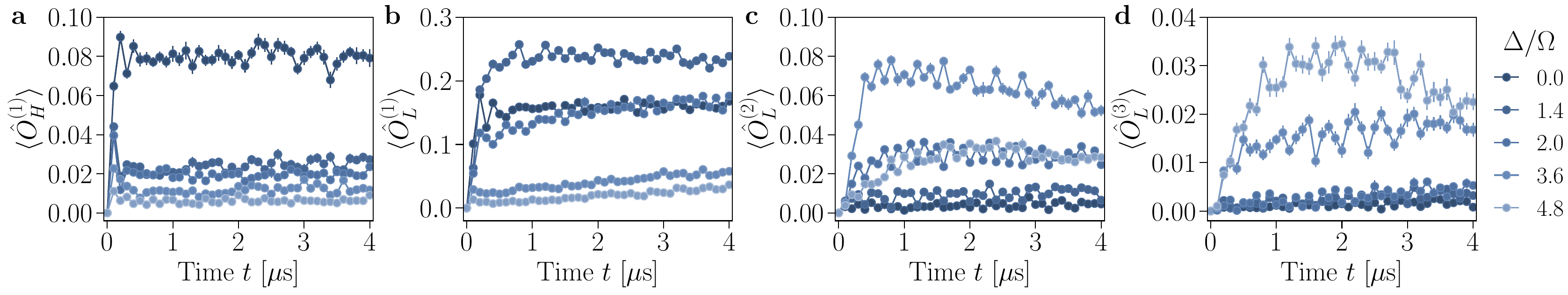}
\caption{
\textbf{High resolution time series for the island observables in the 2D square lattice.} The time evolution of several $k$-island observables, namely \textbf{a.} $xyd$ $1$-islands and \textbf{b}-\textbf{d.} $xy$ $k$-islands with $k = 1,2,3$, for $180$ atoms, on the cut $R_\mathrm{b} / a = 1.4$, and at high time resolution. The quench parameters coincide with the peaks of each of the island operators ($\Delta/\Omega = 0.0, 1.4, 3.6, 4.8$) and the point close to the right exit of the $xy$ 1-island operator region ($\Delta/\Omega = 2.0$), as seen in \Cref{fig:system_size_experiments_2d}a.
}
\label{fig:time_series_2d}
\end{figure*}

Although the time series of local observables can be sampled sparsely to compute reliable time averages, experiments with higher time resolution are required to investigate and identify the presence of finer-scale coherent oscillations. We ran additional experiments on the lattice with $N = 180$ atoms and on the cut $R_\mathrm{b} / a = 1.4$, and examined the more highly resolved time evolution of the island observables for several characteristic points (\Cref{fig:time_series_2d}). In each case, the island observables exhibit rapid oscillations. The oscillations appear to indicate that the amplitudes are largest at the corresponding peaks and diminish away from them, while the frequencies are smallest at the corresponding peaks and increase away from them. However, additional oscillation structure is expected to emerge at higher resolution, where faster oscillations could also be captured. The oscillation amplitude and frequency trends thus qualitatively match the results expected from small numerical simulations (see \Cref{fig:phases_numerics_2d}), despite the experimental noise present in the $N=180$ atom system.

\section{Numerical characterization of dynamical regimes in 2D} 
\label{supp:sec:DPTs_2D_num}

\begin{figure*}[htb]
\centering
\includegraphics[width=1\textwidth]{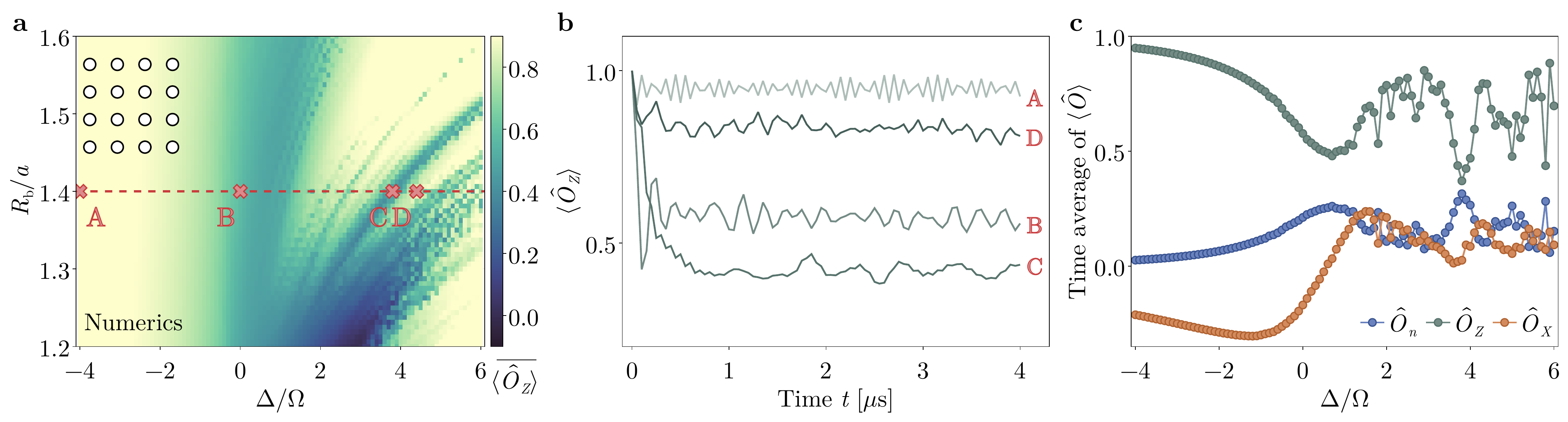}
\caption{
\textbf{One-site observables in numerical simulations for the 2D lattice.} Numerical studies of the dynamical phases for the 2D lattice with $16$ atoms. \textbf{a.} Dynamical phase diagram displaying the time-averaged expectation value of $\hat{O}_Z \equiv \frac{1}{N} \sum_j Z_j$. The atom schematic in the upper left corner shows the geometry. The red dashed horizontal line represents the cut $R_\mathrm{b} / a = 1.4$. \textbf{b.} The time evolution of the expectation values of the one-site observable $\hat{O}_Z$ for characteristic points on the cut.  \textbf{c.} Expectation values of one-site observables, $\hat{O}_Z$, $\hat{O}_n \equiv \frac{1}{N} \sum_j n_j$, and $\hat{O}_X \equiv \frac{1}{N} \sum_j X_j$, averaged over 100 $\mu$s, on the cut. 
}
\label{fig:phases_numerics_2d}
\end{figure*}

We complement the experimental results with numerical simulations of the 2D square lattice using a small system of $N = 16$ atoms, shown in \Cref{fig:phases_numerics_2d}a. The full dynamical phase diagram is constructed from the time-averaged $\hat{O}_Z$ observable. Consistent with experiment, the simulation reveals that the central peak broadens with increasing blockade radius $R_\mathrm{b}/a$, mirroring the broad dynamical response seen in large systems. However, important limitations of the small system are immediately apparent. In particular, the central region of the numerical phase diagram exhibits a series of isolated and discernible resonance peaks, which arise due to finite-size effects. In the large system limit, these features would coalesce into a single smooth central region. Additionally, at small $N$, the right boundary of the central peak is shifted leftward compared to large $N$, due to the limitation on the maximum N{\'e}el cluster size.

The time evolution of $\hat{O}_Z$ expectation values on the experimentally relevant timescales is presented in \Cref{fig:phases_numerics_2d}a. Qualitatively, the dynamics is similar to the 1D dynamics discussed in the main text. At large negative detuning (A), fast low-amplitude oscillations that correspond to the ground state excitation gap are dominant. The mid-central region dynamics (B) exhibit irregular, low-amplitude oscillations. At the central region exit (C), a significant slowdown is observed, signifying the reduction in the number of excitation pathways accessible and the relevance of near-resonant N{\'e}el cluster condition. Dynamics near the resonant islands are similar to those of 1D resonant island dynamics, with more oscillation modes, likely stemming from the enhanced connectivity and the relevance of next-nearest-neighbor interactions. 

The cut at $R_\mathrm{b} / a = 1.4$ is shown in \Cref{fig:phases_numerics_2d}c. Besides the expected main resonances, the observed response includes a large number of irregular weak resonances.  While in 2D, the enhanced spatial connectivity and next-nearest-neighbor interaction effects indeed permit a broader set of distinct resonances, a substantial fraction of the observed resonances originates from boundary effects that are relevant only for small systems.

Taken together, these observations underscore a fundamental contrast between 1D and 2D: while small system simulations in 1D capture the essential features of the dynamics and align well with experimental data, the $N=16$ numerics in 2D are qualitatively inadequate to describe the large-scale experimental response. Key differences include the emergence of resonance peaks in the central region, the shift of the central region exit edge, and the appearance of boundary-driven subleading resonances in the right region of the phase diagram. Although useful for identifying general mechanisms -- such as the role of island excitations in each region -- the small-system simulations fail to accurately reproduce the collective features that emerge only in large-scale 2D systems.

\section{Classical Stoner-Wohlfarth transition}
\label{supp:sec:classical_spin_models}

Stoner and collaborators, in 1948, studied the classical mean-field dynamics of collective magnetization, seeking a microscopic description of the hysteretic behavior of ferromagnets. The appearance of hysteresis in classical magnets followed from the non-analyticity in the magnetization response as a function of the applied magnetic field -- the Stoner-Wohlfarth transition~\cite{Stoner:1948}. Phenomenologically, under a high drive field, the quenched spin will precess close to the axis defined by the applied field for long periods of time; as one reduces the external field strength, a sudden change of behavior (switch) is observed and characterized by a sharp increase and non-analytic behavior of the transversal components of the magnetic susceptibility. At those values of the drive field and just below, the magnetization freely precesses, covering a large portion of its parameter space. As the longitudinal field continues decreasing, the dynamics slowly resettles to small precessions around the field via a gradual crossover. The appearance of hysteresis related to this first-order transition requires a critical value of the effective magnetic anisotropy, with the Stoner-Wohlfarth transition line ending in the critical point at the critical anisotropy.

The Stoner-Wohlfarth dynamical setting has an analog in our quantum quench experiments, and thus motivates us to map our system to equivalent classical collective dynamics. We shift the Rydberg density operators $\hat{n}_i$ in \Cref{eq:hamiltonian} into spin operators $\hat{S}_{z,i}$ and take the classical vector limit of the spins ($S \rightarrow \infty$). The zero-momentum mode, i.e., the total magnetization $M_a = \frac{1}{N}\sum_i S_{a,i}$, is relevant for the all-zero (down) initial state condition, and its dynamics effectively decouples from the rest of the momentum model, giving rise to mean-field magnetization equations of motion:
\begin{equation}
\frac{d\vec{M}}{dt} = \left\lbrace \Omega \, \hat{x} + \left[-\Delta + \frac{1}{2} \left( \sum_{i} z_i V_i \right) (1 + M_z) \right] \hat{z} \right\rbrace \times \vec{M} \, . 
\label{eq:classical_eom}
\end{equation}
Here, $V_i$ are the $i$th neighbor interactions and $z_i$ are the corresponding coordination numbers, each depending on the underlying lattice. Explicitly, $z_i = 2$ for the 1D chain, while $z_i = 4$ for the 2D square lattice. The Rabi frequency plays the role of a transverse drive field; the longitudinal drive is determined by the detuning reduced by the residual interaction strength, while the non-linear interaction-related term represents an effective intrinsic anisotropy.

\begin{figure*}[htb]
\centering
\includegraphics[width=1\textwidth]{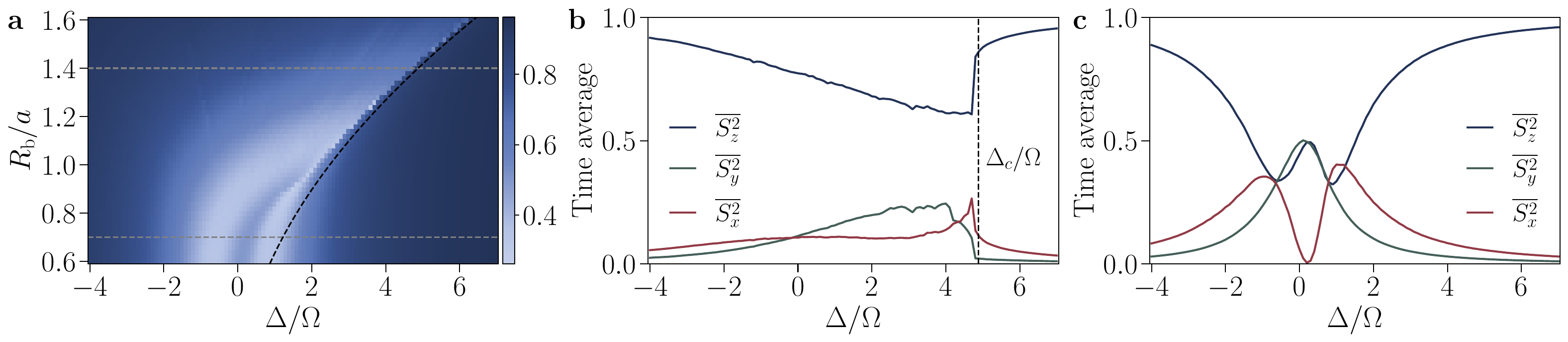}
\caption{
\textbf{Dynamical regimes of the classical spin model for the 2D lattice.} The classical spin dynamics model is described in the Supplementary Information, where the 2D square lattice has coordination numbers $z_1 = z_2 = z_3 = 4$. \textbf{a.} Dynamical phase diagram in terms of the long-time averages of the (squared) spin component $S_z^2$. Panels \textbf{b},\textbf{c} show the three (squared) spin components on \textbf{b.} the cut $R_\mathrm{b} / a = 1.4$ across the phase transition and \textbf{c.} the cut $R_\mathrm{b} / a = 0.7$ across the crossover, each indicated by a grey dashed horizontal line in panel \textbf{a}. The black dashed lines in panels \textbf{a},\textbf{b} show the phase transition from the approximate analytic solution of the Stoner-Wohlfarth astroid.
}
\label{fig:phases_classical_2d}
\end{figure*}

We initialize the state at $\vec{M}(0) = -\hat{z}$, integrate the coupled dynamical equations numerically, and calculate the classical analogs of the relevant local observables $\langle S_a \equiv O_{a}\rangle$, $\langle S_a \rangle^2 \equiv O_{aa} $, $2^{-5}\langle (1-S_z)^4S_z \rangle  \equiv \langle O_L^{(1)}\rangle$ (in 2D), etc., across the Hamiltonian parameter space. \Cref{fig:phases_classical_2d}a showcases the classical dynamical phase diagram based on the $\langle S_z^2\rangle$ observable as a function of detuning and interaction strength for the 2D lattice, i.e., $z_i = 4$. The classical Stoner-Wohlfarth transition is explicit at high detuning fields, as anticipated. The transition eventually ends in a critical point once the intrinsic anisotropy determined by the interaction strength becomes smaller than a critical value. The cut presented in \Cref{fig:phases_classical_2d}b shows a clear non-analyticity in the $\langle S_a^2\rangle$ system response for sufficiently strong detuning, and \Cref{fig:phases_classical_2d}c shows a smooth crossover. 

The critical point for the Stoner-Wohlfarth transition can be obtained analytically. Packaging the different components of \Cref{eq:classical_eom} into generalized variables $h_x$, $h_z$, and $K$ (for the transversal and longitudinal drives, and anisotropy), we evaluate the long-term classical dynamics via the stationary condition for $M_y$,
\begin{equation}
-h_x \cos\theta - h_z \sin\theta + K \sin\theta \cos\theta = 0 \, .
\end{equation}
Let
\begin{equation}
F(\theta; h_x, h_z, V ) := -h_x \cos\theta - h_z \sin\theta + K \sin\theta \cos\theta \, .
\end{equation}
A local analytic branch $\theta (h_x, h_z)$ exists wherever the implicit-function invertibility condition holds, given by
\begin{equation}
\partial_\theta F(\theta; h_x,h_z,V ) \neq 0 \, .
\end{equation}
Therefore, non-analyticity can only occur at parameter values for which
\begin{equation}
F(\theta; h_x, h_z, V ) = 0 \, , \, \partial_\theta F(\theta; h_x, h_z, V ) = 0 \, .
\label{eq:nonanalyticity_conditions}
\end{equation}
Solving the resulting linear system, one obtains
\begin{equation}
h_x = V \sin^3 \theta \, , \, h_z = V \cos^3 \theta \, .
\end{equation}
Eliminating $\theta$ now directly yields the Stoner-Wohlfarth astroid:
\begin{equation}
|\Omega|^{2/3} + \left| \Delta - \sum_i z_i V_i \right|^{2/3} = \left|\sum_i z_i V_i \right|^{2/3} \, .
\label{eq:SW_astroid}
\end{equation}

When $V_i \gg \Omega$, the approximate non-analyticity position (taking the lower branch) is
\begin{equation}
\frac{\Delta_c}{\Omega} \approx \frac{3}{2} \left( \frac{\sum_i z_i V_i}{\Omega} \right)^{1/3} \, .
\label{eq:SW_approximate_solution}
\end{equation}
This expression returns a good agreement with the numerical results in Fig.~\ref{fig:phases_experiments_2d}d and \ref{fig:phases_classical_2d}a,b, slightly overshooting the non-analytic point due to corrections of the order $R_\mathrm{b}^{-2}$. Importantly, it demonstrates a classical scaling of the phase boundary as $\propto (V_1/\Omega)^{1/3} = (R_\mathrm{b}/a)^2$, in contrast with the quantum scaling of $\propto (R_\mathrm{b}/a)^6$, firmly establishing that our quantum phase diagram, while sharing phenomenologies with the Stoner-Wohlfarth dynamics, is independent of the classical limit. 

The non-analyticity itself turns into a crossover for interaction strengths weaker than $\sum_i z_i V_i = \Omega$, as expected from the finite anisotropy threshold for hysteresis. It is likely that the quantum prethermal analog of the Stoner-Wohlfarth transition also becomes a smooth crossover for low interaction strengths. This is, indeed, tentatively supported by the experimental data presented in \Cref{fig:phases_experiments_2d} of the main text, where the sharpness of the central region exit edge is significantly reduced at low $R_\mathrm{b}/a$. A possible crossover scenario at low $R_\mathrm{b}/a$ that is analogous, but not equivalent, to the classical one is further supported by the eventual failure of Floquet-like prethermalization near $V_1 \sim \Omega$. When the prethermal plateau is lost, no thermal transition is accessible, ultimately resulting in a crossover.



\begin{thebibliography}{78}%
\makeatletter
\providecommand \@ifxundefined [1]{%
 \@ifx{#1\undefined}
}%
\providecommand \@ifnum [1]{%
 \ifnum #1\expandafter \@firstoftwo
 \else \expandafter \@secondoftwo
 \fi
}%
\providecommand \@ifx [1]{%
 \ifx #1\expandafter \@firstoftwo
 \else \expandafter \@secondoftwo
 \fi
}%
\providecommand \natexlab [1]{#1}%
\providecommand \enquote  [1]{``#1''}%
\providecommand \bibnamefont  [1]{#1}%
\providecommand \bibfnamefont [1]{#1}%
\providecommand \citenamefont [1]{#1}%
\providecommand \href@noop [0]{\@secondoftwo}%
\providecommand \href [0]{\begingroup \@sanitize@url \@href}%
\providecommand \@href[1]{\@@startlink{#1}\@@href}%
\providecommand \@@href[1]{\endgroup#1\@@endlink}%
\providecommand \@sanitize@url [0]{\catcode `\\12\catcode `\$12\catcode `\&12\catcode `\#12\catcode `\^12\catcode `\_12\catcode `\%12\relax}%
\providecommand \@@startlink[1]{}%
\providecommand \@@endlink[0]{}%
\providecommand \url  [0]{\begingroup\@sanitize@url \@url }%
\providecommand \@url [1]{\endgroup\@href {#1}{\urlprefix }}%
\providecommand \urlprefix  [0]{URL }%
\providecommand \Eprint [0]{\href }%
\providecommand \doibase [0]{https://doi.org/}%
\providecommand \selectlanguage [0]{\@gobble}%
\providecommand \bibinfo  [0]{\@secondoftwo}%
\providecommand \bibfield  [0]{\@secondoftwo}%
\providecommand \translation [1]{[#1]}%
\providecommand \BibitemOpen [0]{}%
\providecommand \bibitemStop [0]{}%
\providecommand \bibitemNoStop [0]{.\EOS\space}%
\providecommand \EOS [0]{\spacefactor3000\relax}%
\providecommand \BibitemShut  [1]{\csname bibitem#1\endcsname}%
\let\auto@bib@innerbib\@empty
\bibitem [{\citenamefont {Polkovnikov}\ \emph {et~al.}(2011)\citenamefont {Polkovnikov}, \citenamefont {Sengupta}, \citenamefont {Silva},\ and\ \citenamefont {Vengalattore}}]{Polkovnikov:2011}%
  \BibitemOpen
  \bibfield  {author} {\bibinfo {author} {\bibfnamefont {A.}~\bibnamefont {Polkovnikov}}, \bibinfo {author} {\bibfnamefont {K.}~\bibnamefont {Sengupta}}, \bibinfo {author} {\bibfnamefont {A.}~\bibnamefont {Silva}},\ and\ \bibinfo {author} {\bibfnamefont {M.}~\bibnamefont {Vengalattore}},\ }\bibfield  {title} {\bibinfo {title} {Colloquium: Nonequilibrium dynamics of closed interacting quantum systems},\ }\href {https://doi.org/10.1103/RevModPhys.83.863} {\bibfield  {journal} {\bibinfo  {journal} {Rev. Mod. Phys.}\ }\textbf {\bibinfo {volume} {83}},\ \bibinfo {pages} {863} (\bibinfo {year} {2011})}\BibitemShut {NoStop}%
\bibitem [{\citenamefont {Mitra}(2018)}]{Mitra:2018}%
  \BibitemOpen
  \bibfield  {author} {\bibinfo {author} {\bibfnamefont {A.}~\bibnamefont {Mitra}},\ }\bibfield  {title} {\bibinfo {title} {Quantum quench dynamics},\ }\href {https://doi.org/https://doi.org/10.1146/annurev-conmatphys-031016-025451} {\bibfield  {journal} {\bibinfo  {journal} {Annual Review of Condensed Matter Physics}\ }\textbf {\bibinfo {volume} {9}},\ \bibinfo {pages} {245} (\bibinfo {year} {2018})}\BibitemShut {NoStop}%
\bibitem [{\citenamefont {Rudner}\ and\ \citenamefont {Lindner}(2020)}]{Rudner:2020}%
  \BibitemOpen
  \bibfield  {author} {\bibinfo {author} {\bibfnamefont {M.~S.}\ \bibnamefont {Rudner}}\ and\ \bibinfo {author} {\bibfnamefont {N.~H.}\ \bibnamefont {Lindner}},\ }\bibfield  {title} {\bibinfo {title} {Band structure engineering and non-equilibrium dynamics in floquet topological insulators},\ }\href {https://doi.org/10.1038/s42254-020-0170-z} {\bibfield  {journal} {\bibinfo  {journal} {Nature Reviews Physics}\ }\textbf {\bibinfo {volume} {2}},\ \bibinfo {pages} {229} (\bibinfo {year} {2020})}\BibitemShut {NoStop}%
\bibitem [{\citenamefont {Vasseur}\ and\ \citenamefont {Moore}(2016)}]{Vasseur:2016}%
  \BibitemOpen
  \bibfield  {author} {\bibinfo {author} {\bibfnamefont {R.}~\bibnamefont {Vasseur}}\ and\ \bibinfo {author} {\bibfnamefont {J.~E.}\ \bibnamefont {Moore}},\ }\bibfield  {title} {\bibinfo {title} {Nonequilibrium quantum dynamics and transport: from integrability to many-body localization},\ }\href {https://doi.org/10.1088/1742-5468/2016/06/064010} {\bibfield  {journal} {\bibinfo  {journal} {Journal of Statistical Mechanics: Theory and Experiment}\ }\textbf {\bibinfo {volume} {2016}},\ \bibinfo {pages} {064010} (\bibinfo {year} {2016})}\BibitemShut {NoStop}%
\bibitem [{\citenamefont {Eisert}\ \emph {et~al.}(2015)\citenamefont {Eisert}, \citenamefont {Friesdorf},\ and\ \citenamefont {Gogolin}}]{Eisert:2015}%
  \BibitemOpen
  \bibfield  {author} {\bibinfo {author} {\bibfnamefont {J.}~\bibnamefont {Eisert}}, \bibinfo {author} {\bibfnamefont {M.}~\bibnamefont {Friesdorf}},\ and\ \bibinfo {author} {\bibfnamefont {C.}~\bibnamefont {Gogolin}},\ }\bibfield  {title} {\bibinfo {title} {Quantum many-body systems out of equilibrium},\ }\href {https://doi.org/10.1038/nphys3215} {\bibfield  {journal} {\bibinfo  {journal} {Nature Physics}\ }\textbf {\bibinfo {volume} {11}},\ \bibinfo {pages} {124} (\bibinfo {year} {2015})}\BibitemShut {NoStop}%
\bibitem [{\citenamefont {Rigol}\ \emph {et~al.}(2008)\citenamefont {Rigol}, \citenamefont {Dunjko},\ and\ \citenamefont {Olshanii}}]{Rigol:2008}%
  \BibitemOpen
  \bibfield  {author} {\bibinfo {author} {\bibfnamefont {M.}~\bibnamefont {Rigol}}, \bibinfo {author} {\bibfnamefont {V.}~\bibnamefont {Dunjko}},\ and\ \bibinfo {author} {\bibfnamefont {M.}~\bibnamefont {Olshanii}},\ }\bibfield  {title} {\bibinfo {title} {Thermalization and its mechanism for generic isolated quantum systems},\ }\href {https://doi.org/10.1038/nature06838} {\bibfield  {journal} {\bibinfo  {journal} {Nature}\ }\textbf {\bibinfo {volume} {452}},\ \bibinfo {pages} {854} (\bibinfo {year} {2008})}\BibitemShut {NoStop}%
\bibitem [{\citenamefont {Srednicki}(1994)}]{Srednicki:1994}%
  \BibitemOpen
  \bibfield  {author} {\bibinfo {author} {\bibfnamefont {M.}~\bibnamefont {Srednicki}},\ }\bibfield  {title} {\bibinfo {title} {Chaos and quantum thermalization},\ }\href {https://doi.org/10.1103/PhysRevE.50.888} {\bibfield  {journal} {\bibinfo  {journal} {Phys. Rev. E}\ }\textbf {\bibinfo {volume} {50}},\ \bibinfo {pages} {888} (\bibinfo {year} {1994})}\BibitemShut {NoStop}%
\bibitem [{\citenamefont {Deutsch}(2018)}]{Deutsch:2018}%
  \BibitemOpen
  \bibfield  {author} {\bibinfo {author} {\bibfnamefont {J.~M.}\ \bibnamefont {Deutsch}},\ }\bibfield  {title} {\bibinfo {title} {Eigenstate thermalization hypothesis},\ }\href {https://doi.org/10.1088/1361-6633/aac9f1} {\bibfield  {journal} {\bibinfo  {journal} {Reports on Progress in Physics}\ }\textbf {\bibinfo {volume} {81}},\ \bibinfo {pages} {082001} (\bibinfo {year} {2018})}\BibitemShut {NoStop}%
\bibitem [{\citenamefont {Abanin}\ \emph {et~al.}(2019)\citenamefont {Abanin}, \citenamefont {Altman}, \citenamefont {Bloch},\ and\ \citenamefont {Serbyn}}]{Abanin:2019}%
  \BibitemOpen
  \bibfield  {author} {\bibinfo {author} {\bibfnamefont {D.~A.}\ \bibnamefont {Abanin}}, \bibinfo {author} {\bibfnamefont {E.}~\bibnamefont {Altman}}, \bibinfo {author} {\bibfnamefont {I.}~\bibnamefont {Bloch}},\ and\ \bibinfo {author} {\bibfnamefont {M.}~\bibnamefont {Serbyn}},\ }\bibfield  {title} {\bibinfo {title} {Colloquium: Many-body localization, thermalization, and entanglement},\ }\href {https://doi.org/10.1103/RevModPhys.91.021001} {\bibfield  {journal} {\bibinfo  {journal} {Rev. Mod. Phys.}\ }\textbf {\bibinfo {volume} {91}},\ \bibinfo {pages} {021001} (\bibinfo {year} {2019})}\BibitemShut {NoStop}%
\bibitem [{\citenamefont {Heyl}(2018)}]{Heyl:2018}%
  \BibitemOpen
  \bibfield  {author} {\bibinfo {author} {\bibfnamefont {M.}~\bibnamefont {Heyl}},\ }\bibfield  {title} {\bibinfo {title} {Dynamical quantum phase transitions: a review},\ }\href {https://doi.org/10.1088/1361-6633/aaaf9a} {\bibfield  {journal} {\bibinfo  {journal} {Reports on Progress in Physics}\ }\textbf {\bibinfo {volume} {81}},\ \bibinfo {pages} {054001} (\bibinfo {year} {2018})}\BibitemShut {NoStop}%
\bibitem [{\citenamefont {Papi{\'{c}}}(2022)}]{Papic:2022}%
  \BibitemOpen
  \bibfield  {author} {\bibinfo {author} {\bibfnamefont {Z.}~\bibnamefont {Papi{\'{c}}}},\ }\bibinfo {title} {Weak ergodicity breaking through the lens of quantum entanglement},\ in\ \href {https://doi.org/10.1007/978-3-031-03998-0_13} {\emph {\bibinfo {booktitle} {Entanglement in Spin Chains: From Theory to Quantum Technology Applications}}},\ \bibinfo {editor} {edited by\ \bibinfo {editor} {\bibfnamefont {A.}~\bibnamefont {Bayat}}, \bibinfo {editor} {\bibfnamefont {S.}~\bibnamefont {Bose}},\ and\ \bibinfo {editor} {\bibfnamefont {H.}~\bibnamefont {Johannesson}}}\ (\bibinfo  {publisher} {Springer International Publishing},\ \bibinfo {address} {Cham},\ \bibinfo {year} {2022})\ pp.\ \bibinfo {pages} {341--395}\BibitemShut {NoStop}%
\bibitem [{\citenamefont {Landau}(1937)}]{Landau:1937obd}%
  \BibitemOpen
  \bibfield  {author} {\bibinfo {author} {\bibfnamefont {L.~D.}\ \bibnamefont {Landau}},\ }\bibfield  {title} {\bibinfo {title} {{On the theory of phase transitions}},\ }\href {https://doi.org/10.1016/B978-0-08-010586-4.50034-1} {\bibfield  {journal} {\bibinfo  {journal} {Zh. Eksp. Teor. Fiz.}\ }\textbf {\bibinfo {volume} {7}},\ \bibinfo {pages} {19} (\bibinfo {year} {1937})}\BibitemShut {NoStop}%
\bibitem [{\citenamefont {Sachdev}(2011)}]{Sachdev:2011}%
  \BibitemOpen
  \bibfield  {author} {\bibinfo {author} {\bibfnamefont {S.}~\bibnamefont {Sachdev}},\ }\href {https://doi.org/10.1017/CBO9780511973765} {\emph {\bibinfo {title} {Quantum Phase Transitions}}},\ \bibinfo {edition} {2nd}\ ed.\ (\bibinfo  {publisher} {Cambridge University Press},\ \bibinfo {year} {2011})\BibitemShut {NoStop}%
\bibitem [{\citenamefont {Ebadi}\ \emph {et~al.}(2021)\citenamefont {Ebadi}, \citenamefont {Wang}, \citenamefont {Levine}, \citenamefont {Keesling}, \citenamefont {Semeghini}, \citenamefont {Omran}, \citenamefont {Bluvstein}, \citenamefont {Samajdar}, \citenamefont {Pichler}, \citenamefont {Ho}, \citenamefont {Choi}, \citenamefont {Sachdev}, \citenamefont {Greiner}, \citenamefont {Vuleti{\'{c}}},\ and\ \citenamefont {Lukin}}]{Ebadi:2021}%
  \BibitemOpen
  \bibfield  {author} {\bibinfo {author} {\bibfnamefont {S.}~\bibnamefont {Ebadi}}, \bibinfo {author} {\bibfnamefont {T.~T.}\ \bibnamefont {Wang}}, \bibinfo {author} {\bibfnamefont {H.}~\bibnamefont {Levine}}, \bibinfo {author} {\bibfnamefont {A.}~\bibnamefont {Keesling}}, \bibinfo {author} {\bibfnamefont {G.}~\bibnamefont {Semeghini}}, \bibinfo {author} {\bibfnamefont {A.}~\bibnamefont {Omran}}, \bibinfo {author} {\bibfnamefont {D.}~\bibnamefont {Bluvstein}}, \bibinfo {author} {\bibfnamefont {R.}~\bibnamefont {Samajdar}}, \bibinfo {author} {\bibfnamefont {H.}~\bibnamefont {Pichler}}, \bibinfo {author} {\bibfnamefont {W.~W.}\ \bibnamefont {Ho}}, \bibinfo {author} {\bibfnamefont {S.}~\bibnamefont {Choi}}, \bibinfo {author} {\bibfnamefont {S.}~\bibnamefont {Sachdev}}, \bibinfo {author} {\bibfnamefont {M.}~\bibnamefont {Greiner}}, \bibinfo {author} {\bibfnamefont {V.}~\bibnamefont {Vuleti{\'{c}}}},\ and\ \bibinfo {author} {\bibfnamefont {M.~D.}\ \bibnamefont {Lukin}},\ }\bibfield  {title} {\bibinfo {title}
  {Quantum phases of matter on a 256-atom programmable quantum simulator},\ }\href {https://doi.org/10.1038/s41586-021-03582-4} {\bibfield  {journal} {\bibinfo  {journal} {Nature}\ }\textbf {\bibinfo {volume} {595}},\ \bibinfo {pages} {227} (\bibinfo {year} {2021})}\BibitemShut {NoStop}%
\bibitem [{\citenamefont {Greiner}\ \emph {et~al.}(2002)\citenamefont {Greiner}, \citenamefont {Mandel}, \citenamefont {Esslinger}, \citenamefont {H{\"a}nsch},\ and\ \citenamefont {Bloch}}]{Greiner:2002}%
  \BibitemOpen
  \bibfield  {author} {\bibinfo {author} {\bibfnamefont {M.}~\bibnamefont {Greiner}}, \bibinfo {author} {\bibfnamefont {O.}~\bibnamefont {Mandel}}, \bibinfo {author} {\bibfnamefont {T.}~\bibnamefont {Esslinger}}, \bibinfo {author} {\bibfnamefont {T.~W.}\ \bibnamefont {H{\"a}nsch}},\ and\ \bibinfo {author} {\bibfnamefont {I.}~\bibnamefont {Bloch}},\ }\bibfield  {title} {\bibinfo {title} {Quantum phase transition from a superfluid to a mott insulator in a gas of ultracold atoms},\ }\href {https://doi.org/10.1038/415039a} {\bibfield  {journal} {\bibinfo  {journal} {Nature}\ }\textbf {\bibinfo {volume} {415}},\ \bibinfo {pages} {39} (\bibinfo {year} {2002})}\BibitemShut {NoStop}%
\bibitem [{\citenamefont {Sengupta}\ \emph {et~al.}(2004)\citenamefont {Sengupta}, \citenamefont {Powell},\ and\ \citenamefont {Sachdev}}]{Sengupta:2004}%
  \BibitemOpen
  \bibfield  {author} {\bibinfo {author} {\bibfnamefont {K.}~\bibnamefont {Sengupta}}, \bibinfo {author} {\bibfnamefont {S.}~\bibnamefont {Powell}},\ and\ \bibinfo {author} {\bibfnamefont {S.}~\bibnamefont {Sachdev}},\ }\bibfield  {title} {\bibinfo {title} {Quench dynamics across quantum critical points},\ }\href {https://doi.org/10.1103/PhysRevA.69.053616} {\bibfield  {journal} {\bibinfo  {journal} {Phys. Rev. A}\ }\textbf {\bibinfo {volume} {69}},\ \bibinfo {pages} {053616} (\bibinfo {year} {2004})}\BibitemShut {NoStop}%
\bibitem [{\citenamefont {Verstraete}\ and\ \citenamefont {Cirac}(2004)}]{Verstraete:2004}%
  \BibitemOpen
  \bibfield  {author} {\bibinfo {author} {\bibfnamefont {F.}~\bibnamefont {Verstraete}}\ and\ \bibinfo {author} {\bibfnamefont {J.~I.}\ \bibnamefont {Cirac}},\ }\href {https://arxiv.org/abs/cond-mat/0407066} {\bibinfo {title} {Renormalization algorithms for quantum-many body systems in two and higher dimensions}} (\bibinfo {year} {2004}),\ \Eprint {https://arxiv.org/abs/cond-mat/0407066} {arXiv:cond-mat/0407066 [cond-mat.str-el]} \BibitemShut {NoStop}%
\bibitem [{\citenamefont {Cazalilla}(2006)}]{Cazalilla:2006}%
  \BibitemOpen
  \bibfield  {author} {\bibinfo {author} {\bibfnamefont {M.~A.}\ \bibnamefont {Cazalilla}},\ }\bibfield  {title} {\bibinfo {title} {Effect of suddenly turning on interactions in the luttinger model},\ }\href {https://doi.org/10.1103/PhysRevLett.97.156403} {\bibfield  {journal} {\bibinfo  {journal} {Phys. Rev. Lett.}\ }\textbf {\bibinfo {volume} {97}},\ \bibinfo {pages} {156403} (\bibinfo {year} {2006})}\BibitemShut {NoStop}%
\bibitem [{\citenamefont {Schollwöck}(2011)}]{Schollwock:2011}%
  \BibitemOpen
  \bibfield  {author} {\bibinfo {author} {\bibfnamefont {U.}~\bibnamefont {Schollwöck}},\ }\bibfield  {title} {\bibinfo {title} {The density-matrix renormalization group in the age of matrix product states},\ }\href {https://doi.org/https://doi.org/10.1016/j.aop.2010.09.012} {\bibfield  {journal} {\bibinfo  {journal} {Annals of Physics}\ }\textbf {\bibinfo {volume} {326}},\ \bibinfo {pages} {96} (\bibinfo {year} {2011})},\ \bibinfo {note} {january 2011 Special Issue}\BibitemShut {NoStop}%
\bibitem [{\citenamefont {Heyl}\ \emph {et~al.}(2013)\citenamefont {Heyl}, \citenamefont {Polkovnikov},\ and\ \citenamefont {Kehrein}}]{Heyl:2013}%
  \BibitemOpen
  \bibfield  {author} {\bibinfo {author} {\bibfnamefont {M.}~\bibnamefont {Heyl}}, \bibinfo {author} {\bibfnamefont {A.}~\bibnamefont {Polkovnikov}},\ and\ \bibinfo {author} {\bibfnamefont {S.}~\bibnamefont {Kehrein}},\ }\bibfield  {title} {\bibinfo {title} {Dynamical quantum phase transitions in the transverse-field ising model},\ }\href {https://doi.org/10.1103/PhysRevLett.110.135704} {\bibfield  {journal} {\bibinfo  {journal} {Phys. Rev. Lett.}\ }\textbf {\bibinfo {volume} {110}},\ \bibinfo {pages} {135704} (\bibinfo {year} {2013})}\BibitemShut {NoStop}%
\bibitem [{\citenamefont {Titum}\ \emph {et~al.}(2019)\citenamefont {Titum}, \citenamefont {Iosue}, \citenamefont {Garrison}, \citenamefont {Gorshkov},\ and\ \citenamefont {Gong}}]{Titum:2019}%
  \BibitemOpen
  \bibfield  {author} {\bibinfo {author} {\bibfnamefont {P.}~\bibnamefont {Titum}}, \bibinfo {author} {\bibfnamefont {J.~T.}\ \bibnamefont {Iosue}}, \bibinfo {author} {\bibfnamefont {J.~R.}\ \bibnamefont {Garrison}}, \bibinfo {author} {\bibfnamefont {A.~V.}\ \bibnamefont {Gorshkov}},\ and\ \bibinfo {author} {\bibfnamefont {Z.-X.}\ \bibnamefont {Gong}},\ }\bibfield  {title} {\bibinfo {title} {Probing ground-state phase transitions through quench dynamics},\ }\href {https://doi.org/10.1103/PhysRevLett.123.115701} {\bibfield  {journal} {\bibinfo  {journal} {Phys. Rev. Lett.}\ }\textbf {\bibinfo {volume} {123}},\ \bibinfo {pages} {115701} (\bibinfo {year} {2019})}\BibitemShut {NoStop}%
\bibitem [{\citenamefont {Titum}\ and\ \citenamefont {Maghrebi}(2020)}]{Titum:2020}%
  \BibitemOpen
  \bibfield  {author} {\bibinfo {author} {\bibfnamefont {P.}~\bibnamefont {Titum}}\ and\ \bibinfo {author} {\bibfnamefont {M.~F.}\ \bibnamefont {Maghrebi}},\ }\bibfield  {title} {\bibinfo {title} {Nonequilibrium criticality in quench dynamics of long-range spin models},\ }\href {https://doi.org/10.1103/PhysRevLett.125.040602} {\bibfield  {journal} {\bibinfo  {journal} {Phys. Rev. Lett.}\ }\textbf {\bibinfo {volume} {125}},\ \bibinfo {pages} {040602} (\bibinfo {year} {2020})}\BibitemShut {NoStop}%
\bibitem [{\citenamefont {Lin}\ \emph {et~al.}(2022)\citenamefont {Lin}, \citenamefont {Zaletel},\ and\ \citenamefont {Pollmann}}]{Lin:2022}%
  \BibitemOpen
  \bibfield  {author} {\bibinfo {author} {\bibfnamefont {S.-H.}\ \bibnamefont {Lin}}, \bibinfo {author} {\bibfnamefont {M.~P.}\ \bibnamefont {Zaletel}},\ and\ \bibinfo {author} {\bibfnamefont {F.}~\bibnamefont {Pollmann}},\ }\bibfield  {title} {\bibinfo {title} {Efficient simulation of dynamics in two-dimensional quantum spin systems with isometric tensor networks},\ }\href {https://doi.org/10.1103/PhysRevB.106.245102} {\bibfield  {journal} {\bibinfo  {journal} {Phys. Rev. B}\ }\textbf {\bibinfo {volume} {106}},\ \bibinfo {pages} {245102} (\bibinfo {year} {2022})}\BibitemShut {NoStop}%
\bibitem [{\citenamefont {Begu\ifmmode \check{s}\else \v{s}\fi{}i\ifmmode~\acute{c}\else \'{c}\fi{}}\ and\ \citenamefont {Chan}(2025)}]{Begusic:2025}%
  \BibitemOpen
  \bibfield  {author} {\bibinfo {author} {\bibfnamefont {T.}~\bibnamefont {Begu\ifmmode \check{s}\else \v{s}\fi{}i\ifmmode~\acute{c}\else \'{c}\fi{}}}\ and\ \bibinfo {author} {\bibfnamefont {G.~K.-L.}\ \bibnamefont {Chan}},\ }\bibfield  {title} {\bibinfo {title} {Real-time operator evolution in two and three dimensions via sparse pauli dynamics},\ }\href {https://doi.org/10.1103/PRXQuantum.6.020302} {\bibfield  {journal} {\bibinfo  {journal} {PRX Quantum}\ }\textbf {\bibinfo {volume} {6}},\ \bibinfo {pages} {020302} (\bibinfo {year} {2025})}\BibitemShut {NoStop}%
\bibitem [{\citenamefont {Park}\ \emph {et~al.}(2025)\citenamefont {Park}, \citenamefont {Gray},\ and\ \citenamefont {Chan}}]{Park:2025}%
  \BibitemOpen
  \bibfield  {author} {\bibinfo {author} {\bibfnamefont {G.}~\bibnamefont {Park}}, \bibinfo {author} {\bibfnamefont {J.}~\bibnamefont {Gray}},\ and\ \bibinfo {author} {\bibfnamefont {G.~K.-L.}\ \bibnamefont {Chan}},\ }\href {https://arxiv.org/abs/2504.07344} {\bibinfo {title} {Simulating quantum dynamics in two-dimensional lattices with tensor network influence functional belief propagation}} (\bibinfo {year} {2025}),\ \Eprint {https://arxiv.org/abs/2504.07344} {arXiv:2504.07344 [quant-ph]} \BibitemShut {NoStop}%
\bibitem [{\citenamefont {Karrasch}\ and\ \citenamefont {Schuricht}(2013)}]{Karrasch:2013}%
  \BibitemOpen
  \bibfield  {author} {\bibinfo {author} {\bibfnamefont {C.}~\bibnamefont {Karrasch}}\ and\ \bibinfo {author} {\bibfnamefont {D.}~\bibnamefont {Schuricht}},\ }\bibfield  {title} {\bibinfo {title} {Dynamical phase transitions after quenches in nonintegrable models},\ }\href {https://doi.org/10.1103/PhysRevB.87.195104} {\bibfield  {journal} {\bibinfo  {journal} {Phys. Rev. B}\ }\textbf {\bibinfo {volume} {87}},\ \bibinfo {pages} {195104} (\bibinfo {year} {2013})}\BibitemShut {NoStop}%
\bibitem [{\citenamefont {Zvyagin}(2016)}]{Zvyagin:2016}%
  \BibitemOpen
  \bibfield  {author} {\bibinfo {author} {\bibfnamefont {A.~A.}\ \bibnamefont {Zvyagin}},\ }\bibfield  {title} {\bibinfo {title} {Dynamical quantum phase transitions (review article)},\ }\href {https://doi.org/10.1063/1.4969869} {\bibfield  {journal} {\bibinfo  {journal} {Low Temperature Physics}\ }\textbf {\bibinfo {volume} {42}},\ \bibinfo {pages} {971} (\bibinfo {year} {2016})}\BibitemShut {NoStop}%
\bibitem [{\citenamefont {Halimeh}\ \emph {et~al.}(2017)\citenamefont {Halimeh}, \citenamefont {Zauner-Stauber}, \citenamefont {McCulloch}, \citenamefont {de~Vega}, \citenamefont {Schollw\"ock},\ and\ \citenamefont {Kastner}}]{Halimeh:2017}%
  \BibitemOpen
  \bibfield  {author} {\bibinfo {author} {\bibfnamefont {J.~C.}\ \bibnamefont {Halimeh}}, \bibinfo {author} {\bibfnamefont {V.}~\bibnamefont {Zauner-Stauber}}, \bibinfo {author} {\bibfnamefont {I.~P.}\ \bibnamefont {McCulloch}}, \bibinfo {author} {\bibfnamefont {I.}~\bibnamefont {de~Vega}}, \bibinfo {author} {\bibfnamefont {U.}~\bibnamefont {Schollw\"ock}},\ and\ \bibinfo {author} {\bibfnamefont {M.}~\bibnamefont {Kastner}},\ }\bibfield  {title} {\bibinfo {title} {Prethermalization and persistent order in the absence of a thermal phase transition},\ }\href {https://doi.org/10.1103/PhysRevB.95.024302} {\bibfield  {journal} {\bibinfo  {journal} {Phys. Rev. B}\ }\textbf {\bibinfo {volume} {95}},\ \bibinfo {pages} {024302} (\bibinfo {year} {2017})}\BibitemShut {NoStop}%
\bibitem [{\citenamefont {\ifmmode \check{Z}\else \v{Z}\fi{}unkovi\ifmmode~\check{c}\else \v{c}\fi{}}\ \emph {et~al.}(2018)\citenamefont {\ifmmode \check{Z}\else \v{Z}\fi{}unkovi\ifmmode~\check{c}\else \v{c}\fi{}}, \citenamefont {Heyl}, \citenamefont {Knap},\ and\ \citenamefont {Silva}}]{Zunkovic:2018}%
  \BibitemOpen
  \bibfield  {author} {\bibinfo {author} {\bibfnamefont {B.}~\bibnamefont {\ifmmode \check{Z}\else \v{Z}\fi{}unkovi\ifmmode~\check{c}\else \v{c}\fi{}}}, \bibinfo {author} {\bibfnamefont {M.}~\bibnamefont {Heyl}}, \bibinfo {author} {\bibfnamefont {M.}~\bibnamefont {Knap}},\ and\ \bibinfo {author} {\bibfnamefont {A.}~\bibnamefont {Silva}},\ }\bibfield  {title} {\bibinfo {title} {Dynamical quantum phase transitions in spin chains with long-range interactions: Merging different concepts of nonequilibrium criticality},\ }\href {https://doi.org/10.1103/PhysRevLett.120.130601} {\bibfield  {journal} {\bibinfo  {journal} {Phys. Rev. Lett.}\ }\textbf {\bibinfo {volume} {120}},\ \bibinfo {pages} {130601} (\bibinfo {year} {2018})}\BibitemShut {NoStop}%
\bibitem [{\citenamefont {Karch}\ \emph {et~al.}(2025)\citenamefont {Karch}, \citenamefont {Bandyopadhyay}, \citenamefont {Sun}, \citenamefont {Impertro}, \citenamefont {Huh}, \citenamefont {Rodríguez}, \citenamefont {Wienand}, \citenamefont {Ketterle}, \citenamefont {Heyl}, \citenamefont {Polkovnikov}, \citenamefont {Bloch},\ and\ \citenamefont {Aidelsburger}}]{Karch:2025}%
  \BibitemOpen
  \bibfield  {author} {\bibinfo {author} {\bibfnamefont {S.}~\bibnamefont {Karch}}, \bibinfo {author} {\bibfnamefont {S.}~\bibnamefont {Bandyopadhyay}}, \bibinfo {author} {\bibfnamefont {Z.-H.}\ \bibnamefont {Sun}}, \bibinfo {author} {\bibfnamefont {A.}~\bibnamefont {Impertro}}, \bibinfo {author} {\bibfnamefont {S.}~\bibnamefont {Huh}}, \bibinfo {author} {\bibfnamefont {I.~P.}\ \bibnamefont {Rodríguez}}, \bibinfo {author} {\bibfnamefont {J.~F.}\ \bibnamefont {Wienand}}, \bibinfo {author} {\bibfnamefont {W.}~\bibnamefont {Ketterle}}, \bibinfo {author} {\bibfnamefont {M.}~\bibnamefont {Heyl}}, \bibinfo {author} {\bibfnamefont {A.}~\bibnamefont {Polkovnikov}}, \bibinfo {author} {\bibfnamefont {I.}~\bibnamefont {Bloch}},\ and\ \bibinfo {author} {\bibfnamefont {M.}~\bibnamefont {Aidelsburger}},\ }\href {https://arxiv.org/abs/2501.16995} {\bibinfo {title} {Probing quantum many-body dynamics using subsystem loschmidt echos}} (\bibinfo {year} {2025}),\ \Eprint {https://arxiv.org/abs/2501.16995} {arXiv:2501.16995
  [cond-mat.quant-gas]} \BibitemShut {NoStop}%
\bibitem [{\citenamefont {Hashizume}\ \emph {et~al.}(2025)\citenamefont {Hashizume}, \citenamefont {Herbort}, \citenamefont {Tindall},\ and\ \citenamefont {Jaksch}}]{Hashizume:2025}%
  \BibitemOpen
  \bibfield  {author} {\bibinfo {author} {\bibfnamefont {T.}~\bibnamefont {Hashizume}}, \bibinfo {author} {\bibfnamefont {F.}~\bibnamefont {Herbort}}, \bibinfo {author} {\bibfnamefont {J.}~\bibnamefont {Tindall}},\ and\ \bibinfo {author} {\bibfnamefont {D.}~\bibnamefont {Jaksch}},\ }\bibfield  {title} {\bibinfo {title} {Dynamical quantum phase transitions on random networks},\ }\href {https://doi.org/10.1088/1367-2630/ade07b} {\bibfield  {journal} {\bibinfo  {journal} {New Journal of Physics}\ }\textbf {\bibinfo {volume} {27}},\ \bibinfo {pages} {064506} (\bibinfo {year} {2025})}\BibitemShut {NoStop}%
\bibitem [{\citenamefont {Moeckel}\ and\ \citenamefont {Kehrein}(2008)}]{Moeckel:2008}%
  \BibitemOpen
  \bibfield  {author} {\bibinfo {author} {\bibfnamefont {M.}~\bibnamefont {Moeckel}}\ and\ \bibinfo {author} {\bibfnamefont {S.}~\bibnamefont {Kehrein}},\ }\bibfield  {title} {\bibinfo {title} {Interaction quench in the hubbard model},\ }\href {https://doi.org/10.1103/PhysRevLett.100.175702} {\bibfield  {journal} {\bibinfo  {journal} {Phys. Rev. Lett.}\ }\textbf {\bibinfo {volume} {100}},\ \bibinfo {pages} {175702} (\bibinfo {year} {2008})}\BibitemShut {NoStop}%
\bibitem [{\citenamefont {Berges}\ \emph {et~al.}(2004)\citenamefont {Berges}, \citenamefont {Bors\'anyi},\ and\ \citenamefont {Wetterich}}]{Berges:2004}%
  \BibitemOpen
  \bibfield  {author} {\bibinfo {author} {\bibfnamefont {J.}~\bibnamefont {Berges}}, \bibinfo {author} {\bibfnamefont {S.}~\bibnamefont {Bors\'anyi}},\ and\ \bibinfo {author} {\bibfnamefont {C.}~\bibnamefont {Wetterich}},\ }\bibfield  {title} {\bibinfo {title} {Prethermalization},\ }\href {https://doi.org/10.1103/PhysRevLett.93.142002} {\bibfield  {journal} {\bibinfo  {journal} {Phys. Rev. Lett.}\ }\textbf {\bibinfo {volume} {93}},\ \bibinfo {pages} {142002} (\bibinfo {year} {2004})}\BibitemShut {NoStop}%
\bibitem [{\citenamefont {Abanin}\ \emph {et~al.}(2017)\citenamefont {Abanin}, \citenamefont {De~Roeck}, \citenamefont {Ho},\ and\ \citenamefont {Huveneers}}]{Abanin:2017}%
  \BibitemOpen
  \bibfield  {author} {\bibinfo {author} {\bibfnamefont {D.}~\bibnamefont {Abanin}}, \bibinfo {author} {\bibfnamefont {W.}~\bibnamefont {De~Roeck}}, \bibinfo {author} {\bibfnamefont {W.~W.}\ \bibnamefont {Ho}},\ and\ \bibinfo {author} {\bibfnamefont {F.}~\bibnamefont {Huveneers}},\ }\bibfield  {title} {\bibinfo {title} {A rigorous theory of many-body prethermalization for periodically driven and closed quantum systems},\ }\href {https://doi.org/10.1007/s00220-017-2930-x} {\bibfield  {journal} {\bibinfo  {journal} {Communications in Mathematical Physics}\ }\textbf {\bibinfo {volume} {354}},\ \bibinfo {pages} {809} (\bibinfo {year} {2017})}\BibitemShut {NoStop}%
\bibitem [{\citenamefont {Mori}\ \emph {et~al.}(2018)\citenamefont {Mori}, \citenamefont {Ikeda}, \citenamefont {Kaminishi},\ and\ \citenamefont {Ueda}}]{Mori:2018}%
  \BibitemOpen
  \bibfield  {author} {\bibinfo {author} {\bibfnamefont {T.}~\bibnamefont {Mori}}, \bibinfo {author} {\bibfnamefont {T.~N.}\ \bibnamefont {Ikeda}}, \bibinfo {author} {\bibfnamefont {E.}~\bibnamefont {Kaminishi}},\ and\ \bibinfo {author} {\bibfnamefont {M.}~\bibnamefont {Ueda}},\ }\bibfield  {title} {\bibinfo {title} {Thermalization and prethermalization in isolated quantum systems: a theoretical overview},\ }\href {https://doi.org/10.1088/1361-6455/aabcdf} {\bibfield  {journal} {\bibinfo  {journal} {Journal of Physics B: Atomic, Molecular and Optical Physics}\ }\textbf {\bibinfo {volume} {51}},\ \bibinfo {pages} {112001} (\bibinfo {year} {2018})}\BibitemShut {NoStop}%
\bibitem [{\citenamefont {Haghshenas}\ \emph {et~al.}(2025)\citenamefont {Haghshenas}, \citenamefont {Chertkov}, \citenamefont {Mills}, \citenamefont {Kadow}, \citenamefont {Lin}, \citenamefont {Chen}, \citenamefont {Cade}, \citenamefont {Niesen}, \citenamefont {Begušić}, \citenamefont {Rudolph}, \citenamefont {Cirstoiu}, \citenamefont {Hemery}, \citenamefont {Keever}, \citenamefont {Lubasch}, \citenamefont {Granet}, \citenamefont {Baldwin}, \citenamefont {Bartolotta}, \citenamefont {Bohn}, \citenamefont {Cline}, \citenamefont {DeCross}, \citenamefont {Dreiling}, \citenamefont {Foltz}, \citenamefont {Francois}, \citenamefont {Gaebler}, \citenamefont {Gilbreth}, \citenamefont {Gray}, \citenamefont {Gresh}, \citenamefont {Hall}, \citenamefont {Hankin}, \citenamefont {Hansen}, \citenamefont {Hewitt}, \citenamefont {Hutson}, \citenamefont {Iqbal}, \citenamefont {Kotibhaskar}, \citenamefont {Lehman}, \citenamefont {Lucchetti}, \citenamefont {Madjarov}, \citenamefont {Mayer}, \citenamefont {Milne}, \citenamefont
  {Moses}, \citenamefont {Neyenhuis}, \citenamefont {Park}, \citenamefont {Ponsioen}, \citenamefont {Schecter}, \citenamefont {Siegfried}, \citenamefont {Stephen}, \citenamefont {Tiemann}, \citenamefont {Urmey}, \citenamefont {Walker}, \citenamefont {Potter}, \citenamefont {Hayes}, \citenamefont {Chan}, \citenamefont {Pollmann}, \citenamefont {Knap}, \citenamefont {Dreyer},\ and\ \citenamefont {Foss-Feig}}]{Haghshenas:2025}%
  \BibitemOpen
  \bibfield  {author} {\bibinfo {author} {\bibfnamefont {R.}~\bibnamefont {Haghshenas}}, \bibinfo {author} {\bibfnamefont {E.}~\bibnamefont {Chertkov}}, \bibinfo {author} {\bibfnamefont {M.}~\bibnamefont {Mills}}, \bibinfo {author} {\bibfnamefont {W.}~\bibnamefont {Kadow}}, \bibinfo {author} {\bibfnamefont {S.-H.}\ \bibnamefont {Lin}}, \bibinfo {author} {\bibfnamefont {Y.-H.}\ \bibnamefont {Chen}}, \bibinfo {author} {\bibfnamefont {C.}~\bibnamefont {Cade}}, \bibinfo {author} {\bibfnamefont {I.}~\bibnamefont {Niesen}}, \bibinfo {author} {\bibfnamefont {T.}~\bibnamefont {Begušić}}, \bibinfo {author} {\bibfnamefont {M.~S.}\ \bibnamefont {Rudolph}}, \bibinfo {author} {\bibfnamefont {C.}~\bibnamefont {Cirstoiu}}, \bibinfo {author} {\bibfnamefont {K.}~\bibnamefont {Hemery}}, \bibinfo {author} {\bibfnamefont {C.~M.}\ \bibnamefont {Keever}}, \bibinfo {author} {\bibfnamefont {M.}~\bibnamefont {Lubasch}}, \bibinfo {author} {\bibfnamefont {E.}~\bibnamefont {Granet}}, \bibinfo {author} {\bibfnamefont {C.~H.}\
  \bibnamefont {Baldwin}}, \bibinfo {author} {\bibfnamefont {J.~P.}\ \bibnamefont {Bartolotta}}, \bibinfo {author} {\bibfnamefont {M.}~\bibnamefont {Bohn}}, \bibinfo {author} {\bibfnamefont {J.}~\bibnamefont {Cline}}, \bibinfo {author} {\bibfnamefont {M.}~\bibnamefont {DeCross}}, \bibinfo {author} {\bibfnamefont {J.~M.}\ \bibnamefont {Dreiling}}, \bibinfo {author} {\bibfnamefont {C.}~\bibnamefont {Foltz}}, \bibinfo {author} {\bibfnamefont {D.}~\bibnamefont {Francois}}, \bibinfo {author} {\bibfnamefont {J.~P.}\ \bibnamefont {Gaebler}}, \bibinfo {author} {\bibfnamefont {C.~N.}\ \bibnamefont {Gilbreth}}, \bibinfo {author} {\bibfnamefont {J.}~\bibnamefont {Gray}}, \bibinfo {author} {\bibfnamefont {D.}~\bibnamefont {Gresh}}, \bibinfo {author} {\bibfnamefont {A.}~\bibnamefont {Hall}}, \bibinfo {author} {\bibfnamefont {A.}~\bibnamefont {Hankin}}, \bibinfo {author} {\bibfnamefont {A.}~\bibnamefont {Hansen}}, \bibinfo {author} {\bibfnamefont {N.}~\bibnamefont {Hewitt}}, \bibinfo {author} {\bibfnamefont {R.~B.}\
  \bibnamefont {Hutson}}, \bibinfo {author} {\bibfnamefont {M.}~\bibnamefont {Iqbal}}, \bibinfo {author} {\bibfnamefont {N.}~\bibnamefont {Kotibhaskar}}, \bibinfo {author} {\bibfnamefont {E.}~\bibnamefont {Lehman}}, \bibinfo {author} {\bibfnamefont {D.}~\bibnamefont {Lucchetti}}, \bibinfo {author} {\bibfnamefont {I.~S.}\ \bibnamefont {Madjarov}}, \bibinfo {author} {\bibfnamefont {K.}~\bibnamefont {Mayer}}, \bibinfo {author} {\bibfnamefont {A.~R.}\ \bibnamefont {Milne}}, \bibinfo {author} {\bibfnamefont {S.~A.}\ \bibnamefont {Moses}}, \bibinfo {author} {\bibfnamefont {B.}~\bibnamefont {Neyenhuis}}, \bibinfo {author} {\bibfnamefont {G.}~\bibnamefont {Park}}, \bibinfo {author} {\bibfnamefont {B.}~\bibnamefont {Ponsioen}}, \bibinfo {author} {\bibfnamefont {M.}~\bibnamefont {Schecter}}, \bibinfo {author} {\bibfnamefont {P.~E.}\ \bibnamefont {Siegfried}}, \bibinfo {author} {\bibfnamefont {D.~T.}\ \bibnamefont {Stephen}}, \bibinfo {author} {\bibfnamefont {B.~G.}\ \bibnamefont {Tiemann}}, \bibinfo {author}
  {\bibfnamefont {M.~D.}\ \bibnamefont {Urmey}}, \bibinfo {author} {\bibfnamefont {J.}~\bibnamefont {Walker}}, \bibinfo {author} {\bibfnamefont {A.~C.}\ \bibnamefont {Potter}}, \bibinfo {author} {\bibfnamefont {D.}~\bibnamefont {Hayes}}, \bibinfo {author} {\bibfnamefont {G.~K.-L.}\ \bibnamefont {Chan}}, \bibinfo {author} {\bibfnamefont {F.}~\bibnamefont {Pollmann}}, \bibinfo {author} {\bibfnamefont {M.}~\bibnamefont {Knap}}, \bibinfo {author} {\bibfnamefont {H.}~\bibnamefont {Dreyer}},\ and\ \bibinfo {author} {\bibfnamefont {M.}~\bibnamefont {Foss-Feig}},\ }\href {https://arxiv.org/abs/2503.20870} {\bibinfo {title} {Digital quantum magnetism at the frontier of classical simulations}} (\bibinfo {year} {2025}),\ \Eprint {https://arxiv.org/abs/2503.20870} {arXiv:2503.20870 [quant-ph]} \BibitemShut {NoStop}%
\bibitem [{\citenamefont {Wurtz}\ \emph {et~al.}(2023)\citenamefont {Wurtz}, \citenamefont {Bylinskii}, \citenamefont {Braverman}, \citenamefont {Amato-Grill}, \citenamefont {Cantu}, \citenamefont {Huber}, \citenamefont {Lukin}, \citenamefont {Liu}, \citenamefont {Weinberg}, \citenamefont {Long}, \citenamefont {Wang}, \citenamefont {Gemelke},\ and\ \citenamefont {Keesling}}]{aquila:2023}%
  \BibitemOpen
  \bibfield  {author} {\bibinfo {author} {\bibfnamefont {J.}~\bibnamefont {Wurtz}}, \bibinfo {author} {\bibfnamefont {A.}~\bibnamefont {Bylinskii}}, \bibinfo {author} {\bibfnamefont {B.}~\bibnamefont {Braverman}}, \bibinfo {author} {\bibfnamefont {J.}~\bibnamefont {Amato-Grill}}, \bibinfo {author} {\bibfnamefont {S.~H.}\ \bibnamefont {Cantu}}, \bibinfo {author} {\bibfnamefont {F.}~\bibnamefont {Huber}}, \bibinfo {author} {\bibfnamefont {A.}~\bibnamefont {Lukin}}, \bibinfo {author} {\bibfnamefont {F.}~\bibnamefont {Liu}}, \bibinfo {author} {\bibfnamefont {P.}~\bibnamefont {Weinberg}}, \bibinfo {author} {\bibfnamefont {J.}~\bibnamefont {Long}}, \bibinfo {author} {\bibfnamefont {S.-T.}\ \bibnamefont {Wang}}, \bibinfo {author} {\bibfnamefont {N.}~\bibnamefont {Gemelke}},\ and\ \bibinfo {author} {\bibfnamefont {A.}~\bibnamefont {Keesling}},\ }\href@noop {} {\bibinfo {title} {Aquila: Quera's 256-qubit neutral-atom quantum computer}} (\bibinfo {year} {2023}),\ \Eprint {https://arxiv.org/abs/2306.11727}
  {arXiv:2306.11727 [quant-ph]} \BibitemShut {NoStop}%
\bibitem [{\citenamefont {Bernien}\ \emph {et~al.}(2017)\citenamefont {Bernien}, \citenamefont {Schwartz}, \citenamefont {Keesling}, \citenamefont {Levine}, \citenamefont {Omran}, \citenamefont {Pichler}, \citenamefont {Choi}, \citenamefont {Zibrov}, \citenamefont {Endres}, \citenamefont {Greiner}, \citenamefont {Vuleti{\'{c}}},\ and\ \citenamefont {Lukin}}]{Bernien:2017}%
  \BibitemOpen
  \bibfield  {author} {\bibinfo {author} {\bibfnamefont {H.}~\bibnamefont {Bernien}}, \bibinfo {author} {\bibfnamefont {S.}~\bibnamefont {Schwartz}}, \bibinfo {author} {\bibfnamefont {A.}~\bibnamefont {Keesling}}, \bibinfo {author} {\bibfnamefont {H.}~\bibnamefont {Levine}}, \bibinfo {author} {\bibfnamefont {A.}~\bibnamefont {Omran}}, \bibinfo {author} {\bibfnamefont {H.}~\bibnamefont {Pichler}}, \bibinfo {author} {\bibfnamefont {S.}~\bibnamefont {Choi}}, \bibinfo {author} {\bibfnamefont {A.~S.}\ \bibnamefont {Zibrov}}, \bibinfo {author} {\bibfnamefont {M.}~\bibnamefont {Endres}}, \bibinfo {author} {\bibfnamefont {M.}~\bibnamefont {Greiner}}, \bibinfo {author} {\bibfnamefont {V.}~\bibnamefont {Vuleti{\'{c}}}},\ and\ \bibinfo {author} {\bibfnamefont {M.~D.}\ \bibnamefont {Lukin}},\ }\bibfield  {title} {\bibinfo {title} {Probing many-body dynamics on a 51-atom quantum simulator},\ }\href {https://doi.org/10.1038/nature24622} {\bibfield  {journal} {\bibinfo  {journal} {Nature}\ }\textbf {\bibinfo {volume}
  {551}},\ \bibinfo {pages} {579} (\bibinfo {year} {2017})}\BibitemShut {NoStop}%
\bibitem [{\citenamefont {Bluvstein}\ \emph {et~al.}(2021)\citenamefont {Bluvstein}, \citenamefont {Omran}, \citenamefont {Levine}, \citenamefont {Keesling}, \citenamefont {Semeghini}, \citenamefont {Ebadi}, \citenamefont {Wang}, \citenamefont {Michailidis}, \citenamefont {Maskara}, \citenamefont {Ho}, \citenamefont {Choi}, \citenamefont {Serbyn}, \citenamefont {Greiner}, \citenamefont {Vuletić},\ and\ \citenamefont {Lukin}}]{Bluvstein:2021}%
  \BibitemOpen
  \bibfield  {author} {\bibinfo {author} {\bibfnamefont {D.}~\bibnamefont {Bluvstein}}, \bibinfo {author} {\bibfnamefont {A.}~\bibnamefont {Omran}}, \bibinfo {author} {\bibfnamefont {H.}~\bibnamefont {Levine}}, \bibinfo {author} {\bibfnamefont {A.}~\bibnamefont {Keesling}}, \bibinfo {author} {\bibfnamefont {G.}~\bibnamefont {Semeghini}}, \bibinfo {author} {\bibfnamefont {S.}~\bibnamefont {Ebadi}}, \bibinfo {author} {\bibfnamefont {T.~T.}\ \bibnamefont {Wang}}, \bibinfo {author} {\bibfnamefont {A.~A.}\ \bibnamefont {Michailidis}}, \bibinfo {author} {\bibfnamefont {N.}~\bibnamefont {Maskara}}, \bibinfo {author} {\bibfnamefont {W.~W.}\ \bibnamefont {Ho}}, \bibinfo {author} {\bibfnamefont {S.}~\bibnamefont {Choi}}, \bibinfo {author} {\bibfnamefont {M.}~\bibnamefont {Serbyn}}, \bibinfo {author} {\bibfnamefont {M.}~\bibnamefont {Greiner}}, \bibinfo {author} {\bibfnamefont {V.}~\bibnamefont {Vuletić}},\ and\ \bibinfo {author} {\bibfnamefont {M.~D.}\ \bibnamefont {Lukin}},\ }\bibfield  {title} {\bibinfo {title}
  {Controlling quantum many-body dynamics in driven rydberg atom arrays},\ }\href {https://doi.org/10.1126/science.abg2530} {\bibfield  {journal} {\bibinfo  {journal} {Science}\ }\textbf {\bibinfo {volume} {371}},\ \bibinfo {pages} {1355} (\bibinfo {year} {2021})}\BibitemShut {NoStop}%
\bibitem [{\citenamefont {González-Cuadra}\ \emph {et~al.}(2025)\citenamefont {González-Cuadra}, \citenamefont {Hamdan}, \citenamefont {Zache}, \citenamefont {Braverman}, \citenamefont {Kornjača}, \citenamefont {Lukin}, \citenamefont {Cantú}, \citenamefont {Liu}, \citenamefont {Wang}, \citenamefont {Keesling}, \citenamefont {Lukin}, \citenamefont {Zoller},\ and\ \citenamefont {Bylinskii}}]{GonzalezCuadra:2024}%
  \BibitemOpen
  \bibfield  {author} {\bibinfo {author} {\bibfnamefont {D.}~\bibnamefont {González-Cuadra}}, \bibinfo {author} {\bibfnamefont {M.}~\bibnamefont {Hamdan}}, \bibinfo {author} {\bibfnamefont {T.~V.}\ \bibnamefont {Zache}}, \bibinfo {author} {\bibfnamefont {B.}~\bibnamefont {Braverman}}, \bibinfo {author} {\bibfnamefont {M.}~\bibnamefont {Kornjača}}, \bibinfo {author} {\bibfnamefont {A.}~\bibnamefont {Lukin}}, \bibinfo {author} {\bibfnamefont {S.~H.}\ \bibnamefont {Cantú}}, \bibinfo {author} {\bibfnamefont {F.}~\bibnamefont {Liu}}, \bibinfo {author} {\bibfnamefont {S.-T.}\ \bibnamefont {Wang}}, \bibinfo {author} {\bibfnamefont {A.}~\bibnamefont {Keesling}}, \bibinfo {author} {\bibfnamefont {M.~D.}\ \bibnamefont {Lukin}}, \bibinfo {author} {\bibfnamefont {P.}~\bibnamefont {Zoller}},\ and\ \bibinfo {author} {\bibfnamefont {A.}~\bibnamefont {Bylinskii}},\ }\bibfield  {title} {\bibinfo {title} {Observation of string breaking on a (2 + 1)d rydberg quantum simulator},\ }\href
  {https://doi.org/10.1038/s41586-025-09051-6} {\bibfield  {journal} {\bibinfo  {journal} {Nature}\ }\textbf {\bibinfo {volume} {642}},\ \bibinfo {pages} {321} (\bibinfo {year} {2025})}\BibitemShut {NoStop}%
\bibitem [{\citenamefont {Keesling}\ \emph {et~al.}(2019)\citenamefont {Keesling}, \citenamefont {Omran}, \citenamefont {Levine}, \citenamefont {Bernien}, \citenamefont {Pichler}, \citenamefont {Choi}, \citenamefont {Samajdar}, \citenamefont {Schwartz}, \citenamefont {Silvi}, \citenamefont {Sachdev}, \citenamefont {Zoller}, \citenamefont {Endres}, \citenamefont {Greiner}, \citenamefont {Vuleti{\'{c}}},\ and\ \citenamefont {Lukin}}]{Keesling:2019}%
  \BibitemOpen
  \bibfield  {author} {\bibinfo {author} {\bibfnamefont {A.}~\bibnamefont {Keesling}}, \bibinfo {author} {\bibfnamefont {A.}~\bibnamefont {Omran}}, \bibinfo {author} {\bibfnamefont {H.}~\bibnamefont {Levine}}, \bibinfo {author} {\bibfnamefont {H.}~\bibnamefont {Bernien}}, \bibinfo {author} {\bibfnamefont {H.}~\bibnamefont {Pichler}}, \bibinfo {author} {\bibfnamefont {S.}~\bibnamefont {Choi}}, \bibinfo {author} {\bibfnamefont {R.}~\bibnamefont {Samajdar}}, \bibinfo {author} {\bibfnamefont {S.}~\bibnamefont {Schwartz}}, \bibinfo {author} {\bibfnamefont {P.}~\bibnamefont {Silvi}}, \bibinfo {author} {\bibfnamefont {S.}~\bibnamefont {Sachdev}}, \bibinfo {author} {\bibfnamefont {P.}~\bibnamefont {Zoller}}, \bibinfo {author} {\bibfnamefont {M.}~\bibnamefont {Endres}}, \bibinfo {author} {\bibfnamefont {M.}~\bibnamefont {Greiner}}, \bibinfo {author} {\bibfnamefont {V.}~\bibnamefont {Vuleti{\'{c}}}},\ and\ \bibinfo {author} {\bibfnamefont {M.~D.}\ \bibnamefont {Lukin}},\ }\bibfield  {title} {\bibinfo {title} {Quantum
  kibble--zurek mechanism and critical dynamics on a programmable rydberg simulator},\ }\href {https://doi.org/10.1038/s41586-019-1070-1} {\bibfield  {journal} {\bibinfo  {journal} {Nature}\ }\textbf {\bibinfo {volume} {568}},\ \bibinfo {pages} {207} (\bibinfo {year} {2019})}\BibitemShut {NoStop}%
\bibitem [{\citenamefont {Daley}\ \emph {et~al.}(2022)\citenamefont {Daley}, \citenamefont {Bloch}, \citenamefont {Kokail}, \citenamefont {Flannigan}, \citenamefont {Pearson}, \citenamefont {Troyer},\ and\ \citenamefont {Zoller}}]{Daley:2022}%
  \BibitemOpen
  \bibfield  {author} {\bibinfo {author} {\bibfnamefont {A.~J.}\ \bibnamefont {Daley}}, \bibinfo {author} {\bibfnamefont {I.}~\bibnamefont {Bloch}}, \bibinfo {author} {\bibfnamefont {C.}~\bibnamefont {Kokail}}, \bibinfo {author} {\bibfnamefont {S.}~\bibnamefont {Flannigan}}, \bibinfo {author} {\bibfnamefont {N.}~\bibnamefont {Pearson}}, \bibinfo {author} {\bibfnamefont {M.}~\bibnamefont {Troyer}},\ and\ \bibinfo {author} {\bibfnamefont {P.}~\bibnamefont {Zoller}},\ }\bibfield  {title} {\bibinfo {title} {Practical quantum advantage in quantum simulation},\ }\href {https://doi.org/10.1038/s41586-022-04940-6} {\bibfield  {journal} {\bibinfo  {journal} {Nature}\ }\textbf {\bibinfo {volume} {607}},\ \bibinfo {pages} {667} (\bibinfo {year} {2022})}\BibitemShut {NoStop}%
\bibitem [{\citenamefont {Smale}\ \emph {et~al.}(2019)\citenamefont {Smale}, \citenamefont {He}, \citenamefont {Olsen}, \citenamefont {Jackson}, \citenamefont {Sharum}, \citenamefont {Trotzky}, \citenamefont {Marino}, \citenamefont {Rey},\ and\ \citenamefont {Thywissen}}]{Smale:2019}%
  \BibitemOpen
  \bibfield  {author} {\bibinfo {author} {\bibfnamefont {S.}~\bibnamefont {Smale}}, \bibinfo {author} {\bibfnamefont {P.}~\bibnamefont {He}}, \bibinfo {author} {\bibfnamefont {B.~A.}\ \bibnamefont {Olsen}}, \bibinfo {author} {\bibfnamefont {K.~G.}\ \bibnamefont {Jackson}}, \bibinfo {author} {\bibfnamefont {H.}~\bibnamefont {Sharum}}, \bibinfo {author} {\bibfnamefont {S.}~\bibnamefont {Trotzky}}, \bibinfo {author} {\bibfnamefont {J.}~\bibnamefont {Marino}}, \bibinfo {author} {\bibfnamefont {A.~M.}\ \bibnamefont {Rey}},\ and\ \bibinfo {author} {\bibfnamefont {J.~H.}\ \bibnamefont {Thywissen}},\ }\bibfield  {title} {\bibinfo {title} {Observation of a transition between dynamical phases in a quantum degenerate fermi gas},\ }\href {https://doi.org/10.1126/sciadv.aax1568} {\bibfield  {journal} {\bibinfo  {journal} {Science Advances}\ }\textbf {\bibinfo {volume} {5}},\ \bibinfo {pages} {eaax1568} (\bibinfo {year} {2019})}\BibitemShut {NoStop}%
\bibitem [{\citenamefont {Muniz}\ \emph {et~al.}(2020)\citenamefont {Muniz}, \citenamefont {Barberena}, \citenamefont {Lewis-Swan}, \citenamefont {Young}, \citenamefont {Cline}, \citenamefont {Rey},\ and\ \citenamefont {Thompson}}]{Muniz:2020}%
  \BibitemOpen
  \bibfield  {author} {\bibinfo {author} {\bibfnamefont {J.~A.}\ \bibnamefont {Muniz}}, \bibinfo {author} {\bibfnamefont {D.}~\bibnamefont {Barberena}}, \bibinfo {author} {\bibfnamefont {R.~J.}\ \bibnamefont {Lewis-Swan}}, \bibinfo {author} {\bibfnamefont {D.~J.}\ \bibnamefont {Young}}, \bibinfo {author} {\bibfnamefont {J.~R.~K.}\ \bibnamefont {Cline}}, \bibinfo {author} {\bibfnamefont {A.~M.}\ \bibnamefont {Rey}},\ and\ \bibinfo {author} {\bibfnamefont {J.~K.}\ \bibnamefont {Thompson}},\ }\bibfield  {title} {\bibinfo {title} {Exploring dynamical phase transitions with cold atoms in an optical  cavity},\ }\href {https://doi.org/10.1038/s41586-020-2224-x} {\bibfield  {journal} {\bibinfo  {journal} {Nature}\ }\textbf {\bibinfo {volume} {580}},\ \bibinfo {pages} {602} (\bibinfo {year} {2020})}\BibitemShut {NoStop}%
\bibitem [{\citenamefont {Chu}\ \emph {et~al.}(2020)\citenamefont {Chu}, \citenamefont {Will}, \citenamefont {Arlt}, \citenamefont {Klempt},\ and\ \citenamefont {Rey}}]{Chu:2020}%
  \BibitemOpen
  \bibfield  {author} {\bibinfo {author} {\bibfnamefont {A.}~\bibnamefont {Chu}}, \bibinfo {author} {\bibfnamefont {J.}~\bibnamefont {Will}}, \bibinfo {author} {\bibfnamefont {J.}~\bibnamefont {Arlt}}, \bibinfo {author} {\bibfnamefont {C.}~\bibnamefont {Klempt}},\ and\ \bibinfo {author} {\bibfnamefont {A.~M.}\ \bibnamefont {Rey}},\ }\bibfield  {title} {\bibinfo {title} {Simulation of $xxz$ spin models using sideband transitions in trapped bosonic gases},\ }\href {https://doi.org/10.1103/PhysRevLett.125.240504} {\bibfield  {journal} {\bibinfo  {journal} {Phys. Rev. Lett.}\ }\textbf {\bibinfo {volume} {125}},\ \bibinfo {pages} {240504} (\bibinfo {year} {2020})}\BibitemShut {NoStop}%
\bibitem [{\citenamefont {Zhang}\ \emph {et~al.}(2017)\citenamefont {Zhang}, \citenamefont {Pagano}, \citenamefont {Hess}, \citenamefont {Kyprianidis}, \citenamefont {Becker}, \citenamefont {Kaplan}, \citenamefont {Gorshkov}, \citenamefont {Gong},\ and\ \citenamefont {Monroe}}]{Zhang:2017}%
  \BibitemOpen
  \bibfield  {author} {\bibinfo {author} {\bibfnamefont {J.}~\bibnamefont {Zhang}}, \bibinfo {author} {\bibfnamefont {G.}~\bibnamefont {Pagano}}, \bibinfo {author} {\bibfnamefont {P.~W.}\ \bibnamefont {Hess}}, \bibinfo {author} {\bibfnamefont {A.}~\bibnamefont {Kyprianidis}}, \bibinfo {author} {\bibfnamefont {P.}~\bibnamefont {Becker}}, \bibinfo {author} {\bibfnamefont {H.}~\bibnamefont {Kaplan}}, \bibinfo {author} {\bibfnamefont {A.~V.}\ \bibnamefont {Gorshkov}}, \bibinfo {author} {\bibfnamefont {Z.-X.}\ \bibnamefont {Gong}},\ and\ \bibinfo {author} {\bibfnamefont {C.}~\bibnamefont {Monroe}},\ }\bibfield  {title} {\bibinfo {title} {Observation of a many-body dynamical phase transition with a 53-qubit quantum simulator},\ }\href {https://doi.org/10.1038/nature24654} {\bibfield  {journal} {\bibinfo  {journal} {Nature}\ }\textbf {\bibinfo {volume} {551}},\ \bibinfo {pages} {601} (\bibinfo {year} {2017})}\BibitemShut {NoStop}%
\bibitem [{\citenamefont {De}\ \emph {et~al.}(2025)\citenamefont {De}, \citenamefont {Cook}, \citenamefont {Ali}, \citenamefont {Collins}, \citenamefont {Morong}, \citenamefont {Paz}, \citenamefont {Titum}, \citenamefont {Pagano}, \citenamefont {Gorshkov}, \citenamefont {Maghrebi},\ and\ \citenamefont {Monroe}}]{De:2025}%
  \BibitemOpen
  \bibfield  {author} {\bibinfo {author} {\bibfnamefont {A.}~\bibnamefont {De}}, \bibinfo {author} {\bibfnamefont {P.}~\bibnamefont {Cook}}, \bibinfo {author} {\bibfnamefont {M.}~\bibnamefont {Ali}}, \bibinfo {author} {\bibfnamefont {K.}~\bibnamefont {Collins}}, \bibinfo {author} {\bibfnamefont {W.}~\bibnamefont {Morong}}, \bibinfo {author} {\bibfnamefont {D.}~\bibnamefont {Paz}}, \bibinfo {author} {\bibfnamefont {P.}~\bibnamefont {Titum}}, \bibinfo {author} {\bibfnamefont {G.}~\bibnamefont {Pagano}}, \bibinfo {author} {\bibfnamefont {A.~V.}\ \bibnamefont {Gorshkov}}, \bibinfo {author} {\bibfnamefont {M.}~\bibnamefont {Maghrebi}},\ and\ \bibinfo {author} {\bibfnamefont {C.}~\bibnamefont {Monroe}},\ }\bibfield  {title} {\bibinfo {title} {Non-equilibrium critical scaling and universality in a quantum simulator},\ }\href {https://doi.org/10.1038/s41467-025-63398-y} {\bibfield  {journal} {\bibinfo  {journal} {Nature Communications}\ }\textbf {\bibinfo {volume} {16}},\ \bibinfo {pages} {7939} (\bibinfo {year}
  {2025})}\BibitemShut {NoStop}%
\bibitem [{\citenamefont {Xu}\ \emph {et~al.}(2020)\citenamefont {Xu}, \citenamefont {Sun}, \citenamefont {Liu}, \citenamefont {Zhang}, \citenamefont {Li}, \citenamefont {Dong}, \citenamefont {Ren}, \citenamefont {Zhang}, \citenamefont {Nori}, \citenamefont {Zheng}, \citenamefont {Fan},\ and\ \citenamefont {Wang}}]{Xu:2020}%
  \BibitemOpen
  \bibfield  {author} {\bibinfo {author} {\bibfnamefont {K.}~\bibnamefont {Xu}}, \bibinfo {author} {\bibfnamefont {Z.-H.}\ \bibnamefont {Sun}}, \bibinfo {author} {\bibfnamefont {W.}~\bibnamefont {Liu}}, \bibinfo {author} {\bibfnamefont {Y.-R.}\ \bibnamefont {Zhang}}, \bibinfo {author} {\bibfnamefont {H.}~\bibnamefont {Li}}, \bibinfo {author} {\bibfnamefont {H.}~\bibnamefont {Dong}}, \bibinfo {author} {\bibfnamefont {W.}~\bibnamefont {Ren}}, \bibinfo {author} {\bibfnamefont {P.}~\bibnamefont {Zhang}}, \bibinfo {author} {\bibfnamefont {F.}~\bibnamefont {Nori}}, \bibinfo {author} {\bibfnamefont {D.}~\bibnamefont {Zheng}}, \bibinfo {author} {\bibfnamefont {H.}~\bibnamefont {Fan}},\ and\ \bibinfo {author} {\bibfnamefont {H.}~\bibnamefont {Wang}},\ }\bibfield  {title} {\bibinfo {title} {Probing dynamical phase transitions with a superconducting quantum simulator},\ }\href {https://doi.org/10.1126/sciadv.aba4935} {\bibfield  {journal} {\bibinfo  {journal} {Science Advances}\ }\textbf {\bibinfo {volume} {6}},\
  \bibinfo {pages} {eaba4935} (\bibinfo {year} {2020})}\BibitemShut {NoStop}%
\bibitem [{\citenamefont {Jaksch}\ \emph {et~al.}(2000)\citenamefont {Jaksch}, \citenamefont {Cirac}, \citenamefont {Zoller}, \citenamefont {Rolston}, \citenamefont {C\^ot\'e},\ and\ \citenamefont {Lukin}}]{Jaksch:2000}%
  \BibitemOpen
  \bibfield  {author} {\bibinfo {author} {\bibfnamefont {D.}~\bibnamefont {Jaksch}}, \bibinfo {author} {\bibfnamefont {J.~I.}\ \bibnamefont {Cirac}}, \bibinfo {author} {\bibfnamefont {P.}~\bibnamefont {Zoller}}, \bibinfo {author} {\bibfnamefont {S.~L.}\ \bibnamefont {Rolston}}, \bibinfo {author} {\bibfnamefont {R.}~\bibnamefont {C\^ot\'e}},\ and\ \bibinfo {author} {\bibfnamefont {M.~D.}\ \bibnamefont {Lukin}},\ }\bibfield  {title} {\bibinfo {title} {Fast quantum gates for neutral atoms},\ }\href {https://doi.org/10.1103/PhysRevLett.85.2208} {\bibfield  {journal} {\bibinfo  {journal} {Phys. Rev. Lett.}\ }\textbf {\bibinfo {volume} {85}},\ \bibinfo {pages} {2208} (\bibinfo {year} {2000})}\BibitemShut {NoStop}%
\bibitem [{\citenamefont {Lukin}\ \emph {et~al.}(2001)\citenamefont {Lukin}, \citenamefont {Fleischhauer}, \citenamefont {Cote}, \citenamefont {Duan}, \citenamefont {Jaksch}, \citenamefont {Cirac},\ and\ \citenamefont {Zoller}}]{Lukin:2001}%
  \BibitemOpen
  \bibfield  {author} {\bibinfo {author} {\bibfnamefont {M.~D.}\ \bibnamefont {Lukin}}, \bibinfo {author} {\bibfnamefont {M.}~\bibnamefont {Fleischhauer}}, \bibinfo {author} {\bibfnamefont {R.}~\bibnamefont {Cote}}, \bibinfo {author} {\bibfnamefont {L.~M.}\ \bibnamefont {Duan}}, \bibinfo {author} {\bibfnamefont {D.}~\bibnamefont {Jaksch}}, \bibinfo {author} {\bibfnamefont {J.~I.}\ \bibnamefont {Cirac}},\ and\ \bibinfo {author} {\bibfnamefont {P.}~\bibnamefont {Zoller}},\ }\bibfield  {title} {\bibinfo {title} {Dipole blockade and quantum information processing in mesoscopic atomic ensembles},\ }\href {https://doi.org/10.1103/PhysRevLett.87.037901} {\bibfield  {journal} {\bibinfo  {journal} {Phys. Rev. Lett.}\ }\textbf {\bibinfo {volume} {87}},\ \bibinfo {pages} {037901} (\bibinfo {year} {2001})}\BibitemShut {NoStop}%
\bibitem [{\citenamefont {Haldar}\ \emph {et~al.}(2021)\citenamefont {Haldar}, \citenamefont {Mallayya}, \citenamefont {Heyl}, \citenamefont {Pollmann}, \citenamefont {Rigol},\ and\ \citenamefont {Das}}]{Haldar:2021}%
  \BibitemOpen
  \bibfield  {author} {\bibinfo {author} {\bibfnamefont {A.}~\bibnamefont {Haldar}}, \bibinfo {author} {\bibfnamefont {K.}~\bibnamefont {Mallayya}}, \bibinfo {author} {\bibfnamefont {M.}~\bibnamefont {Heyl}}, \bibinfo {author} {\bibfnamefont {F.}~\bibnamefont {Pollmann}}, \bibinfo {author} {\bibfnamefont {M.}~\bibnamefont {Rigol}},\ and\ \bibinfo {author} {\bibfnamefont {A.}~\bibnamefont {Das}},\ }\bibfield  {title} {\bibinfo {title} {Signatures of quantum phase transitions after quenches in quantum chaotic one-dimensional systems},\ }\href {https://doi.org/10.1103/PhysRevX.11.031062} {\bibfield  {journal} {\bibinfo  {journal} {Phys. Rev. X}\ }\textbf {\bibinfo {volume} {11}},\ \bibinfo {pages} {031062} (\bibinfo {year} {2021})}\BibitemShut {NoStop}%
\bibitem [{\citenamefont {Surace}\ \emph {et~al.}(2020)\citenamefont {Surace}, \citenamefont {Mazza}, \citenamefont {Giudici}, \citenamefont {Lerose}, \citenamefont {Gambassi},\ and\ \citenamefont {Dalmonte}}]{Surace:2020}%
  \BibitemOpen
  \bibfield  {author} {\bibinfo {author} {\bibfnamefont {F.~M.}\ \bibnamefont {Surace}}, \bibinfo {author} {\bibfnamefont {P.~P.}\ \bibnamefont {Mazza}}, \bibinfo {author} {\bibfnamefont {G.}~\bibnamefont {Giudici}}, \bibinfo {author} {\bibfnamefont {A.}~\bibnamefont {Lerose}}, \bibinfo {author} {\bibfnamefont {A.}~\bibnamefont {Gambassi}},\ and\ \bibinfo {author} {\bibfnamefont {M.}~\bibnamefont {Dalmonte}},\ }\bibfield  {title} {\bibinfo {title} {Lattice gauge theories and string dynamics in rydberg atom quantum simulators},\ }\href {https://doi.org/10.1103/PhysRevX.10.021041} {\bibfield  {journal} {\bibinfo  {journal} {Phys. Rev. X}\ }\textbf {\bibinfo {volume} {10}},\ \bibinfo {pages} {021041} (\bibinfo {year} {2020})}\BibitemShut {NoStop}%
\bibitem [{\citenamefont {Turner}\ \emph {et~al.}(2018)\citenamefont {Turner}, \citenamefont {Michailidis}, \citenamefont {Abanin}, \citenamefont {Serbyn},\ and\ \citenamefont {Papi{\'{c}}}}]{Turner:2018}%
  \BibitemOpen
  \bibfield  {author} {\bibinfo {author} {\bibfnamefont {C.~J.}\ \bibnamefont {Turner}}, \bibinfo {author} {\bibfnamefont {A.~A.}\ \bibnamefont {Michailidis}}, \bibinfo {author} {\bibfnamefont {D.~A.}\ \bibnamefont {Abanin}}, \bibinfo {author} {\bibfnamefont {M.}~\bibnamefont {Serbyn}},\ and\ \bibinfo {author} {\bibfnamefont {Z.}~\bibnamefont {Papi{\'{c}}}},\ }\bibfield  {title} {\bibinfo {title} {Weak ergodicity breaking from quantum many-body scars},\ }\href {https://doi.org/10.1038/s41567-018-0137-5} {\bibfield  {journal} {\bibinfo  {journal} {Nature Physics}\ }\textbf {\bibinfo {volume} {14}},\ \bibinfo {pages} {745} (\bibinfo {year} {2018})}\BibitemShut {NoStop}%
\bibitem [{Mar(2025)}]{Mark:2025}%
  \BibitemOpen
  \href@noop {} {\bibinfo {title} {{Private communication with Daniel K. Mark, Soonwon Choi, and Manuel Endres. To appear concurrently.}}} (\bibinfo {year} {2025})\BibitemShut {NoStop}%
\bibitem [{\citenamefont {Stoner}\ and\ \citenamefont {Wohlfarth}(1948)}]{Stoner:1948}%
  \BibitemOpen
  \bibfield  {author} {\bibinfo {author} {\bibfnamefont {E.~C.}\ \bibnamefont {Stoner}}\ and\ \bibinfo {author} {\bibfnamefont {E.~P.}\ \bibnamefont {Wohlfarth}},\ }\bibfield  {title} {\bibinfo {title} {A mechanism of magnetic hysteresis in heterogeneous alloys},\ }\href {https://doi.org/10.1098/rsta.1948.0007} {\bibfield  {journal} {\bibinfo  {journal} {Philosophical Transactions of the Royal Society of London. Series A, Mathematical and Physical Sciences}\ }\textbf {\bibinfo {volume} {240}},\ \bibinfo {pages} {599} (\bibinfo {year} {1948})}\BibitemShut {NoStop}%
\bibitem [{\citenamefont {Liu}\ \emph {et~al.}(2022)\citenamefont {Liu}, \citenamefont {Yang}, \citenamefont {Bienias}, \citenamefont {Iadecola},\ and\ \citenamefont {Gorshkov}}]{Liu:2022}%
  \BibitemOpen
  \bibfield  {author} {\bibinfo {author} {\bibfnamefont {F.}~\bibnamefont {Liu}}, \bibinfo {author} {\bibfnamefont {Z.-C.}\ \bibnamefont {Yang}}, \bibinfo {author} {\bibfnamefont {P.}~\bibnamefont {Bienias}}, \bibinfo {author} {\bibfnamefont {T.}~\bibnamefont {Iadecola}},\ and\ \bibinfo {author} {\bibfnamefont {A.~V.}\ \bibnamefont {Gorshkov}},\ }\bibfield  {title} {\bibinfo {title} {Localization and criticality in antiblockaded two-dimensional rydberg atom arrays},\ }\href {https://doi.org/10.1103/PhysRevLett.128.013603} {\bibfield  {journal} {\bibinfo  {journal} {Phys. Rev. Lett.}\ }\textbf {\bibinfo {volume} {128}},\ \bibinfo {pages} {013603} (\bibinfo {year} {2022})}\BibitemShut {NoStop}%
\bibitem [{\citenamefont {Bluvstein}\ \emph {et~al.}(2022)\citenamefont {Bluvstein}, \citenamefont {Levine}, \citenamefont {Semeghini}, \citenamefont {Wang}, \citenamefont {Ebadi}, \citenamefont {Kalinowski}, \citenamefont {Keesling}, \citenamefont {Maskara}, \citenamefont {Pichler}, \citenamefont {Greiner}, \citenamefont {Vuleti{\'{c}}},\ and\ \citenamefont {Lukin}}]{Bluvstein:2022}%
  \BibitemOpen
  \bibfield  {author} {\bibinfo {author} {\bibfnamefont {D.}~\bibnamefont {Bluvstein}}, \bibinfo {author} {\bibfnamefont {H.}~\bibnamefont {Levine}}, \bibinfo {author} {\bibfnamefont {G.}~\bibnamefont {Semeghini}}, \bibinfo {author} {\bibfnamefont {T.~T.}\ \bibnamefont {Wang}}, \bibinfo {author} {\bibfnamefont {S.}~\bibnamefont {Ebadi}}, \bibinfo {author} {\bibfnamefont {M.}~\bibnamefont {Kalinowski}}, \bibinfo {author} {\bibfnamefont {A.}~\bibnamefont {Keesling}}, \bibinfo {author} {\bibfnamefont {N.}~\bibnamefont {Maskara}}, \bibinfo {author} {\bibfnamefont {H.}~\bibnamefont {Pichler}}, \bibinfo {author} {\bibfnamefont {M.}~\bibnamefont {Greiner}}, \bibinfo {author} {\bibfnamefont {V.}~\bibnamefont {Vuleti{\'{c}}}},\ and\ \bibinfo {author} {\bibfnamefont {M.~D.}\ \bibnamefont {Lukin}},\ }\bibfield  {title} {\bibinfo {title} {A quantum processor based on coherent transport of entangled atom arrays},\ }\href {https://doi.org/10.1038/s41586-022-04592-6} {\bibfield  {journal} {\bibinfo  {journal} {Nature}\
  }\textbf {\bibinfo {volume} {604}},\ \bibinfo {pages} {451} (\bibinfo {year} {2022})}\BibitemShut {NoStop}%
\bibitem [{\citenamefont {Swingle}\ \emph {et~al.}(2016)\citenamefont {Swingle}, \citenamefont {Bentsen}, \citenamefont {Schleier-Smith},\ and\ \citenamefont {Hayden}}]{Swingle:2016}%
  \BibitemOpen
  \bibfield  {author} {\bibinfo {author} {\bibfnamefont {B.}~\bibnamefont {Swingle}}, \bibinfo {author} {\bibfnamefont {G.}~\bibnamefont {Bentsen}}, \bibinfo {author} {\bibfnamefont {M.}~\bibnamefont {Schleier-Smith}},\ and\ \bibinfo {author} {\bibfnamefont {P.}~\bibnamefont {Hayden}},\ }\bibfield  {title} {\bibinfo {title} {Measuring the scrambling of quantum information},\ }\href {https://doi.org/10.1103/PhysRevA.94.040302} {\bibfield  {journal} {\bibinfo  {journal} {Phys. Rev. A}\ }\textbf {\bibinfo {volume} {94}},\ \bibinfo {pages} {040302} (\bibinfo {year} {2016})}\BibitemShut {NoStop}%
\bibitem [{\citenamefont {Daniel}\ \emph {et~al.}(2023)\citenamefont {Daniel}, \citenamefont {Hallam}, \citenamefont {Desaules}, \citenamefont {Hudomal}, \citenamefont {Su}, \citenamefont {Halimeh},\ and\ \citenamefont {Papi\ifmmode~\acute{c}\else \'{c}\fi{}}}]{Daniel:2023}%
  \BibitemOpen
  \bibfield  {author} {\bibinfo {author} {\bibfnamefont {A.}~\bibnamefont {Daniel}}, \bibinfo {author} {\bibfnamefont {A.}~\bibnamefont {Hallam}}, \bibinfo {author} {\bibfnamefont {J.-Y.}\ \bibnamefont {Desaules}}, \bibinfo {author} {\bibfnamefont {A.}~\bibnamefont {Hudomal}}, \bibinfo {author} {\bibfnamefont {G.-X.}\ \bibnamefont {Su}}, \bibinfo {author} {\bibfnamefont {J.~C.}\ \bibnamefont {Halimeh}},\ and\ \bibinfo {author} {\bibfnamefont {Z.}~\bibnamefont {Papi\ifmmode~\acute{c}\else \'{c}\fi{}}},\ }\bibfield  {title} {\bibinfo {title} {Bridging quantum criticality via many-body scarring},\ }\href {https://doi.org/10.1103/PhysRevB.107.235108} {\bibfield  {journal} {\bibinfo  {journal} {Phys. Rev. B}\ }\textbf {\bibinfo {volume} {107}},\ \bibinfo {pages} {235108} (\bibinfo {year} {2023})}\BibitemShut {NoStop}%
\bibitem [{\citenamefont {Hudomal}\ \emph {et~al.}(2022)\citenamefont {Hudomal}, \citenamefont {Desaules}, \citenamefont {Mukherjee}, \citenamefont {Su}, \citenamefont {Halimeh},\ and\ \citenamefont {Papi\ifmmode~\acute{c}\else \'{c}\fi{}}}]{Hudomal:2022}%
  \BibitemOpen
  \bibfield  {author} {\bibinfo {author} {\bibfnamefont {A.}~\bibnamefont {Hudomal}}, \bibinfo {author} {\bibfnamefont {J.-Y.}\ \bibnamefont {Desaules}}, \bibinfo {author} {\bibfnamefont {B.}~\bibnamefont {Mukherjee}}, \bibinfo {author} {\bibfnamefont {G.-X.}\ \bibnamefont {Su}}, \bibinfo {author} {\bibfnamefont {J.~C.}\ \bibnamefont {Halimeh}},\ and\ \bibinfo {author} {\bibfnamefont {Z.}~\bibnamefont {Papi\ifmmode~\acute{c}\else \'{c}\fi{}}},\ }\bibfield  {title} {\bibinfo {title} {Driving quantum many-body scars in the pxp model},\ }\href {https://doi.org/10.1103/PhysRevB.106.104302} {\bibfield  {journal} {\bibinfo  {journal} {Phys. Rev. B}\ }\textbf {\bibinfo {volume} {106}},\ \bibinfo {pages} {104302} (\bibinfo {year} {2022})}\BibitemShut {NoStop}%
\bibitem [{\citenamefont {Chertkov}\ \emph {et~al.}(2023)\citenamefont {Chertkov}, \citenamefont {Cheng}, \citenamefont {Potter}, \citenamefont {Gopalakrishnan}, \citenamefont {Gatterman}, \citenamefont {Gerber}, \citenamefont {Gilmore}, \citenamefont {Gresh}, \citenamefont {Hall}, \citenamefont {Hankin}, \citenamefont {Matheny}, \citenamefont {Mengle}, \citenamefont {Hayes}, \citenamefont {Neyenhuis}, \citenamefont {Stutz},\ and\ \citenamefont {Foss-Feig}}]{Chertkov:2023}%
  \BibitemOpen
  \bibfield  {author} {\bibinfo {author} {\bibfnamefont {E.}~\bibnamefont {Chertkov}}, \bibinfo {author} {\bibfnamefont {Z.}~\bibnamefont {Cheng}}, \bibinfo {author} {\bibfnamefont {A.~C.}\ \bibnamefont {Potter}}, \bibinfo {author} {\bibfnamefont {S.}~\bibnamefont {Gopalakrishnan}}, \bibinfo {author} {\bibfnamefont {T.~M.}\ \bibnamefont {Gatterman}}, \bibinfo {author} {\bibfnamefont {J.~A.}\ \bibnamefont {Gerber}}, \bibinfo {author} {\bibfnamefont {K.}~\bibnamefont {Gilmore}}, \bibinfo {author} {\bibfnamefont {D.}~\bibnamefont {Gresh}}, \bibinfo {author} {\bibfnamefont {A.}~\bibnamefont {Hall}}, \bibinfo {author} {\bibfnamefont {A.}~\bibnamefont {Hankin}}, \bibinfo {author} {\bibfnamefont {M.}~\bibnamefont {Matheny}}, \bibinfo {author} {\bibfnamefont {T.}~\bibnamefont {Mengle}}, \bibinfo {author} {\bibfnamefont {D.}~\bibnamefont {Hayes}}, \bibinfo {author} {\bibfnamefont {B.}~\bibnamefont {Neyenhuis}}, \bibinfo {author} {\bibfnamefont {R.}~\bibnamefont {Stutz}},\ and\ \bibinfo {author} {\bibfnamefont
  {M.}~\bibnamefont {Foss-Feig}},\ }\bibfield  {title} {\bibinfo {title} {Characterizing a non-equilibrium phase transition on a quantum computer},\ }\href {https://doi.org/10.1038/s41567-023-02199-w} {\bibfield  {journal} {\bibinfo  {journal} {Nature Physics}\ }\textbf {\bibinfo {volume} {19}},\ \bibinfo {pages} {1799} (\bibinfo {year} {2023})}\BibitemShut {NoStop}%
\bibitem [{\citenamefont {Iadecola}\ \emph {et~al.}(2025)\citenamefont {Iadecola}, \citenamefont {Wilson},\ and\ \citenamefont {Pixley}}]{Iadecola:2025}%
  \BibitemOpen
  \bibfield  {author} {\bibinfo {author} {\bibfnamefont {T.}~\bibnamefont {Iadecola}}, \bibinfo {author} {\bibfnamefont {J.~H.}\ \bibnamefont {Wilson}},\ and\ \bibinfo {author} {\bibfnamefont {J.}~\bibnamefont {Pixley}},\ }\bibfield  {title} {\bibinfo {title} {Concomitant entanglement and control criticality driven by collective measurements},\ }\href {https://doi.org/10.1103/PRXQuantum.6.010351} {\bibfield  {journal} {\bibinfo  {journal} {PRX Quantum}\ }\textbf {\bibinfo {volume} {6}},\ \bibinfo {pages} {010351} (\bibinfo {year} {2025})}\BibitemShut {NoStop}%
\bibitem [{Blo(2023)}]{BloqadeJulia:2023}%
  \BibitemOpen
  \href {https://github.com/QuEraComputing/Bloqade.jl/} {\bibinfo {title} {Bloqade.jl: {P}ackage for the quantum computation and quantum simulation based on the neutral-atom architecture.}} (\bibinfo {year} {2023})\BibitemShut {NoStop}%
\bibitem [{\citenamefont {Weinberg}\ \emph {et~al.}(2024)\citenamefont {Weinberg}, \citenamefont {Wu}, \citenamefont {Long},\ and\ \citenamefont {Luo}}]{bloqadepython:2024}%
  \BibitemOpen
  \bibfield  {author} {\bibinfo {author} {\bibfnamefont {P.}~\bibnamefont {Weinberg}}, \bibinfo {author} {\bibfnamefont {K.-H.}\ \bibnamefont {Wu}}, \bibinfo {author} {\bibfnamefont {J.}~\bibnamefont {Long}},\ and\ \bibinfo {author} {\bibfnamefont {X.-z.~R.}\ \bibnamefont {Luo}},\ }\href {https://doi.org/10.5281/zenodo.11114110} {\bibinfo {title} {Queracomputing/bloqade-python: v0.15.11}} (\bibinfo {year} {2024})\BibitemShut {NoStop}%
\bibitem [{per(2023)}]{perlmutter:2023}%
  \BibitemOpen
  \href {https://www.nersc.gov/systems/perlmutter} {\bibinfo {title} {Perlmutter.}} (\bibinfo {year} {2023})\BibitemShut {NoStop}%
\bibitem [{\citenamefont {Beugeling}\ \emph {et~al.}(2014)\citenamefont {Beugeling}, \citenamefont {Moessner},\ and\ \citenamefont {Haque}}]{Beugeling:2014}%
  \BibitemOpen
  \bibfield  {author} {\bibinfo {author} {\bibfnamefont {W.}~\bibnamefont {Beugeling}}, \bibinfo {author} {\bibfnamefont {R.}~\bibnamefont {Moessner}},\ and\ \bibinfo {author} {\bibfnamefont {M.}~\bibnamefont {Haque}},\ }\bibfield  {title} {\bibinfo {title} {Finite-size scaling of eigenstate thermalization},\ }\href {https://doi.org/10.1103/PhysRevE.89.042112} {\bibfield  {journal} {\bibinfo  {journal} {Phys. Rev. E}\ }\textbf {\bibinfo {volume} {89}},\ \bibinfo {pages} {042112} (\bibinfo {year} {2014})}\BibitemShut {NoStop}%
\bibitem [{\citenamefont {Deutsch}(1991)}]{Deutsch:1991}%
  \BibitemOpen
  \bibfield  {author} {\bibinfo {author} {\bibfnamefont {J.~M.}\ \bibnamefont {Deutsch}},\ }\bibfield  {title} {\bibinfo {title} {Quantum statistical mechanics in a closed system},\ }\href {https://doi.org/10.1103/PhysRevA.43.2046} {\bibfield  {journal} {\bibinfo  {journal} {Phys. Rev. A}\ }\textbf {\bibinfo {volume} {43}},\ \bibinfo {pages} {2046} (\bibinfo {year} {1991})}\BibitemShut {NoStop}%
\bibitem [{\citenamefont {Sakurai}(1994)}]{Sakurai:1994}%
  \BibitemOpen
  \bibfield  {author} {\bibinfo {author} {\bibfnamefont {J.~J.}\ \bibnamefont {Sakurai}},\ }\href@noop {} {\emph {\bibinfo {title} {Modern Quantum Mechanics}}}\ (\bibinfo  {publisher} {Addison Wesley},\ \bibinfo {year} {1994})\BibitemShut {NoStop}%
\bibitem [{\citenamefont {Darbha}\ \emph {et~al.}(2024{\natexlab{a}})\citenamefont {Darbha}, \citenamefont {Kornja\ifmmode~\check{c}\else \v{c}\fi{}a}, \citenamefont {Liu}, \citenamefont {Balewski}, \citenamefont {Hirsbrunner}, \citenamefont {Lopes}, \citenamefont {Wang}, \citenamefont {Van~Beeumen}, \citenamefont {Camps},\ and\ \citenamefont {Klymko}}]{Darbha:2024a}%
  \BibitemOpen
  \bibfield  {author} {\bibinfo {author} {\bibfnamefont {S.}~\bibnamefont {Darbha}}, \bibinfo {author} {\bibfnamefont {M.}~\bibnamefont {Kornja\ifmmode~\check{c}\else \v{c}\fi{}a}}, \bibinfo {author} {\bibfnamefont {F.}~\bibnamefont {Liu}}, \bibinfo {author} {\bibfnamefont {J.}~\bibnamefont {Balewski}}, \bibinfo {author} {\bibfnamefont {M.~R.}\ \bibnamefont {Hirsbrunner}}, \bibinfo {author} {\bibfnamefont {P.~L.~S.}\ \bibnamefont {Lopes}}, \bibinfo {author} {\bibfnamefont {S.-T.}\ \bibnamefont {Wang}}, \bibinfo {author} {\bibfnamefont {R.}~\bibnamefont {Van~Beeumen}}, \bibinfo {author} {\bibfnamefont {D.}~\bibnamefont {Camps}},\ and\ \bibinfo {author} {\bibfnamefont {K.}~\bibnamefont {Klymko}},\ }\bibfield  {title} {\bibinfo {title} {False vacuum decay and nucleation dynamics in neutral atom systems},\ }\href {https://doi.org/10.1103/PhysRevB.110.155103} {\bibfield  {journal} {\bibinfo  {journal} {Phys. Rev. B}\ }\textbf {\bibinfo {volume} {110}},\ \bibinfo {pages} {155103} (\bibinfo {year}
  {2024}{\natexlab{a}})}\BibitemShut {NoStop}%
\bibitem [{\citenamefont {Darbha}\ \emph {et~al.}(2024{\natexlab{b}})\citenamefont {Darbha}, \citenamefont {Kornja\ifmmode~\check{c}\else \v{c}\fi{}a}, \citenamefont {Liu}, \citenamefont {Balewski}, \citenamefont {Hirsbrunner}, \citenamefont {Lopes}, \citenamefont {Wang}, \citenamefont {Van~Beeumen}, \citenamefont {Klymko},\ and\ \citenamefont {Camps}}]{Darbha:2024b}%
  \BibitemOpen
  \bibfield  {author} {\bibinfo {author} {\bibfnamefont {S.}~\bibnamefont {Darbha}}, \bibinfo {author} {\bibfnamefont {M.}~\bibnamefont {Kornja\ifmmode~\check{c}\else \v{c}\fi{}a}}, \bibinfo {author} {\bibfnamefont {F.}~\bibnamefont {Liu}}, \bibinfo {author} {\bibfnamefont {J.}~\bibnamefont {Balewski}}, \bibinfo {author} {\bibfnamefont {M.~R.}\ \bibnamefont {Hirsbrunner}}, \bibinfo {author} {\bibfnamefont {P.~L.~S.}\ \bibnamefont {Lopes}}, \bibinfo {author} {\bibfnamefont {S.-T.}\ \bibnamefont {Wang}}, \bibinfo {author} {\bibfnamefont {R.}~\bibnamefont {Van~Beeumen}}, \bibinfo {author} {\bibfnamefont {K.}~\bibnamefont {Klymko}},\ and\ \bibinfo {author} {\bibfnamefont {D.}~\bibnamefont {Camps}},\ }\bibfield  {title} {\bibinfo {title} {Long-lived oscillations of metastable states in neutral atom systems},\ }\href {https://doi.org/10.1103/PhysRevB.110.155114} {\bibfield  {journal} {\bibinfo  {journal} {Phys. Rev. B}\ }\textbf {\bibinfo {volume} {110}},\ \bibinfo {pages} {155114} (\bibinfo {year}
  {2024}{\natexlab{b}})}\BibitemShut {NoStop}%
\bibitem [{\citenamefont {Manovitz}\ \emph {et~al.}(2025)\citenamefont {Manovitz}, \citenamefont {Li}, \citenamefont {Ebadi}, \citenamefont {Samajdar}, \citenamefont {Geim}, \citenamefont {Evered}, \citenamefont {Bluvstein}, \citenamefont {Zhou}, \citenamefont {Koyluoglu}, \citenamefont {Feldmeier}, \citenamefont {Dolgirev}, \citenamefont {Maskara}, \citenamefont {Kalinowski}, \citenamefont {Sachdev}, \citenamefont {Huse}, \citenamefont {Greiner}, \citenamefont {Vuleti{\'{c}}},\ and\ \citenamefont {Lukin}}]{Manovitz:2025}%
  \BibitemOpen
  \bibfield  {author} {\bibinfo {author} {\bibfnamefont {T.}~\bibnamefont {Manovitz}}, \bibinfo {author} {\bibfnamefont {S.~H.}\ \bibnamefont {Li}}, \bibinfo {author} {\bibfnamefont {S.}~\bibnamefont {Ebadi}}, \bibinfo {author} {\bibfnamefont {R.}~\bibnamefont {Samajdar}}, \bibinfo {author} {\bibfnamefont {A.~A.}\ \bibnamefont {Geim}}, \bibinfo {author} {\bibfnamefont {S.~J.}\ \bibnamefont {Evered}}, \bibinfo {author} {\bibfnamefont {D.}~\bibnamefont {Bluvstein}}, \bibinfo {author} {\bibfnamefont {H.}~\bibnamefont {Zhou}}, \bibinfo {author} {\bibfnamefont {N.~U.}\ \bibnamefont {Koyluoglu}}, \bibinfo {author} {\bibfnamefont {J.}~\bibnamefont {Feldmeier}}, \bibinfo {author} {\bibfnamefont {P.~E.}\ \bibnamefont {Dolgirev}}, \bibinfo {author} {\bibfnamefont {N.}~\bibnamefont {Maskara}}, \bibinfo {author} {\bibfnamefont {M.}~\bibnamefont {Kalinowski}}, \bibinfo {author} {\bibfnamefont {S.}~\bibnamefont {Sachdev}}, \bibinfo {author} {\bibfnamefont {D.~A.}\ \bibnamefont {Huse}}, \bibinfo {author} {\bibfnamefont
  {M.}~\bibnamefont {Greiner}}, \bibinfo {author} {\bibfnamefont {V.}~\bibnamefont {Vuleti{\'{c}}}},\ and\ \bibinfo {author} {\bibfnamefont {M.~D.}\ \bibnamefont {Lukin}},\ }\bibfield  {title} {\bibinfo {title} {Quantum coarsening and collective dynamics on a programmable simulator},\ }\href {https://doi.org/10.1038/s41586-024-08353-5} {\bibfield  {journal} {\bibinfo  {journal} {Nature}\ }\textbf {\bibinfo {volume} {638}},\ \bibinfo {pages} {86} (\bibinfo {year} {2025})}\BibitemShut {NoStop}%
\bibitem [{\citenamefont {Mattioli}\ \emph {et~al.}(2015)\citenamefont {Mattioli}, \citenamefont {Glätzle},\ and\ \citenamefont {Lechner}}]{Mattioli:2015}%
  \BibitemOpen
  \bibfield  {author} {\bibinfo {author} {\bibfnamefont {M.}~\bibnamefont {Mattioli}}, \bibinfo {author} {\bibfnamefont {A.~W.}\ \bibnamefont {Glätzle}},\ and\ \bibinfo {author} {\bibfnamefont {W.}~\bibnamefont {Lechner}},\ }\bibfield  {title} {\bibinfo {title} {From classical to quantum non-equilibrium dynamics of rydberg excitations in optical lattices},\ }\href {https://doi.org/10.1088/1367-2630/17/11/113039} {\bibfield  {journal} {\bibinfo  {journal} {New Journal of Physics}\ }\textbf {\bibinfo {volume} {17}},\ \bibinfo {pages} {113039} (\bibinfo {year} {2015})}\BibitemShut {NoStop}%
\bibitem [{\citenamefont {Marcuzzi}\ \emph {et~al.}(2017)\citenamefont {Marcuzzi}, \citenamefont {Min\'a\ifmmode~\check{r}\else \v{r}\fi{}}, \citenamefont {Barredo}, \citenamefont {de~L\'es\'eleuc}, \citenamefont {Labuhn}, \citenamefont {Lahaye}, \citenamefont {Browaeys}, \citenamefont {Levi},\ and\ \citenamefont {Lesanovsky}}]{Marcuzzi:2017}%
  \BibitemOpen
  \bibfield  {author} {\bibinfo {author} {\bibfnamefont {M.}~\bibnamefont {Marcuzzi}}, \bibinfo {author} {\bibfnamefont {J.~c.~v.}\ \bibnamefont {Min\'a\ifmmode~\check{r}\else \v{r}\fi{}}}, \bibinfo {author} {\bibfnamefont {D.}~\bibnamefont {Barredo}}, \bibinfo {author} {\bibfnamefont {S.}~\bibnamefont {de~L\'es\'eleuc}}, \bibinfo {author} {\bibfnamefont {H.}~\bibnamefont {Labuhn}}, \bibinfo {author} {\bibfnamefont {T.}~\bibnamefont {Lahaye}}, \bibinfo {author} {\bibfnamefont {A.}~\bibnamefont {Browaeys}}, \bibinfo {author} {\bibfnamefont {E.}~\bibnamefont {Levi}},\ and\ \bibinfo {author} {\bibfnamefont {I.}~\bibnamefont {Lesanovsky}},\ }\bibfield  {title} {\bibinfo {title} {Facilitation dynamics and localization phenomena in rydberg lattice gases with position disorder},\ }\href {https://doi.org/10.1103/PhysRevLett.118.063606} {\bibfield  {journal} {\bibinfo  {journal} {Phys. Rev. Lett.}\ }\textbf {\bibinfo {volume} {118}},\ \bibinfo {pages} {063606} (\bibinfo {year} {2017})}\BibitemShut {NoStop}%
\bibitem [{\citenamefont {Nandkishore}\ and\ \citenamefont {Hermele}(2019)}]{Nandkishore:2019}%
  \BibitemOpen
  \bibfield  {author} {\bibinfo {author} {\bibfnamefont {R.~M.}\ \bibnamefont {Nandkishore}}\ and\ \bibinfo {author} {\bibfnamefont {M.}~\bibnamefont {Hermele}},\ }\bibfield  {title} {\bibinfo {title} {Fractons},\ }\href {https://doi.org/https://doi.org/10.1146/annurev-conmatphys-031218-013604} {\bibfield  {journal} {\bibinfo  {journal} {Annual Review of Condensed Matter Physics}\ }\textbf {\bibinfo {volume} {10}},\ \bibinfo {pages} {295} (\bibinfo {year} {2019})}\BibitemShut {NoStop}%
\bibitem [{\citenamefont {Chamon}(2005)}]{Chamon:2005}%
  \BibitemOpen
  \bibfield  {author} {\bibinfo {author} {\bibfnamefont {C.}~\bibnamefont {Chamon}},\ }\bibfield  {title} {\bibinfo {title} {Quantum glassiness in strongly correlated clean systems: An example of topological overprotection},\ }\href {https://doi.org/10.1103/PhysRevLett.94.040402} {\bibfield  {journal} {\bibinfo  {journal} {Phys. Rev. Lett.}\ }\textbf {\bibinfo {volume} {94}},\ \bibinfo {pages} {040402} (\bibinfo {year} {2005})}\BibitemShut {NoStop}%
\bibitem [{\citenamefont {Hirsbrunner}\ \emph {et~al.}(2025)\citenamefont {Hirsbrunner}, \citenamefont {Kornjača}, \citenamefont {Samajdar}, \citenamefont {Darbha}, \citenamefont {Hamdan}, \citenamefont {Balewski}, \citenamefont {Rrapaj}, \citenamefont {Wang}, \citenamefont {Camps}, \citenamefont {Liu}, \citenamefont {Lopes},\ and\ \citenamefont {Klymko}}]{Hirsbrunner:2025}%
  \BibitemOpen
  \bibfield  {author} {\bibinfo {author} {\bibfnamefont {M.~R.}\ \bibnamefont {Hirsbrunner}}, \bibinfo {author} {\bibfnamefont {M.}~\bibnamefont {Kornjača}}, \bibinfo {author} {\bibfnamefont {R.}~\bibnamefont {Samajdar}}, \bibinfo {author} {\bibfnamefont {S.}~\bibnamefont {Darbha}}, \bibinfo {author} {\bibfnamefont {M.}~\bibnamefont {Hamdan}}, \bibinfo {author} {\bibfnamefont {J.}~\bibnamefont {Balewski}}, \bibinfo {author} {\bibfnamefont {E.}~\bibnamefont {Rrapaj}}, \bibinfo {author} {\bibfnamefont {S.-T.}\ \bibnamefont {Wang}}, \bibinfo {author} {\bibfnamefont {D.}~\bibnamefont {Camps}}, \bibinfo {author} {\bibfnamefont {F.}~\bibnamefont {Liu}}, \bibinfo {author} {\bibfnamefont {P.~L.~S.}\ \bibnamefont {Lopes}},\ and\ \bibinfo {author} {\bibfnamefont {K.}~\bibnamefont {Klymko}},\ }\href {https://arxiv.org/abs/2508.05737} {\bibinfo {title} {Quantum criticality and nonequilibrium dynamics on a lieb lattice of rydberg atoms}} (\bibinfo {year} {2025}),\ \Eprint {https://arxiv.org/abs/2508.05737}
  {arXiv:2508.05737 [cond-mat.quant-gas]} \BibitemShut {NoStop}%
\bibitem [{\citenamefont {Castelnovo}\ \emph {et~al.}(2012)\citenamefont {Castelnovo}, \citenamefont {Moessner},\ and\ \citenamefont {Sondhi}}]{Castelnovo:2012}%
  \BibitemOpen
  \bibfield  {author} {\bibinfo {author} {\bibfnamefont {C.}~\bibnamefont {Castelnovo}}, \bibinfo {author} {\bibfnamefont {R.}~\bibnamefont {Moessner}},\ and\ \bibinfo {author} {\bibfnamefont {S.}~\bibnamefont {Sondhi}},\ }\bibfield  {title} {\bibinfo {title} {Spin ice, fractionalization, and topological order},\ }\href {https://doi.org/https://doi.org/10.1146/annurev-conmatphys-020911-125058} {\bibfield  {journal} {\bibinfo  {journal} {Annual Review of Condensed Matter Physics}\ }\textbf {\bibinfo {volume} {3}},\ \bibinfo {pages} {35} (\bibinfo {year} {2012})}\BibitemShut {NoStop}%
\bibitem [{\citenamefont {Kornjača}\ \emph {et~al.}(2024)\citenamefont {Kornjača}, \citenamefont {Hu}, \citenamefont {Zhao}, \citenamefont {Wurtz}, \citenamefont {Weinberg}, \citenamefont {Hamdan}, \citenamefont {Zhdanov}, \citenamefont {Cantu}, \citenamefont {Zhou}, \citenamefont {Bravo}, \citenamefont {Bagnall}, \citenamefont {Basham}, \citenamefont {Campo}, \citenamefont {Choukri}, \citenamefont {DeAngelo}, \citenamefont {Frederick}, \citenamefont {Haines}, \citenamefont {Hammett}, \citenamefont {Hsu}, \citenamefont {Hu}, \citenamefont {Huber}, \citenamefont {Jepsen}, \citenamefont {Jia}, \citenamefont {Karolyshyn}, \citenamefont {Kwon}, \citenamefont {Long}, \citenamefont {Lopatin}, \citenamefont {Lukin}, \citenamefont {Macrì}, \citenamefont {Marković}, \citenamefont {Martínez-Martínez}, \citenamefont {Meng}, \citenamefont {Ostroumov}, \citenamefont {Paquette}, \citenamefont {Robinson}, \citenamefont {Rodriguez}, \citenamefont {Singh}, \citenamefont {Sinha}, \citenamefont {Thoreen}, \citenamefont
  {Wan}, \citenamefont {Waxman-Lenz}, \citenamefont {Wong}, \citenamefont {Wu}, \citenamefont {Lopes}, \citenamefont {Boger}, \citenamefont {Gemelke}, \citenamefont {Kitagawa}, \citenamefont {Keesling}, \citenamefont {Gao}, \citenamefont {Bylinskii}, \citenamefont {Yelin}, \citenamefont {Liu},\ and\ \citenamefont {Wang}}]{Kornjaca:2024}%
  \BibitemOpen
  \bibfield  {author} {\bibinfo {author} {\bibfnamefont {M.}~\bibnamefont {Kornjača}}, \bibinfo {author} {\bibfnamefont {H.-Y.}\ \bibnamefont {Hu}}, \bibinfo {author} {\bibfnamefont {C.}~\bibnamefont {Zhao}}, \bibinfo {author} {\bibfnamefont {J.}~\bibnamefont {Wurtz}}, \bibinfo {author} {\bibfnamefont {P.}~\bibnamefont {Weinberg}}, \bibinfo {author} {\bibfnamefont {M.}~\bibnamefont {Hamdan}}, \bibinfo {author} {\bibfnamefont {A.}~\bibnamefont {Zhdanov}}, \bibinfo {author} {\bibfnamefont {S.~H.}\ \bibnamefont {Cantu}}, \bibinfo {author} {\bibfnamefont {H.}~\bibnamefont {Zhou}}, \bibinfo {author} {\bibfnamefont {R.~A.}\ \bibnamefont {Bravo}}, \bibinfo {author} {\bibfnamefont {K.}~\bibnamefont {Bagnall}}, \bibinfo {author} {\bibfnamefont {J.~I.}\ \bibnamefont {Basham}}, \bibinfo {author} {\bibfnamefont {J.}~\bibnamefont {Campo}}, \bibinfo {author} {\bibfnamefont {A.}~\bibnamefont {Choukri}}, \bibinfo {author} {\bibfnamefont {R.}~\bibnamefont {DeAngelo}}, \bibinfo {author} {\bibfnamefont {P.}~\bibnamefont
  {Frederick}}, \bibinfo {author} {\bibfnamefont {D.}~\bibnamefont {Haines}}, \bibinfo {author} {\bibfnamefont {J.}~\bibnamefont {Hammett}}, \bibinfo {author} {\bibfnamefont {N.}~\bibnamefont {Hsu}}, \bibinfo {author} {\bibfnamefont {M.-G.}\ \bibnamefont {Hu}}, \bibinfo {author} {\bibfnamefont {F.}~\bibnamefont {Huber}}, \bibinfo {author} {\bibfnamefont {P.~N.}\ \bibnamefont {Jepsen}}, \bibinfo {author} {\bibfnamefont {N.}~\bibnamefont {Jia}}, \bibinfo {author} {\bibfnamefont {T.}~\bibnamefont {Karolyshyn}}, \bibinfo {author} {\bibfnamefont {M.}~\bibnamefont {Kwon}}, \bibinfo {author} {\bibfnamefont {J.}~\bibnamefont {Long}}, \bibinfo {author} {\bibfnamefont {J.}~\bibnamefont {Lopatin}}, \bibinfo {author} {\bibfnamefont {A.}~\bibnamefont {Lukin}}, \bibinfo {author} {\bibfnamefont {T.}~\bibnamefont {Macrì}}, \bibinfo {author} {\bibfnamefont {O.}~\bibnamefont {Marković}}, \bibinfo {author} {\bibfnamefont {L.~A.}\ \bibnamefont {Martínez-Martínez}}, \bibinfo {author} {\bibfnamefont {X.}~\bibnamefont {Meng}},
  \bibinfo {author} {\bibfnamefont {E.}~\bibnamefont {Ostroumov}}, \bibinfo {author} {\bibfnamefont {D.}~\bibnamefont {Paquette}}, \bibinfo {author} {\bibfnamefont {J.}~\bibnamefont {Robinson}}, \bibinfo {author} {\bibfnamefont {P.~S.}\ \bibnamefont {Rodriguez}}, \bibinfo {author} {\bibfnamefont {A.}~\bibnamefont {Singh}}, \bibinfo {author} {\bibfnamefont {N.}~\bibnamefont {Sinha}}, \bibinfo {author} {\bibfnamefont {H.}~\bibnamefont {Thoreen}}, \bibinfo {author} {\bibfnamefont {N.}~\bibnamefont {Wan}}, \bibinfo {author} {\bibfnamefont {D.}~\bibnamefont {Waxman-Lenz}}, \bibinfo {author} {\bibfnamefont {T.}~\bibnamefont {Wong}}, \bibinfo {author} {\bibfnamefont {K.-H.}\ \bibnamefont {Wu}}, \bibinfo {author} {\bibfnamefont {P.~L.~S.}\ \bibnamefont {Lopes}}, \bibinfo {author} {\bibfnamefont {Y.}~\bibnamefont {Boger}}, \bibinfo {author} {\bibfnamefont {N.}~\bibnamefont {Gemelke}}, \bibinfo {author} {\bibfnamefont {T.}~\bibnamefont {Kitagawa}}, \bibinfo {author} {\bibfnamefont {A.}~\bibnamefont {Keesling}},
  \bibinfo {author} {\bibfnamefont {X.}~\bibnamefont {Gao}}, \bibinfo {author} {\bibfnamefont {A.}~\bibnamefont {Bylinskii}}, \bibinfo {author} {\bibfnamefont {S.~F.}\ \bibnamefont {Yelin}}, \bibinfo {author} {\bibfnamefont {F.}~\bibnamefont {Liu}},\ and\ \bibinfo {author} {\bibfnamefont {S.-T.}\ \bibnamefont {Wang}},\ }\href {https://arxiv.org/abs/2407.02553} {\bibinfo {title} {Large-scale quantum reservoir learning with an analog quantum computer}} (\bibinfo {year} {2024}),\ \Eprint {https://arxiv.org/abs/2407.02553} {arXiv:2407.02553 [quant-ph]} \BibitemShut {NoStop}%
\end{thebibliography}
\end{document}